\documentclass[12pt]{article}
\pdfoutput=1
\usepackage{cite,amsmath,amssymb,mathrsfs,collref,graphicx,epstopdf,hyperref,feynmp,ifpdf}

\graphicspath{{./}
              {./Figures/}
              {./Figs/}
              }

\ifpdf
\fi

\makeatletter
\def\endfmffile{%
  \fmfcmd{\p@rcent\space the end.^^J%
          end.^^J%
          endinput;}%
  \if@fmfio
    \immediate\closeout\@outfmf
  \fi
  \IfFileExists{\thefmffile.mp}{\immediate\write18{mpost \thefmffile}}{}
  \let\thefmffile\relax
}
\makeatother

\newcommand{\lag}{\mathscr{L}}
\newcommand{\dd}{\text{d}}
\newcommand{\abs}[1]{\left\lvert #1 \right\rvert}

\newcommand{\hx}{\ensuremath{h_1^x}}
\newcommand{\epsY}{\ensuremath{\epsilon_Y}}

\newcommand{\ccdot}{\!\cdot\!}

\newcommand{\gev}{\text{GeV}}
\newcommand{\mev}{\text{MeV}}
\newcommand{\beq}{\begin{equation}}
\newcommand{\eeq}{\end{equation}}
\newcommand{\bea}{\begin{eqnarray}}
\newcommand{\eea}{\end{eqnarray}}
\newcommand{\nnmb}{\nonumber}
\newcommand{\lrf}[2]{\left(\frac{#1}{#2}\right)}
\newcommand{\met}{/\hspace{-0.3cm}E_T}
\newcommand{\modeqref}[1]{Eq.~\eqref{#1}}
\newcommand{\twoeqref}[2]{Eqs.~\eqref{#1} and~\eqref{#2}}
\newcommand{\threeqref}[3]{Eqs.~\eqref{#1}, \eqref{#2} and~\eqref{#3}}
\newcommand{\foureqref}[4]{Eqs.~\eqref{#1}, \eqref{#2}, \eqref{#3} and~\eqref{#4}}
\newcommand{\figref}[1]{Fig.~\ref{#1}}
\newcommand{\figrefs}[2]{Figs.~\ref{#1} and~\ref{#2}}
\newcommand{\figfigrefs}[3]{Figs.~\ref{#1}, \ref{#2} and~\ref{#3}}

\newcommand{\thrufigrefs}[2]{Figs.~\ref{#1}--\ref{#2}}
\newcommand{\tabref}[1]{Table~\ref{#1}}
\newcommand{\tabrefs}[2]{Tables~\ref{#1} and~\ref{#2}}
\newcommand{\refcite}[1]{Ref.~\cite{#1}}
\newcommand{\refscite}[1]{Refs.~\cite{#1}}
\newcommand{\secref}[1]{Section~\ref{#1}}
\newcommand{\secrefs}[2]{Sections~\ref{#1} and~\ref{#2}}
\newcommand{\appref}[1]{Appendix~\ref{#1}}

\newcommand{\threeapprefs}[3]{Appendices~\ref{#1}, \ref{#2} and~\ref{#3}}
\newcommand{\order}[1]{\mathcal{O}(#1)}

\begin{document}

\begin{titlepage}
\noindent
\vspace{1cm}
\begin{center}
  \begin{Large}
    \begin{bf}
New Limits on Light Hidden Sectors\vspace{0.15cm}\\from Fixed-Target Experiments
     \end{bf}
  \end{Large}
\end{center}
\vspace{0.2cm}
\begin{center}
\begin{large}
David E. Morrissey$^{(a)}$ and Andrew P. Spray$^{(b)}$
\end{large}
\vspace{1cm}\\
\begin{it}
(a) TRIUMF, 4004 Wesbrook Mall, Vancouver, BC V6T 2A3, Canada\vspace{0.2cm}\\
(b) ARC Centre of Excellence for Particle Physics at the Terascale,\\ 
School of Physics, University of Melbourne, Victoria 3010, Australia
\vspace{0.5cm}\\
email: \emph{\texttt{dmorri@triumf.ca}}, \emph{\texttt{andrew.spray@coepp.org.au}}
\vspace{0.2cm}
\end{it}
\end{center}
\center{\today}

\begin{abstract}

 New physics can be light if it is hidden, coupling very weakly
to the Standard Model.  In this work we investigate the discovery
prospects of Abelian hidden sectors in lower-energy
fixed-target and high-precision experiments.  We focus on a minimal
supersymmetric realization consisting of an Abelian vector multiplet, 
coupled to hypercharge by kinetic mixing, and a pair of chiral Higgs multiplets.
This simple theory can give rise to a broad range of experimental 
signals, 
including both commonly-studied patterns of hidden vector decay 
as well as new and distinctive hidden sector cascades.  
We find limits from the production of hidden states other than 
the vector itself.  In particular, we show that if the 
hidden Abelian symmetry is higgsed, and the corresponding hidden 
Higgs boson has visible decays, it severely restricts the ability 
of the hidden sector to explain the anomalous muon magnetic moment.

\end{abstract}

\end{titlepage}

\setcounter{page}{2}


\section{Introduction}\label{sec:intro}
New physics beyond the Standard Model~(SM) could take many forms.  
To be consistent with existing experiments, any new particles must either 
be heavier than the electroweak scale or they must interact very weakly 
with us~\cite{Essig:2013lka}.  Lower-energy experiments
with a very high precision or intensity are particularly well-suited to 
discovering the second possibility when the characteristic mass of the 
new physics lies below the electroweak 
scale~\cite{Jaeckel:2010ni,Hewett:2012ns}.  
The results of such experiments may also provide guidance for 
future searches at high energies.

  An interesting variety of new physics below the electroweak scale 
is an exotic Abelian gauge force $U(1)_x$ that couples to the SM 
exclusively through a small kinetic mixing with 
hypercharge~\cite{Holdom:1985ag,Foot:1991kb}
\beq
  \lag \supset -\frac{\epsY}{2}X_{\mu\nu}B^{\mu\nu} \ ,
  \label{eq:vecport}
\eeq  
where $X_{\mu\nu} = (\partial_{\mu}X_{\nu}\!-\!\partial_{\nu}X_{\mu})$ 
and $X^{\mu}$ is the $U(1)_x$ gauge boson.  
If the only new particle is the massless $X^{\mu}$, 
it can be rotated with the photon such that one linear 
combination decouples from the theory~\cite{Holdom:1985ag}.  
However, if there are new matter fields charged under $U(1)_x$ 
they will develop millicharges under this rotation~\cite{Holdom:1985ag}.  
Since light millicharged matter is very strongly constrained by 
direct searches and astrophysical 
observations~\cite{Davidson:2000hf,Arias:2012az}, we
will focus on a theory where $U(1)_x$ is spontaneously broken.

  The simplest way to break $U(1)_x$ is with a hidden Higgs field.  
In this work we study a minimal supersymmetric theory in which the new gauge
force is broken well below the weak scale by a pair of hidden Higgs bosons.
With supersymmetry, the relatively low breaking scale can be
natural if the hidden sector feels superymmetry breaking less strongly
than the visible SM sector and $\epsilon$ is sufficiently 
small~\cite{ArkaniHamed:2008qn,ArkaniHamed:2008qp,Cheung:2009qd,Katz:2009qq,Morrissey:2009ur,Goodsell:2009xc,Essig:2010ye,Andreas:2011in}.
The hidden sector will also contain a dark matter candidate 
in the presence of $R$-parity.  While naturalness and dark matter
are attractive features, our primary motivation for investigating 
this particular model is that it provides a simple but non-minimal
theory of a light hidden sector.

  Light hidden vectors interacting with the SM through the kinetic 
mixing interaction of \modeqref{eq:vecport} have been studied 
extensively in recent years.
The two scenarios that have received the most attention are minimal models
where the new massive vector $Z^x$ decays primarily to 
the SM~\cite{Pospelov:2008zw,Batell:2009yf,Reece:2009un,Bjorken:2009mm} 
or to a pair of dark matter particles~\cite{Batell:2009di,
deNiverville:2011it,Izaguirre:2013uxa,Diamond:2013oda,Essig:2013vha}.  
These decay channels can also occur in the non-minimal theory we study here, 
but other decay channels can be dominant as well. 
In particular, we investigate phases of the theory where the leading
vector decays are $Z^x\to h_1^xA^x$, where $h_1^x$ and $A^x$ are 
hidden Higgs bosons, and $Z^x\to \chi_1^x\chi_2^x$, 
where $\chi_1^x$ and $\chi_2^x$ are hidden neutralinos.
The experimental signals of these additional channels have not received
as much attention as the purely visible or invisible channels
(although see Ref.~\cite{Schuster:2009au}).
We find that the $h_1^x$, $A^x$, and $\chi_2^x$ decay products 
can be metastable, and that fixed-target experiments are particularly 
well-suited to finding them.

  In this paper, we study the sensitivity of low-energy and high-precision 
experiments to a minimal supersymmetric Abelian hidden sector, 
with a focus on fixed-target experiments.  
We begin in \secref{sec:theory} by defining the theory and 
characterizing the mass eigenstates and their decay modes.  
We also formulate four distinct sets of benchmark parameters 
to be studied in the sections to come.
In \secref{sec:precision}, we investigate the existing limits on the theory 
from non-fixed-target experiments, including precision tests, meson factories, 
and cosmology.  Next, in \secrefs{sec:ebeam}{sec:hbeam} 
we study the sensitivities of current and future electron and hadron 
fixed target experiments to the theory in its various phases.
The implications of our results for the LHC are discussed in \secref{sec:lhc},
and we conclude in \secref{sec:conc}.  Some technical details about the theory
and our calculation of vector production and detection in fixed target 
experiments are collected in 
the \threeapprefs{app:hidden}{app:edetails}{app:scatter}.

\section{A Minimal Supersymmetric Hidden Sector}\label{sec:theory}
We consider the minimal supersymmetric hidden sector formulated in 
\refscite{Morrissey:2009ur,Chan:2011aa} 
consisting of a $U(1)_x$ vector multiplet $X$ together 
with a pair of chiral multiplets $H$ and $H'$ with charges $x_{H,H'}=\pm 1$ 
that develop vacuum expectation values~(VEVs).  
In addition to a minimal K\"ahler potential for $H$ and $H'$, 
the interactions in the hidden sector are taken to be:
\beq
  \lag_{hid} = \lag_{V}+\lag_{W}+\lag_{soft} \ ,
\eeq
with the vector terms
\beq
  \lag_V = \int d^2 \theta \, \biggl( \frac{1}{4} B^\alpha B_\alpha + \frac{1}{4} X^\alpha X_\alpha + \frac{1}{2} \frac{\epsilon}{\cos \theta_W} \, B^\alpha X_\alpha \biggr) + \text{h.c.} \ ,
  \label{eq:svecport}
\eeq
the superpotential terms
\beq
  \lag_W = \int d^2\theta\;\mu'HH' \ ,
\eeq
and the soft terms
\beq
  -\lag_{soft} = m_H^2|H|^2+m_{H'}^2|H'|^2- (b'HH'+\text{h.c.}) - M_x\tilde{X}^2 \ .
\eeq
Note that the kinetic mixing has been written in terms of 
$\epsilon \equiv \epsY \cos \theta_W$, where $\theta_W$ is the Weinberg angle, 
to agree with the most common convention.  
This is a different normalisation to that made in 
\refcite{Chan:2011aa} (which took $\epsilon \equiv \epsY$).  
We assume $\epsilon\ll 1$ and take the dimensionful $\mu'$ and 
soft terms to be at or below a GeV.  
This choice is consistent with supersymmetric naturalness 
for sufficiently small $\epsilon$~\cite{Morrissey:2009ur,Chan:2011aa}, 
and predicts a hidden vector mass of the same order.  
The visible sector is assumed to be the minimal 
supersymmetric standard model~(MSSM).

\subsection{Hidden Masses}

  For appropriate values of the model parameters, the scalar components of
$H$ and $H'$ develop VEVs,
\beq
  \langle H\rangle = \eta\,\sin\zeta,~~~~~
\langle H'\rangle = \eta\,\cos\zeta \ ,
\eeq
with $\zeta \in [0,\pi/2]$.  
As a result, the hidden vector $X_{\mu}$ obtains a mass equal to
\beq
  m_x = \sqrt{2}\, g_x\eta \ .
\eeq
We will denote this state $Z^x$.

  Hidden symmetry breaking also gives rise to mass mixing among 
the scalar and fermion sectors of the theory.  
The remaining bosonic mass eigenstates are a pair of
real scalars $h_1^x$ and $h_2^x$ with masses $m_{h_1^x} \leq m_{h_2^x}$, 
and a pseudoscalar $A^x$ of mass $m_{A^x}$.\footnote{
We neglect CP violation in the present work.}
The fermionic mass eigenstates are mixtures of the $U(1)_x$ gaugino 
and the hidden higgsinos:
$\chi_{1}^x,\,\chi_{2}^x,\,\chi_{3}^x$ with masses 
$|m_{\chi_1^x}| < |m_{\chi_2^x}| < |m_{\chi_3^x}|$.  
A more detailed exposition of hidden-sector mass mixing can be found 
in \refcite{Chan:2011aa} and \appref{app:hidden}.
  
  When symmetry breaking occurs in the hidden sector, 
a convenient basis of parameters for the theory is 
$\{g_x,\mu',m_x,m_{A^x},M_x,\tan\zeta,\epsilon\}$.  
Note as well that the structure of the hidden Higgs sector mirrors 
that of the neutral Higgs of the MSSM~\cite{Martin:1997ns}.  
As such, we obtain $m_{h_1^x} < m_x,\,m_{A^x} < m_{h_2^x}$ at tree-level,
and we expect this ordering to persist at loop level provided $g_x$ is small
and $\mu',\,M_x \sim m_x$.  We will also assume that $R$-parity 
is conserved and that the gravitino is moderately heavy, 
so that $\chi_1^x$ is a stable lightest superpartner~(LSP).

\subsection{Hidden Couplings and Decays}

   Particles in the hidden sector interact with each other through 
the gauge coupling $g_x$ and with the visible (MSSM) sector via the 
kinetic mixing parameter $\epsilon$.
For $e\,\epsilon \ll g_x$, as we assume here, 
hidden states nearly always decay to other hidden states 
when kinematically allowed.  However, when the hidden channels 
have been exhausted, decays to SM can dominate giving rise
to new signals in low- and high-energy experiments.

  The main connector to the SM is the hidden vector $Z^x$, and it can
decay in various ways depending on the mass spectrum of the hidden sector.  
The experimental signatures of the theory can be classified according 
to the most probable decay mode of the vector. 
We will consider the four most important possibilities:
\beq
Z^x ~~\to~~ \left\{
\begin{array}{l}
\text{SM}\!+\!\text{SM}\\
\chi_1^x+\chi_1^x,\\
h_1^x+A^{x},\\
\chi_i^x+\chi_j^x,~i > 1
\end{array}
\right.
\eeq
Below, we will define four benchmark parameter slopes that realize each 
of these four cases.

  An interesting feature of this theory is that the lightest scalar 
$h_1^x$ is always lighter than the vector and all the other scalars 
(at weak coupling).  The only possible hidden channels are
\beq
h_1^x\to \chi_i^x\chi_j^x \ .
\eeq
When all these channels are forbidden, $m_{h_1^x} < 2m_{\chi_1^x}$, 
the $h_1^x$ state will decay to the SM.  The two main contributions 
to the decay come from mass mixing with the MSSM Higgs bosons 
and from a loop of hidden vectors connecting to a pair of SM fermions.  
The effective Higgs mixing angle goes like $\epsilon\,(m_x/m_{h^0})$,
where $m_{h^0}$ is the mass of the lighter MSSM Higgs boson,
while the vector loop is proportional to $\epsilon^2$.
In computing the $h_1^x$ decay width,
we find that the Higgs mixing contribution tends to dominate over
the loop contribution above the dimuon threshold.
However, both contributions have significant suppressions, 
and the $h_1^x$ state is nearly always long-lived on collider time scales.
More details can be found in \appref{app:hidden}.

  Among the heavier scalars, there are a number of possible decay channels:
\bea
  A^x &\to& h_1^xZ^{x(*)},\,\chi_i^x\chi_j^{x} \ ,\nnmb\\
  h_2^x&\to& A^xZ^{x\,(*)},\,Z^x Z^{x(*)},\, h_1^xh_1^{x\,(*)},\, \chi_i^x\chi_j^{x} \ ,
\eea
where $(*)$ denotes that the decay product is potentially off-shell.
These decays tend to be prompt when they are two-body, but can become slow
when they are forced to be three-body with a SM final state
(\emph{e.g.} $A^x\to h_1^xZ^{x\,*}\to h_1^x(f\bar{f})$ where $f$ is 
a charged SM fermion).
  
  With our assumption of $R$-parity conservation, 
the lightest hidden neutralino is the stable LSP 
(in the absence of a light gravitino).  The heavier hidden 
neutralinos decay according to
\beq
  \chi_i^x \to \chi_j^x+\left\{Z^{x(*)},\,A^{x\,(*)},\,h_{1,2}^{x\,(*)}\right\} 
\ .
\eeq
Again, the branching fractions of these decays depend on the spectrum 
and mixing of hidden states.  Note as well that the lightest MSSM
superpartner can decay to the hidden sector through gauge kinetic mixing
(\emph{e.g.} $\chi_1^0 \to \chi_1^x+\{Z^x,\,h_1^x,\,h_2^x,\,A^x\}$).

\subsection{Benchmark Slopes}

  This minimal supersymmetric hidden sector can give rise
to a broad range of experimental signatures.  
To illustrate this range, we investigate the signals of 
four sample slopes corresponding to four different primary 
decay modes of the hidden vector $Z^x$.  
For each slope, we set $\tan\zeta = 3$ and $\alpha_x = \alpha$, 
and we take the remaining model parameters to be fixed ratios of 
the vector mass $m_x$.  
These ratios and the resulting mass spectra (in units of $m_x$) 
are listed in \tabref{tab:cases}.

  Each of the four slopes corresponds to a different dominant decay channel
for the hidden vector boson.  They are:
\begin{itemize}
\item[\textbf{A.}] $\mathbf{Z^x\to SM\!+\!SM}$\\
  The vector has no two-body decays within the hidden sector, and as a result
decays almost entirely through gauge kinetic mixing with the photon.
This scenario occurs readily for $m_x < M_x,\,\mu',m_{A^x}$,
and has been studied in great detail~\cite{Essig:2013lka}. 
Limits on this scenario are dominated by direct production of the vector
followed by its decay to visible SM final states.  We find some additional limits at fixed-target experiments based on the production of the lightest Higgs \hx\ through an \emph{off-shell} vector, followed by observing either the decay or the scattering of the scalar.  Decays of the 
lighter scalar $h_1^x$ are addressed above and in \appref{app:hidden}.
\item[\textbf{B.}] $\mathbf{Z^x\to \chi_1^x\chi_1^x}$\\
  This channel is now the only two-body hidden mode, 
and it dominates over the $SM\!+\!SM$ channel (for $e\,\epsilon \ll g_x$).  
It is likely to occur for $M_x < m_x < m_{A^x},\,\mu'$.
The final state is invisible, and has been searched for in two ways:
missing energy signatures in meson factories; and elastic scattering
of $\chi_1^x$ particles in the detector of fixed-target 
experiments~\cite{Essig:2013lka}.  As for the previous benchmark, we find some new limits based on the decay of the lightest Higgs \hx, produced through an off-shell vector.
\item[\textbf{C.}] $\mathbf{Z^x\to h_1^xA^x}$\\
The initial vector decay is prompt, but the $h_1^x$ and $A^x$ products 
decay further to the SM and are typically delayed.  
The pseudoscalar decays via $A^x\to h_1^x+Z^{x\,*}$ with 
$Z^{x\,*} \to SM\!+\!SM$, and it can be slow as well. 
An ordering $m_{A^x} < m_x < \mu',\,M_x$ is likely to produce this
decay channel. 
Depending on their decay lengths, the production of long-lived $A^x$ and $h_1^x$
states via the vector can lead to missing energy signatures in meson factories.  It can also lead to either highly displaced visible decays or elastic scattering signals in the detectors of fixed-target experiments. 
This case has only been studied in Ref.~\cite{Schuster:2009au} 
within a simplified model of the hidden sector. 
\item[\textbf{D.}] $\mathbf{Z^x\to \chi_1^x+\chi_2^x}$\\
In this case, the vector has a pair of two-body hidden decay channels
to neutralinos with branching fractions
$BR(\chi_1^x\chi_2^x)\simeq 0.94$ and $BR(\chi_1^x\chi_1^x)\simeq 0.06$.
The presence of two lighter hidden neutralinos occurs readily when  
$\mu' < m_x < M_x,\,m_{A^x}$.  The heavier of the two states tends 
to have a larger gaugino fraction, and this favours the 
$\chi_1^x\chi_2^x$ channel over the $\chi_1^x\chi_1^x$.  
The $\chi_2^x$ produced in the vector decay
goes to $\chi_2^x\to \chi_1^x+Z^{x^*}$ with $Z^{x^*} \to SM\!+\!SM$,
and can be long-lived.  This can lead to visible decays in 
fixed-target experiments.  We also have elastic scattering signals at these experiments from $\chi_1^x$ and (for long enough lifetimes) from $\chi_2^x$ .
We do not know of any previous studies of this decay channel.

\end{itemize}
In the sections to follow, we will investigate the current experimental
bounds on these four scenarios.

\begin{table}
  \centering
  \begin{tabular}{cccccccccc}
~~Case~~&~~$M_x$~~&~~$\mu'$~~&~~$m_{A^x}$~~&~~$m_{h_1^x}$~~&~~$m_{h_2^x}$~~&~~$m_{\chi_1^x}$~~&~~$m_{\chi_2^x}$~~&~~$m_{\chi_3^x}$~~\\
\hline
A&3.0&4.0&2.0&0.76&2.10&2.50&-4.02&4.53\\
B&1.0&1.5&1.5&0.73&1.65&0.39&-1.59&2.20\\
C&1.5&2.0&0.5&0.38&1.05&0.86&-2.06&2.69\\
D&3.0&0.5&1.5&0.73&1.65&0.24&0.57&3.33\\
\hline
\end{tabular}
\caption{Sample parameters and particle masses for the four benchmark
slopes.  All quantities are listed in units of the vector mass $m_x$,
and we also fix $\alpha_x = \alpha$ and $\tan\zeta = 3$.}
\label{tab:cases}
\end{table}

\section{Non-Fixed Target Limits}\label{sec:precision}
  To begin, we collect limits on the four scenarios from precision
tests, meson factories, and cosmology.

\subsection{Model-Independent Limits}\label{sec:modind}

  Precision electroweak observables and lepton magnetic moments provide
important limits on the mass and kinetic mixing of the hidden vector
that are independent of how it decays.  We review here
the model-independent constraints that apply to all four cases.

  Kinetic mixing ties the hidden vector to the hypercharge 
vector, and this induces a mixing between $Z^x$ and the $Z^0$ after 
electroweak and hidden symmetry 
breaking~\cite{Chang:2006fp,Chun:2010ve}.  A global fit to precision 
data in the presence of this mixing was performed in \refcite{Hook:2010tw}, 
and the limit for $m_x \lesssim 10\,\gev$ was found to be
\beq
\epsilon < (0.03)c_W \simeq 0.026 \ .
\eeq

  The leading effects of a kinetically-mixed hidden vector on the 
anomalous magnetic moments $a_{\ell} = (g\!-\!2)_{\ell}$ of the electron 
and muon were computed in \refcite{Pospelov:2008zw}.  
The current experimental status 
for the muon is~\cite{Beringer:1900zz}
\bea
\Delta a_{\mu} &=& a_{\mu}^{obs}-a_{\mu}^{SM} = (287\pm 80)\times 10^{-11} \ .
\eea
We demand that $0< \Delta a_{\mu} < 447\times 10^{-11}$, so as to lie
within $2\sigma$ of the observed value while also allowing for consistency
with the SM.  The additional contribution due to the hidden vector
tends to increase the predicted value of $\Delta a_{\mu}$, 
and an interesting possibility is that a hidden vector is
the source of the apparent ($3.6\sigma$) discrepancy 
with the SM~\cite{Pospelov:2008zw}.

  For the anomalous magnetic moment of the electron, 
we use~\cite{Davoudiasl:2012ig,Giudice:2012ms,Endo:2012hp}
\beq
\Delta a_e = (-10.5\pm 8.1)\times 10^{-13} \ ,
\eeq
and we impose the $3\sigma$ limit $-34.8 < \Delta a_e\cdot 10^{13} < 13.8$.
This range uses the updated direct measurement of 
$a_e$~\cite{Hanneke:2008tm} compared with the SM prediction~\cite{Aoyama:2012wj}
computed with the value of of $\alpha$ derived from atomic measurements in 
rubidium~\cite{Bouchendira:2010es}.
The combined precision electroweak and magnetic moment bounds are
shown in \figrefs{fig:benchC}{fig:benchD}.

\subsection{Meson Factories}\label{sec:mesfact}

  New light states have been searched for in a diverse range of meson
factories, including BaBar, Belle, KLOE, and in rare Kaon decays
at Brookhaven.  We examine the bounds they imply for the theory.

\paragraph{Case A}
The direct signals of the vector in this case have been studied previously.
For $2m_{\mu} < m_x < 10.335\,\gev$, there are strong limits from the BaBar
search for an exotic pseudoscalar $a^0$ via $\Upsilon(3s,2s) \to \gamma\,a^0$
with $a^0\to \mu^+\mu^-$~\cite{Aubert:2009au,Aubert:2009cp} 
when applied to the continuum process $e^+e^-\to \gamma\,Z^x$ 
with $Z^x\to \mu^+\mu^-$~\cite{Borodatchenkova:2005ct,Bjorken:2009mm}.  
For $m_x \in [100, 400]\,\mev$, the strongest bounds come from the KLOE search
for $\phi\to \eta\,Z^x$ with 
$Z^x\to e^+e^-$~\cite{Archilli:2011zc,Babusci:2012cr}.  When $m_x < 100\,\mev$, WASA-at-COSY is most constraining based on $\pi^0 \to Z^x \gamma$ with prompt $Z^x \to e^+ e^-$~\cite{Adlarson:2013eza}.
In all cases, the bound is approximately $\epsilon \lesssim 3\times 10^{-3}$.

  Our supersymmetric realization of this scenario will also induce 
new decay channels for mesons through the mixing of $h_1^x$ and $h_2^x$ 
with the MSSM Higgs bosons.
Limits on this mixing were studied in 
\refscite{Batell:2009di,Batell:2009jf,Clarke:2013aya},
where the strongest bounds were found to come from $K^+\to\pi^+h_1^x$ 
and $B^+\to K^+h_1^x$ with $h_1^x\to \mu^+\mu^-,\,invisible$.
The effective Higgs mixing angle in this theory is proportional 
to $\epsilon\,(m_x/m_{h^0})$, where $m_{h^0}$ is the SM-like Higgs boson mass
(see \appref{app:hidden}).
We find that the model-independent and direct limits on $\epsilon$
and $m_x$ from the effects of the vector are stronger than those from
decays through Higgs mixing for $m_x < 10\,\gev$.
In the same way, no useful limits are found from leptonic 
heavy meson decays~\cite{Aditya:2012ay}.

\paragraph{Case B} 
Direct bounds on an invisibly decaying hidden vector were investigated
in \refscite{Izaguirre:2013uxa,Essig:2013vha}.  
The BaBar search for an invisible pseudoscalar
in $\Upsilon(3s)\to \gamma\,a^0$ with $a^0\to invisible$~\cite{Aubert:2008as} 
can be reinterpreted as a limit on $e^+e^-\to \gamma\,Z^x$ 
with $Z^x\to invisible$~\cite{Borodatchenkova:2005ct}, 
and imply $\epsilon \lesssim (1\!-\!4)\times 10^{-3}$ depending on the 
masses $m_x$ ($< 8\,\gev$) and $m_{\chi_1^x}$.  
The Brookhaven E787 and E949 searches for 
$K^+\to \pi^+Z^x$ with $Z^x \to invisible$~\cite{Adler:2004hp,Artamonov:2009sz} 
also limit $\epsilon \lesssim 1\times 10^{-3}$ in the regions 
$m_x \in [0,120]\cup[160,240]\,\mev$~\cite{Izaguirre:2013uxa,Essig:2013vha}.  
No additional bounds are obtained from rare meson decays through Higgs mixing.

\begin{figure}
  \centering
  \includegraphics[width=0.8\textwidth]{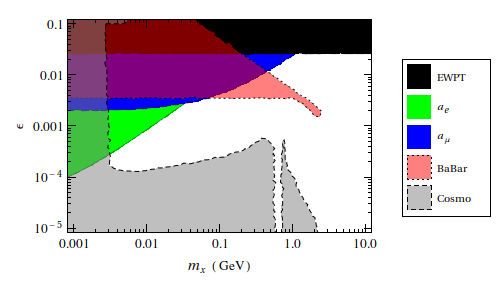}
  \caption{Non-fixed target limits on Case C.  The black, green and blue regions show the model-independent exclusions of \secref{sec:modind}; the red regions the BaBar constraints discussed in \secref{sec:mesfact}; and the grey regions the cosmological limits from \secref{sec:cosmo}.  The cosmological limits are less robust, as discussed in \secref{sec:cosmo}.}
\label{fig:benchC}\end{figure}

\paragraph{Case C}
The vector decay $Z^x\to h_1^xA^x$ has not been studied 
in as much detail as the other cases~\cite{Schuster:2009au}.
Both products can have long lifetimes, and the limits found 
for Case B can often be applied to this case as well.  
The BaBar $\Upsilon(3s)\to \gamma+invisible$ search imposed
a veto on additional activity in the electromagnetic calorimeter
and on hits in the muon chamber opposite 
to the photon direction~\cite{Aubert:2008as}.
To approximate the exclusion implied by this search, we apply the bounds
found in Refs.~~\cite{Izaguirre:2013uxa,Essig:2013vha} 
with the addtional condition that $\overline{\gamma{v}}_T > 100\,\text{cm}$ 
for both $h_1^x$ and $A^x$, where $\overline{\gamma{v}}_T$ is the mean 
boosted transverse velocity in the lab frame.  
The value of $100\,\text{cm}$ is the approximate transverse radius 
of the BaBar electromagnetic calorimeter 
and muon chamber~\cite{Harrison:1998yr}. 
We estimate $\overline{\gamma{v}}_T = \bar{p}_T/m$ using
a simple Monte Carlo simulation that takes into account the angular
distribution of the production cross section~\cite{Essig:2009nc}
and the angular acceptance of photons in the search~\cite{Aubert:2008as}.
The exclusions derived in this way are shown in \figref{fig:benchC}
together with the model-independent bounds.
We also find that the exclusions from $K^+\to \pi^++inv$
apply here for $\epsilon \lesssim 10^{-2}$, although we do not
show them in the figures.
  
  In contrast to searches for invisible decay modes, the existing 
meson-factory searches for the visible decays of a hidden vector 
presented in \refscite{Aubert:2009au,Aubert:2009cp,Archilli:2011zc,
Babusci:2012cr,Aubert:2009af,Lees:2012ra,Adlarson:2013eza} do not appear to give any relevant 
limits on this scenario.  These searches tend to focus on leptonic decays, 
they typically look for one or more resonances,
and they often demand that the energies of the visible products reconstruct 
the mass of the decaying meson or the beam energy.  Vector decays in case C 
will typically have two visible products and missing energy 
(for $h_1^x$ long-lived), or six visible products (for $h_1^x$ short-lived).  
When there is missing energy, the energies of the visible products will not 
reconstruct the total beam or decaying meson energy, as required 
in \refscite{Aubert:2009au,Aubert:2009cp,Archilli:2011zc,
Babusci:2012cr,Aubert:2009af}.  Missing energy will be absent when 
$h_1^x$ decays promptly.  However, this only occurs when 
$m_{h_1^x} \gtrsim \gev$ where the hidden Higgs has only a very small 
branching fraction to leptons.  This implies that the search of \refcite{Adlarson:2013eza} using pion decays, as well as the inclusive leptonic 
searches of \refcite{Lees:2012ra},  are insensitive.  Finally, note that
the pseudoscalar $A^x$ decay involves an off-shell vector, and its decay
products will not reconstruct a sharp resonance.

\begin{figure}
  \centering
  \includegraphics[width=0.8\textwidth]{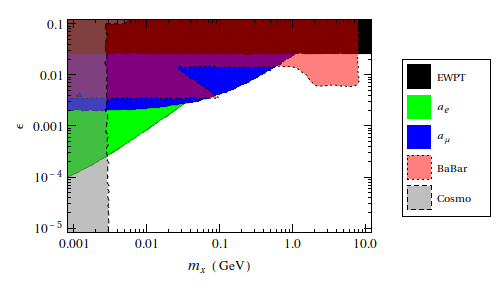}
  \caption{Non-fixed target limits on Case D, in the same format as \figref{fig:benchC}.}
\label{fig:benchD}\end{figure}

\paragraph{Case D}
The vector now decays primarily to $Z^x\to \chi_1^x\chi_2^x$ 
with $BR_{12} \simeq 0.94$, but also to $Z^x\to \chi_1^x\chi_1^x$ 
with $BR_{11} \simeq 0.06$.  In the former channel, the subsequent
decay of $\chi_2^x$ to $\chi_1^x$ and visible matter via an off-shell $Z^x$
can be delayed.  As in Case~C, we find that meson factory searches for
visible decays do not give any useful bounds.  However, the BaBar
$\Upsilon(3s)\to \gamma+invisible$ search of \refcite{Aubert:2008as} is applicable here.  
When the $\chi_2^x$ decay occurs relatively
promptly, $\overline{\gamma{v}_T}\tau < 100\,\text{cm}$ (with the mean
value of $\overline{\gamma{v}_T} = \bar{p}_T/m_{\chi_2^x}$ estimated
with a simple Monte Carlo simulation), only the $Z^x\to\chi_1^x\chi_1^x$
channel contributes significantly. In this situation we rescale the limits on 
$\epsilon$ obtained \refscite{Izaguirre:2013uxa,Essig:2013vha} 
by $1/\sqrt{BR_{11}}$.
For $\overline{\gamma{v}_T}\tau > 100\,\text{cm}$, no such rescaling is needed.
The exclusions derived in this way are shown in \figref{fig:benchD}.

\subsection{Cosmology and Astrophysics}\label{sec:cosmo}

  This theory can also be constrained by cosmological observations 
given a specific assumption for the evolution of the Universe.  
An obvious requirement for all cases is that the relic density of 
the lightest hidden neutralino $\chi_1^x$ not be too large.  
The relic density calculations of \refscite{Pospelov:2007mp,Pospelov:2008jd} 
are applicable here.  In general, the thermal relic density tends to 
be well below critical for $m_{\chi_1^x} > m_x$ unless $\alpha_x$ is also
very small, but can become unacceptably large when $m_{\chi_1^x} < m_x$.

  If $\chi_1^x$ does make up the dark matter, it is typically 
consistent with existing limits from direct detection.  Scattering
can occur through the exchange of a hidden vector (via kinetic mixing)
or a hidden Higgs boson (via Higgs portal mixing).  Since the lightest
hidden neutralino is Majorana, vector-mediated scattering
corresponds to the effective operator 
$(\bar{\chi}_1^x\gamma^{\mu}\gamma^5\chi)(\bar{f}\gamma_{\mu}f)$, 
which leads to spin-independent~(SI) scattering suppressed by two powers 
of the dark matter velocity~\cite{Chang:2009yt}, 
in addition to a factor of $\epsilon^2$.  The exchange of a hidden Higgs
boson will also contribute to SI scattering, now with a suppression factor on
the order of $\epsilon^2(m_xm_Z/m_{h_1^x}^2)^2$ relative to SI scattering
mediated by the SM Higgs boson.  These suppressions lead to
direct-detection cross sections well below existing limits 
from nuclear-~\cite{Cushman:2013zza} and 
electron-recoil~\cite{Essig:2011nj,Essig:2012yx} 
analyses once the precision bounds on $\epsilon$ are applied.

  Additional bounds from cosmology can arise if some of the states in
the theory are long-lived.  Lifetimes greater
than $\tau \simeq 0.05\,\text{s}$ will lead to decays after the onset
of big-bang nucleosynthesis~(BBN) and can potentially alter the
abundances of light elements~\cite{Kawasaki:2004qu,Jedamzik:2006xz}.  
Late decays can also lead to spectral distortions in the cosmic
microwave background radiation~\cite{Hu:1993gc} or an excess
in gamma rays~\cite{Essig:2013goa}.  

  In cases~A--C, the longest-lived state is nearly always the $h_1^x$ scalar.
It decays mainly through mixing with the MSSM Higgs bosons to pairs of
SM particles.  When $m_{h_1^x} > 2m_{\pi}$, a significant fraction of these
decays will be hadronic and will ultimately create light pions.  
When $\tau > 0.05\,\text{s}$, these pions can scatter with protons 
to create more neutrons than would otherwise be 
present~\cite{Kawasaki:2004qu,Jedamzik:2006xz,Pospelov:2010cw}.
Below the pion threshold, all decays will be purely electromagnetic,
producing muons, electrons, or photons.  These particles thermalize
efficiently with the plasma, and for $\tau \lesssim 10^4\,\text{s}$,
they do so before potentially destroying 
light nuclei~\cite{Kawasaki:2004qu,Jedamzik:2006xz,Pospelov:2010cw}.
Thus, longer lifetimes are safe below the pion threshold.
In \figref{fig:benchC} we show the regions where $h_1^x$ decay
could be dangerous for nucleosynthesis (or the CMB) based 
on these considerations. 
These regions are not necessarily ruled out if the yield $Y = n/s$ 
of $h_1^x$ is sufficiently small at the time of decay.
However, in a thermal scenario where $h_1^x$ begins with an equilibrium
density and freezes out, the freeze-out process relies on the three-body
annihilation mode $h_1^xh_1^x\to Z^xZ^{x\,*}$ and tends to produce 
a relatively large yield.

  The long-lived state in case~D is the next-to-lightest neutralino $\chi_2^x$,
which decays through an off-shell vector.  
For $\Delta m_{\chi} = m_{\chi_2^x}-m_{\chi_1^x} > 2m_{\pi}$, a significant
fraction of these decays will be hadronic.  For smaller mass differences
the decays will be electromagntic.  In Fig.~\ref{fig:benchD} we
show the regions where this decay could potentially affect nucleosynthesis.
Note, however, that the transfer reaction $\chi_2^x\chi_1^x\to\chi_1^x\chi_1^x$
is generally efficient, and the thermal yield of $\chi_2^x$ is typically
very small.

  Further constraints on all four cases can be derived from  
the observed cooling rate of supernova~(SN) 
1987A~\cite{Bjorken:2009mm,Dent:2012mx,Dreiner:2013mua}.  
The hidden vector can be created by thermal scattering within a SN and 
generate new energy loss mechanisms.  In case~A, this vector can be 
long-lived and 
escape the SN before it decays to the SM,
and small values of $\epsilon \lesssim 10^{-7}$ are excluded for 
$m_{x} \lesssim 100\,\mev$~\cite{Bjorken:2009mm}.  
In cases~B--D, the vector will decay promptly to long-lived hidden states 
such as $h_1^x$ or $\chi_1^x$ which can carry energy away if they leave 
the SN before decaying or scattering.  For these cases, we expect limits 
similar to those found in \refcite{Dreiner:2013mua}, 
which studied the bounds on hidden dark matter in the form of 
a Dirac fermion or a complex scalar charged under $U(1)_x$.
A potential important difference in the present case is that the metastable
states interact differently with the vector boson, and may be less likely
to scatter before exiting the SN.  We defer an analysis of this effect
to a future work.

\section{Electron Fixed Target Experiments}\label{sec:ebeam}
Fixed target experiments using electron beams have played an important role in studies of hidden vectors.  
Significant limits have been set when the gauge bosons decay directly to the SM~\cite{Bjorken:2009mm,Andreas:2012mt,Merkel:2011ze,Abrahamyan:2011gv,Andreas:2012rm}.  Additionally, a number of experiments have been proposed to extend the search reach based on both visible~\cite{Essig:2010xa,Freytsis:2009bh,HPS,2013arXiv1301.1103H} and invisible~\cite{Izaguirre:2013uxa} decays.  It is therefore natural to examine how they constrain the more general hidden sectors we consider here.

\begin{figure}[ttt]
  \centering
  \includegraphics[width=0.85\textwidth]{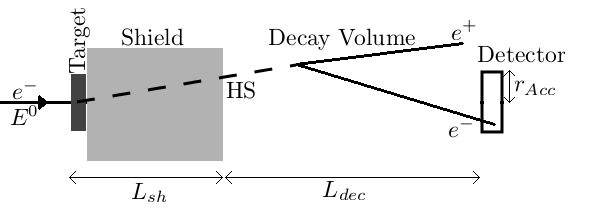}
  \caption{Sketch of the experimental configuration for fixed target searches for visible decays.  The electron beam strikes the target, producing a metastable hidden sector state (dashed line); this decays to visible sector particles within the decay volume after passing through the shield.  A signal requires that at least one of the visible decay products strike the detector.}
\label{fig:beamdump}\end{figure}

Most searches require hidden states decaying to the SM, so we begin by considering that case.  The experimental geometry is sketched in \figref{fig:beamdump}.  A signal requires the production of a metastable state at the interaction point, which decays to visible sector particles (typically required to be leptons) within the decay volume.  In previous work, the long-lived state was the hidden vector itself; the lifetime requirement leads to limits being set at small kinetic mixing and/or low mass.

However, as already discussed in \secref{sec:theory}, our model displays a broader range of phenomenology.  Of our four benchmarks, only in Case~A do the standard limits apply.  In all other Cases, the vector $Z^x$ decays promptly to the hidden sector.  Instead, limits may be derived from other metastable particles: the lightest scalar \hx\ (Cases~A--C), pseudoscalar $A^x$ (Case~C) and next-to-lightest fermion $\chi_2^x$ (Case~D).  The subsequent limits at electron beam dumps have not been previously studied.

\begin{figure}[ttt]
  \centering
  \includegraphics[width=0.6\textwidth]{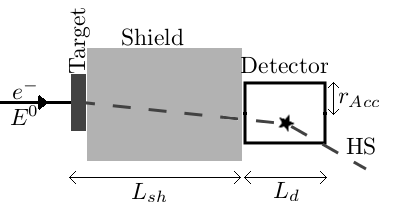}
  \caption{Sketch of the experimental configuration for fixed target searches for invisible final states.  The electron beam strikes the target, producing a stable or long-lived hidden sector state (dashed line).  This then scatters inside the detector at the starred point.}
\label{fig:beaminvis}\end{figure}

Another class of searches look for the production of invisible states at fixed target experiments.  These have been proposed as a way to search for light dark matter~\cite{Izaguirre:2013uxa}.  The experimental geometry is sketched in \figref{fig:beaminvis}.  Production of hidden sector states proceeds as before, producing either stable or sufficiently long-lived hidden particles.  The experiment would seek to observe the quasi-elastic scattering of these states with nuclei within the detector, similar to searches for the direct detection of dark matter.  Such a search is highly relevant in Cases~B and~D, when the $Z^x$ can decay to the stable lightest fermion $\chi_1^x$.

After some general comments, we discuss the production of the hidden sector through an on-shell (off-shell) $Z^x$ in \secref{sec:onX} (\ref{sec:offX}).  We relate the cross sections to the expected signals in \secref{sec:sigmatoN}, and discuss relevant experiments in \secref{sec:Experiments}.  The theoretical framework for the production of a hidden vector at electron fixed target experiments is well established~\cite{Kim:1973he,Bjorken:2009mm,Andreas:2012mt,Beranek:2013nqa} (see also \refcite{Izaguirre:2013uxa}), and we outline the main points here to be self-contained and extend the formalism in some points for the analysis at hand.
Additional technical details are presented in \appref{app:edetails}.  The reader who is interested 
primarily in results should therefore skip to \secref{sec:limits}, where we present the combined limits from electron beam dumps for our four benchmark scenarios.

\subsection{Production Cross Section: On-Shell Vector}\label{sec:onX}

\begin{figure}[ttt]
  \begin{center}
    \begin{fmffile}{epepX1}
      \begin{fmfgraph*}(120,60)
        \fmfstraight
        \fmfleft{i1,ia,i2,i3} \fmfright{o0,oa,o1,o2}
        \fmf{fermion,label=$p_e$}{i2,v1}
        \fmf{dbl_plain,label=$\overrightarrow{p_N}$,label.side=right}{i1,v3}
        \fmf{dbl_plain,label=$\overrightarrow{k_{N'}}$,label.side=right}{v3,o0}
        \fmfblob{.08w}{v3}
        \fmf{fermion,label=$q_a$,label.side=left}{v1,v2}
        \fmf{fermion,label=$k_e$,label.side=right}{v2,o1}
        \fmf{boson,label=$q\uparrow$,label.side=left,tension=0.4}{v3,v1}
        \fmffreeze
        \fmf{boson,label=$k_x$,label.side=left}{v2,o2}
        \fmflabel{$Z^x$}{o2}
      \end{fmfgraph*}
    \end{fmffile}
    \begin{fmffile}{epepX2}
      \begin{fmfgraph*}(120,60)
        \fmfstraight
        \fmfleft{i1,ia,i2,i3} \fmfright{o0,oa,o1,o2}
        \fmf{fermion,label=$p_e$}{i2,v1}
        \fmf{fermion,label=$q_b$}{v1,v2}
        \fmf{fermion,label=$k_e$}{v2,o1}
        \fmf{dbl_plain,label=$\overrightarrow{p_N}$,label.side=right}{i1,v3}
        \fmf{dbl_plain,label=$\overrightarrow{k_{N'}}$}{v3,o0}
        \fmfblob{.08w}{v3}
        \fmf{boson,label=$\uparrow q$,label.side=right,tension=0.4}{v3,v2}
        \fmffreeze
        \fmf{boson,label=$k_x$,label.side=left}{v1,o2}
        \fmflabel{$Z^x$}{o2}
      \end{fmfgraph*}
    \end{fmffile}
  \end{center}
  \caption{The two leading Feynman diagrams for production of on-shell massive vector in electron-fixed target scattering.  The double lines represent the initial and final atomic/nuclear states $N$ and $N'$.}\label{fig:epepX}
\end{figure}

We first consider the production of an on-shell hidden vector.  This has been the usual focus of attention, with a long-lived vector decaying to leptons; that is relevant in Case A.  Our interest is on a vector that decays promptly to hidden sector scalars (Case C) or fermions (Cases B and D).

\begin{figure}[ttt]
  \centering
  \includegraphics[width=0.6\textwidth]{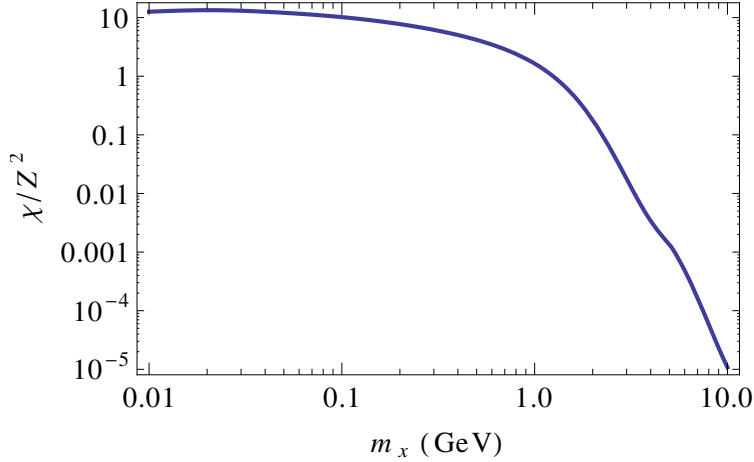}
  \caption{The effective flux $\chi$ for the E137 experiment, normalised to the target atomic number $Z = 13$.}\label{fig:effflux}
\end{figure}

There are two Feynman diagrams at leading order, shown in \figref{fig:epepX}.  The improved Weizsacker-Williams (WW) approximation~\cite{Kim:1973he} lets us factor out the dependence on the structure of the target into an integral over form factors.  For an electron of energy $E_e$, this approximation is valid when~\cite{Andreas:2012mt}
\begin{equation}
  m_e \ll m_x \ll E_e \quad \text{and} \quad E_x \theta_x^2 \ll E_e ,
\label{eq:WWcons}\end{equation}
where $E_x$, $\theta_x$ are respectively the energy and scattering angle of the hidden vector.  The production differential cross section for the full process is then related to that for electron-photon scattering:
\begin{equation}
  \frac{d \sigma\, (eN \to eZ^xN')}{d (p_e \cdot k_x) \, d (p_N \cdot k_x)} = \frac{\alpha}{\pi} \, \frac{d \sigma\, (e\gamma \to eZ^x)}{d (p_e \cdot k_x)}\biggr\rvert_{q = q^\ast} \, \frac{\chi}{p_N \cdot k_e} \, .
\label{eq:wwonshell}\end{equation}
The two-to-two cross section is evaluated at the special kinematics $q=q^\ast$, where $\vec{q^\ast}$ is parallel to $\vec{p}_e - \vec{k}_x$.  The effective flux $\chi$ contains all details of the target structure; it is also a function of the electron energy $E_e$ and vector mass $m_x$.  We plot $\chi/Z^2$ in \figref{fig:effflux} for the E137 experiment~\cite{Riordan:1987aw} (see \secref{sec:Experiments}), where the target atomic number $Z = 13$ (compare Fig.~10 of \refcite{Bjorken:2009mm}).  It is $\order{1\text{--}100}$ for $m_x < 1$~GeV, and drops sharply above that point.  More details about this function are provided in \appref{app:effflux}.

Introducing $x_e \equiv E_x/E_e$, the total differential cross section is~\cite{Bjorken:2009mm}
\begin{equation}
  \frac{d \sigma}{d x_e \, d \cos \theta_x} = 8 \, \alpha^3 \epsilon^2 E_e^2 \, x_e \  \chi \, \sqrt{1 - \frac{m_x^2}{E_e^2}} \, f(x_e, m_x^2) \,,
\label{eq:dsigdxdth}\end{equation}
where for later convenience we have defined
\begin{equation}
  f(x_e, m_x^2) = \frac{1 - x_e + \frac{1}{2} x_e^2}{U^2} - \frac{(1 - x_e) \, x_e \, m_x^2}{U^3} + \frac{(1 - x_e)^2 m_x^4}{U^4} \,.
\end{equation}
$U$ is the virtuality of the intermediate electron in the second diagram of \figref{fig:epepX}, and is a function of both $x_e$ and $\theta_x$:
\begin{equation}
  U \approx E_e^2 \, x_e \, \theta_x^2 + m_x^2 \, \frac{1 - x_e}{x_e} + m_e^2 \, x_e .
\label{eq:Udef}\end{equation}
This cross section is sharply peaked at $\theta_x \approx 0$ and $x_e \approx 1$.  Indeed, we see from Eqs.~\eqref{eq:dsigdxdth} through \eqref{eq:Udef} that there is a singularity in that limit that is only regulated by the electron mass.  The WW approximation can be understood as focusing on these singular terms.

We integrate \modeqref{eq:dsigdxdth} over the angular variable $\theta_x$.  This gives us the differential cross section with respect to the electron energy,
\begin{equation}
  \frac{d\sigma}{dx_e} = 4 \, \alpha^3 \epsilon^2 \, \chi \, \sqrt{1 - \frac{m_x^2}{E_e^2}} \ \frac{1 - x_e + \frac{1}{3} x_e^2}{m_x^2 \frac{1-x_e}{x_e} + m_e^2 x_e} \,.
\label{eq:dsigdx}\end{equation}
The total cross section for production of on-shell hidden vectors at the E137 and proposed JLab experiments are shown in \figref{fig:exsec}.

\subsection{Production Cross Section: Off-Shell Vector}\label{sec:offX}

We next consider the situation where we enter the hidden sector through an \emph{off-shell} vector.  This is the natural generalisation of the previous scenario.  It is particularly relevant to Case B, where it is the only way to get a visible signal from the hidden sector; and Case A, where it is the only way to produce hidden states other than the hidden vector.

\begin{figure}
  \begin{center}
    \begin{fmffile}{epepXstar1}
      \begin{fmfgraph*}(150,60)
        \fmfstraight
        \fmfleft{i1,ia,i2,i3} \fmfright{o0,oa,o1,o2,o3}
        \fmf{fermion,label=$p_e$}{i2,v1}
        \fmf{dbl_plain,label=$\overrightarrow{p_N}$,label.side=right}{i1,v3}
        \fmf{dbl_plain,label=$\overrightarrow{k_{N'}}$,label.side=right}{v3,o0}
        \fmfblob{.08w}{v3}
        \fmf{fermion,label=$q_a$,label.side=left}{v1,v2}
        \fmf{fermion,label=$k_e$,label.side=right}{v2,o1}
        \fmf{boson,label=$q\uparrow$,label.side=left,tension=0.4}{v3,v1}
        \fmffreeze
        \fmf{boson,label=$k_x$,label.side=left}{v2,v4}
        \fmf{dots}{o2,v4,o3}
      \end{fmfgraph*}
    \end{fmffile}
    \begin{fmffile}{epepXstar2}
      \begin{fmfgraph*}(150,60)
        \fmfstraight
        \fmfleft{i1,ia,i2,i3} \fmfright{o0,oa,o1,o2,o3}
        \fmf{fermion,label=$p_e$}{i2,v1}
        \fmf{fermion,label=$q_b$}{v1,v2}
        \fmf{fermion,label=$k_e$}{v2,o1}
        \fmf{dbl_plain,label=$\overrightarrow{p_N}$,label.side=right}{i1,v3}
        \fmf{dbl_plain,label=$\overrightarrow{k_{N'}}$}{v3,o0}
        \fmfblob{.08w}{v3}
        \fmf{boson,label=$\uparrow q$,label.side=right,tension=0.4}{v3,v2}
        \fmffreeze
        \fmf{boson,label=$k_x$,label.side=left}{v1,v4}
        \fmf{dots}{o2,v4,o3}
      \end{fmfgraph*}
    \end{fmffile}
  \end{center}
  \caption{The two leading Feynman diagrams for production of off-shell massive vector in electron-fixed target scattering.  The dotted lines represent the on-shell hidden states; we integrate over the phase space of these particles in deriving \modeqref{eq:offshell}.}\label{fig:epepXoffshell}
\end{figure}

\begin{figure}
  \centering
  \includegraphics[width=0.7\textwidth]{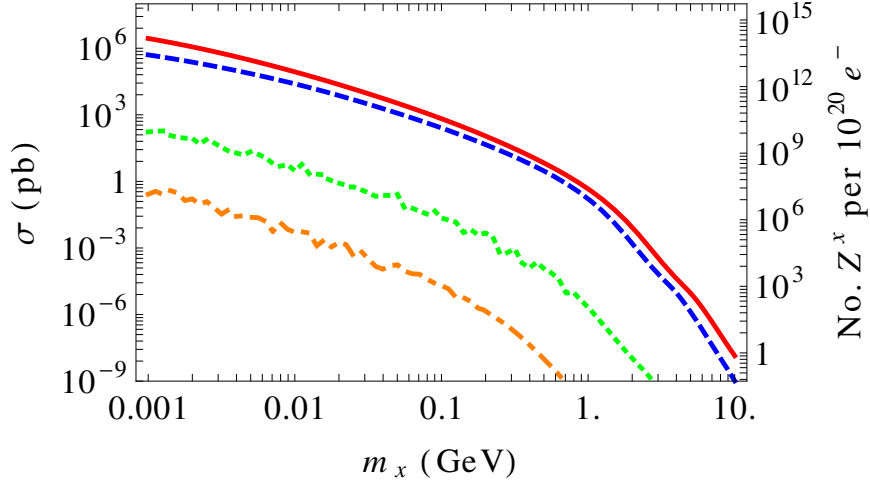}
  \caption{Production cross sections and hidden sector thick-target yields for $\epsilon = 10^{-3}$  at electron beam dump experiments.  The red, solid (blue, dashed) lines are for on-shell production at E137 (JLab).  The orange, dot-dashed (green, dotted) lines are for off-shell production at E137 in the scalar (Higgsstrahlung) channels in Case A; these are the extremal off-shell cross sections for our model.}\label{fig:exsec}
\end{figure}

We focus on the production of two on-shell hidden states through the diagrams shown in \figref{fig:epepXoffshell}.  The dotted lines represent either $h^x_1 Z^x$ (Higgsstrahlung channel), $h^x_1 A^x$ (scalar channel) or $\chi_1^x \chi_{1,2}^x$ (fermion channel).  Of these, the first is usually most important on kinematic grounds.  Before fully generalising the WW approximation, we note that by integrating over the hidden state phase space and using the QED Ward identity, we can obtain a simple extension of \modeqref{eq:wwonshell} (see \refcite{Izaguirre:2013uxa} and \appref{app:offshell}):
\begin{equation}
  \frac{d \sigma \, (eN \to e Z^{x(\ast)}N')}{d k_x^2 \, d (p_e \cdot k_x) \, d (p_N \cdot k_x)} = \biggl( \frac{1}{\pi} \, \frac{\sqrt{k_x^2} \, \Gamma_{HS} (k_x^2)}{(k_x^2 - m_x^2)^2} \biggr) \frac{\alpha}{\pi} \, \frac{d \sigma\, (e\gamma \to eZ^x)}{d (p_e \cdot k_x)}\biggr\rvert_{q^\ast} \, \frac{\chi}{p_N \cdot k_e} \, .\label{eq:offshell}\end{equation}
This expression is valid under the conditions of \modeqref{eq:WWcons} with the replacement $m_x \to \sqrt{k_x^2}$.  The partial width $\Gamma_{HS}$ is for a vector of mass $m_x^2 = k_x^2$ to decay to the relevant final state; explicit expressions are given in \appref{app:vecwidth}.  Note that if we replace the numerator with a Breit-Wigner propagator and allow the vector to go on-shell, then in the narrow-width approximation the bracketed term in \modeqref{eq:offshell} reduces to a delta function, reproducing \modeqref{eq:wwonshell}.  We show the production cross sections for off-shell hidden vectors in \figref{fig:exsec} for two cases that are extremal among those we consider.

The full generalisation of the WW approximation is most conveniently expressed in terms of integrated matrix elements, due to non-trivial phase space factors.  We denote the electron-target and electron-photon amplitudes by $\mathcal{M}_{eN}$ and $\mathcal{M}_{e\gamma}$ respectively, where the latter is evaluated at the special kinematics $q = q^\ast$.  Then the WW approximation is
\begin{equation}
  \int d \Pi_{n+2} \, \frac{1}{2N_i} \sum \abs{\mathcal{M}_{eN}}^2 = \int \biggl( \prod_{i=1}^n \frac{d^3k_i}{(2\pi)^3 2E_i} \biggr) \, \frac{1}{4} \sum \abs{\mathcal{M}_{e\gamma}}^2 \frac{M_i}{\lvert\vec{p}_e - \vec{k}_x\rvert} \, \frac{\alpha \, \chi}{t_{aux}} \,.
\label{eq:WWgenoff}\end{equation}
Here, $M_i$ is the mass of the target, $N_i$ its internal degrees of freedom, the vector magnitude $\lvert \vec{p}_e - \vec{k}_x\rvert$ is evaluated in the lab frame, and $t_{aux}\equiv (p_e \cdot k_X - \frac{1}{2} m_x^2)/(E_e - E_x)$ is roughly the minimum momentum transfer between the electron and the target.  There are $n$ hidden sector particles with momenta $k_i$, $i = 1 \ldots n$, and $d\Pi_l$ represents the $l$-dimensional phase space integral including the delta function that enforces overall conservation of momentum.  More details are presented in \appref{app:offshell}.

For the case $n = 2$ of interest, the hidden sector kinematics may be expressed in terms of the momentum of the off-shell vector $k_x$, and $k_\Delta \equiv k_1 - k_2$.  Requiring that the final states be on-shell imposes
\begin{equation}
  \begin{split}
    k_\Delta^2 & = 2m_1^2 + 2m_2^2 - k_x^2 \,, \\
    k_\Delta \cdot k_x & = m_1^2 - m_2^2 \,.
  \end{split}
\label{eq:kDelcon}\end{equation}
The first constraint can be used to fix $\lvert\vec{k}_\Delta\rvert$, and the second to fix the angle between $\vec{k}_x$ and $\vec{k}_\Delta$.  $k_\Delta$ is then defined by $k_\Delta^0$ and $\phi_\Delta$, with the latter being the azimuthal angle for the vector $\vec{k}_\Delta$ relative to $\vec{k}_x$.  The lab-frame cross sections we desire are, in the Higgsstrahlung channel:
\begin{multline}
  \frac{\dd \sigma}{\dd k_x^2 \, \dd k_x^0 \, \dd \cos\theta_x \, \dd k_\Delta^0 \, \dd \phi_\Delta} = \frac{\alpha^3 \alpha_x \epsilon^2 \chi}{4\pi^2} \, N_{H,a}^2 \, \frac{1}{(k_x^2 - m_x^2)^2} \\
  \times \, \bigl[ \bigl(k_x^2 - 2 m_{h^x_1}^2 + 6 m_x^2 \bigr) \, f(x_e, k_x^2) + 2 \Omega(k_\Delta) \bigr] ,
\label{eq:higgsstr}\end{multline}
for the production of a pair of scalars:
\begin{multline}
  \frac{\dd \sigma}{\dd k_x^2 \, \dd k_x^0 \, \dd \cos\theta_x \, \dd k_\Delta^0 \, \dd \phi_\Delta} = \frac{\alpha^3 \alpha_x \epsilon^2 \chi}{4\pi^2} \, N_{S,a}^2 \, \frac{1}{(k_x^2 - m_x^2)^2} \\
  \times \bigl[ \bigl(k_x^2 - 2 m_1^2 - 2 m_2^2 \bigr) \, f(x_e, k_x^2) + 2 \Omega(k_\Delta) \bigr] ,
\label{eq:dsigOSS}\end{multline}
and for a pair of fermions
\begin{multline}
  \frac{\dd \sigma}{\dd k_x^2 \, \dd k_x^0 \, \dd \cos\theta_x \, \dd k_\Delta^0 \, \dd \phi_\Delta} = \frac{\alpha^3 \alpha_x \epsilon^2 \chi}{2\pi^2} \, \frac{1}{(k_x^2 - m_x^2)^2} \\
  \times \bigl[ \bigl( \abs{N_{F,ij}}^2 k_x^2 - 4 m_1 m_2 (\Re \, N_{F,ij}^2) \bigr) \, f(x_e, k_x^2) - 2 \abs{N_{F,ij}}^2\Omega(k_\Delta) \bigr] \,.
\label{eq:dsigOSF}\end{multline}
The various $N$s are the combinations of mixing matrices that appear in the three-point vertices $X^\mu$--$X^\nu$--$h_a^x$, $X^\mu$--$h_a^x$--$A^x$ and $X^\mu$--$\chi_i^x$--$\chi_j^x$, respectively.  In practice we need only consider $a=1$ and $i=1$, $j\in \{1, 2\}$.  They are explicitly given by  
\begin{equation}
  \begin{split}
    N_{H,a} & = \sin\zeta \, R_{1a} + \cos\zeta , R_{2a} \,, \\
    N_{S,a} & = \cos \zeta \, R_{1a} - \sin \zeta \, R_{2a} \,, \\
    N_{F,ij} & = P_{i1} P_{j1}^\ast - P_{i2} P_{j2}^\ast \,.
  \end{split}
\label{eq:couplings}\end{equation}
$R$ is the matrix that relates the real hidden scalar mass and gauge bases, as defined in \modeqref{eq:hmix}.  $P$ is the equivalent matrix for the hidden fermions~\cite{Chan:2011aa}.  We have used $x_e$ and $U$ defined as in the on-shell case with $m_x^2 \to k_x^2$.  The $k_\Delta$-dependence comes through the function
\begin{multline}
  \Omega(k_\Delta) = \frac{(1-x_e)}{U^3} \, (m_1^2 - m_2^2) \, \bigl( (1 - x_e) \, p_e \cdot k_\Delta + (p_e - q) \cdot k_\Delta - (m_1^2 - m_2^2) \bigr) \\
  - \frac{(1 - x_e)^2}{U^4} \, k_x^2 \, \Bigl( (p_e \cdot k_\Delta)^2 + \bigl( (p_e - q) \cdot k_\Delta \bigr)^2 - (m_1^2 - m_2^2) \, (2p_e - q) \cdot k_\Delta \Bigr) .
\end{multline}
In the limit that the WW approximation is valid, $(p_e - q) \cdot k_\Delta \approx p_e \cdot k_\Delta$; while
\begin{equation}
  p_e \cdot k_\Delta = E_e \bigl( k_\Delta^0 - \lvert\vec{k}_\Delta\rvert (\cos \theta_x \cos\theta_\Delta - \sin\theta_x \sin\theta_\Delta \cos\phi_\Delta ) \bigr) \,,
\end{equation}
with $\theta_\Delta$ the angle between $\vec{k}_x$ and $\vec{k}_\Delta$.

\subsection{Expected Number of Signal Events}\label{sec:sigmatoN}

In the previous two sections we gave the differential cross sections for the production of hidden sector states at beam dump experiments.  Here we relate those expressions to the actual experimental observables, specifically the number of events seen in the detectors.  We first relate the cross sections to the number of hidden sector particles actually produced, and then discuss the acceptance factors that give the number of signals per hidden sector event.

For a fixed target experiment, we must account for interactions throughout the full depth of the target material.  This must also include energy loss from the electron beam as it travels through the target.  In particular, the electron energy in \modeqref{eq:dsigdxdth} and similar expressions satisfies $E_e \leq E_0$, the incident beam energy.  This effect is handled by convolving the production cross section with an attenuation function.  We have the following expression for the distribution of the number of hidden vectors produced:
\begin{equation}
  \frac{d N}{d x_0} = N_e \, \frac{N_0 X_0}{A} \int_{E_x + m_e}^{E_0} dE_e \int_0^T ds \, I(E_0, E_e, s) \, \frac{E_0}{E_e} \, \frac{d\sigma}{dx_e} \, .
\label{eq:diffNe}\end{equation}
Here, $N_e$ is the total number of electrons delivered to the target; $N_0$ is Avogadro's number; $A$ is the atomic number and $X_0$ the unit radiation length (M/L$^3$) of the target material; $s$ is the depth in radiation lengths within the target, of total thickness $T$; and $x_{0,e} = E_x/E_{0,e}$.

The attenuation function of \modeqref{eq:diffNe} gives the energy distribution of an initiallly monochromatic electron beam after travelling $s$ radiation lengths through the target material.  We follow \refscite{Tsai:1986tx,Andreas:2012mt} in taking\footnote{For an alternate but similar parameterisation, see \refscite{Bjorken:2009mm,Izaguirre:2013uxa}.}
\begin{equation}
  I(E_0, E_e, s) = \frac{1}{E_0} \frac{\biggl[ \ln \biggl( \frac{E_0}{E_e} \biggr) \biggr]^{bs - 1}}{\Gamma (b s)} ,
\end{equation}
with $\Gamma$ the Gamma function and $b = \frac{4}{3}$.  Since $\sigma \propto E_e^2$ (see \modeqref{eq:dsigdxdth}), production is dominated from electrons with the initial beam energy; and from $I$, this mostly occurs in the first few radiation lengths of the target.

We define an average acceptance number that is the ratio of the number of signal events to the total number of hidden sector events computed above.  There are two steps in deriving these factors.  First, we need to infer the kinematic distributions  for the stable or metastable states from the differential cross sections of \secrefs{sec:onX}{sec:offX}.  
Next we must determine either what fraction of metastable states' decay products will give visible signals; or what fraction of hidden sector particles will scatter in the detector, depending on the nature of the experiment.

We compute these using a simple Monte Carlo simulation. First we generate events drawn from the appropriate one of the distributions in \foureqref{eq:dsigdxdth}{eq:higgsstr}{eq:dsigOSS}{eq:dsigOSF}.  For this purpose, an event is defined by the four-momenta of one or two on-shell hidden states.  When the hidden vector has on-shell decays to the final states of interest, we generate its kinematics using \modeqref{eq:dsigdxdth}; otherwise we use the appropriate one of \threeqref{eq:higgsstr}{eq:dsigOSS}{eq:dsigOSF} and generate both hidden states' four-momenta.

For on-shell processes, we select the $Z^x$ energy fraction $x_e$ from \modeqref{eq:dsigdx} first, then the polar angle $\theta_x$ given $x_e$ from \modeqref{eq:dsigdxdth}.  To select the variables, we generate candidate values randomly within the physically allowed range, then keep those values with probability given by the ratio of the kinematic distribution to its maximum value.  For example, a candidate value for $x_e$ is kept with probability
\begin{equation}
  P^{keep} = \left. \frac{d\sigma}{dx_e}(x^{cand}_e) \middle/ \frac{d\sigma}{dx_e} (x^{max}_e) \right. .
\end{equation}
When choosing to keep or reject $\theta_x$, we maximise \modeqref{eq:dsigdxdth} for fixed $x_e$.  The azimuthal angle is selected randomly, as the spin-averaged cross sections do not depend on it.

For off-shell processes, we first generate the hidden vector four-momenta, then use the appropriate one of \threeqref{eq:higgsstr}{eq:dsigOSS}{eq:dsigOSF} to simultaneously generate the remaining two non-trivial kinematic variables $k_\Delta^0$ and $\phi_\Delta$.  Generating $k_x$ is done as for on-shell processes, except we must first select the unconstrained $k_x^2$ using the distribution (derivable from \twoeqref{eq:wwonshell}{eq:offshell})
\begin{equation}
  P(k_x^2) \propto \frac{\sqrt{k_x^2}\,\Gamma_{HS} (k_x^2)}{(k_x^2 - m_x^2)^2} \, \sigma(k_x^2) \,.
\end{equation}
Again, there is no dependence on the $k_x$ azimuthal angle, and other kinematics are constrained by requiring final states by on-shell, \modeqref{eq:kDelcon}.

We then decay through the hidden sector till we reach the stable and metastable states.  In doing so we use the narrow width approximation, treating all decays as prompt and isotropic in the parent rest frame.  We also track all branching ratios as necessary, though in most cases these are trivial.  This effectively gives us the (meta-)stable particle kinematics.

The subsequent steps then depend on the nature of the experiment.  For searches based on visible sector particles, there is another branching ratio; the experiments we will discuss triggered only on electrons, suppressing sensitivity for hidden sector masses above the muon threshold.  The acceptance factor is the product of all relevant branching ratios, the probability that the metastable state decays within the decay volume, and that its visible daughter particles hit the detector with sufficient energy to pass the threshold.  We compute this on an event-by-event basis.  The decay probability is 
\begin{align}
  P_{dec} & = \exp\biggl[ - \frac{L_{sh}}{\gamma v_z \tau} \biggr] \, \biggl( 1 -  \exp\biggl[ - \frac{L_{dec}}{\gamma v_z \tau} \biggr] \biggr) \notag \\
  & = \exp\biggl[ - \frac{L_{sh}}{\tau} \, \frac{m}{p_z}\biggr] \, \biggl( 1 -  \exp\biggl[ - \frac{L_{dec}}{\tau} \, \frac{m}{p_z}\biggr] \biggr) \,, \label{eq:pdec}
\end{align}
with $L_{sh}$ and $L_{dec}$ as shown in \figref{fig:beamdump}, $v_z$ and $p_z$ are the parent's velocity and momentum along the beam direction, $m$ is its mass and $\tau$ its lifetime.  The angular acceptance probability is found by extending our Monte Carlo, decaying the metastable states within the decay volume and counting what fraction of daughter particles pass the experimental cuts.  To this end, we model the experiments as cylinders of radius $r_{Acc}$ with hard energy threshold cuts at $E_{thr}$.  The final number of signal events is
\begin{equation}
  N_{sig} = N \times \frac{1}{EV} \sum_{i \in HIT} \text{Br}_{tot}^i \times P_{dec}^i \,,
\end{equation}
where $EV$ is the total number of events we generate, and $HIT$ the set of metastable states with daughter particles observed in the detector.

For the proposed search of \refcite{Izaguirre:2013uxa} based on invisible particles, the remaining factors are much simpler.  The angular acceptance is directly found from the kinematic distribution of metastable states, asking what fraction strike the detector (including, where necessary, a survival probability).  This is multiplied by the event-by-event probability that the hidden sector state scatter within the detector:
\begin{equation}
  P_{scat} = \frac{\rho}{m_N} \, L_d \, \sigma_{\chi N} \,,
\label{eq:pscat}\end{equation}
with $\rho$ the target density, $m_N$ the nucleon mass, $L_d = 1$~m the detector depth as in \figref{fig:beaminvis} and $\sigma_{\chi N}$ the total hidden sector-nucleon scattering cross section summed over final states.  This assumes that multiple scattering and attenuation of the dark sector beam is negligible ($P_{scat} \ll 1$).  Scattering to different hidden sector states often dominates due to mixing matrix suppression of the elastic process.  For this reason we neglect scattering from electrons, where the inelastic channels are usually kinematically forbidden.

Full details on the differential scattering cross sections are given in \appref{app:scatter}.  When the incident particle is relativistic we may approximate~\cite{Izaguirre:2013uxa}
\begin{equation}
  \sigma_{\chi N} \sim 4\,\pi\, \epsilon^2 \, \alpha \, \alpha_x \, N_I^2 \, \frac{Z}{A} \, \frac{1}{\mu^2} \,,
\label{eq:scatxsec}\end{equation}
with
\begin{equation}
  \frac{1}{\mu^2} = 
  \begin{cases}
    (Q_{max}^2-Q_{min}^2)/m_x^4 & m_x^2 \gg Q_{max}^2 \,, \\
    1/m_x^2 & Q_{min}^2 \lesssim m_x^2 \lesssim Q_{max}^2 \,, \\
    1/Q_{min}^2 & m_x^2 \ll Q_{min}^2 \,.
  \end{cases}
\end{equation}
$Q^2$ is the momentum transfer in the scattering process, with $Q_{min} \sim 140$~MeV and $Q_{max} \sim 1$~GeV (for the JLab experiment).  $N_I$ is the mixing matrix factors for the three-point coupling of the hidden sector state to $X_\mu$, summed over possible final states.  In the case when all hidden sector states are accessible, we have
\begin{equation}
  N_I^2 = 
  \begin{cases}
    1 & \text{ for } A^x, \ h_a^x \,, \\
    \abs{P_{i1}}^2 + \abs{P_{i2}}^2  & \text{ for } \chi_i^x \,.
  \end{cases}
\label{eq:scatmixmat}\end{equation}
We take the signal yield as 
\begin{equation}
  N_{sig} = N \times \frac{1}{EV} \sum_{i \in HIT} \text{Br}_{tot}^i \times P_{scat}^i \times P_{surv}^i \,,
\label{eq:neinv}\end{equation}
with $EV$ the number of simulated particles, $HIT$ the set of hidden particles with trajectories intersecting the detector, and $P_{surv}$ the probability of that state not decaying before passing through it.

\subsection{Current and Prospective Experiments}\label{sec:Experiments}

Let us first briefly review the experiments that have been used or proposed to set limits on low energy hidden sectors.

\paragraph{Previous Beam Dumps} Several 
previous
beam dump experiments have been used to set limits on hidden vectors decaying directly to leptons~\cite{Andreas:2012mt,Bjorken:2009mm}.   These include E137~\cite{Bjorken:1988as} and E141~\cite{Riordan:1987aw} at SLAC, E774~\cite{Bross:1989mp} at Fermilab, KEK~\cite{Konaka:1986cb}, and Orsay~\cite{Davier:1989wz}.  We use the experimental parameters as given in \tabref{tab:ebeamexp}.  The energy thresholds are taken from the relevant experimental papers; they were usually ignored in previous work, as they are trivial for searches based on $Z^x \to e^+ e^-$.  For our more complex decay chains they must be included.  Note that all these experiments were only sensititve to $e^\pm$, except for E141 which was only sensitive to positrons.  We demand that the number of expected events be less than $N_{95\%}$.

\begin{table}
  \centering
  \begin{tabular}{ccrrrrrrr}
    Experiment & Target & $E_0$ & $N_e$ & $L_{sh}$ & $L_{dec}$ & $E_{thr}$ & $r_{Acc}$ & $N_{95\%}$ \\
    \hline
    E137 & Al & 20 & $1.87 \times 10^{20}$ & 179 & 204 & 2 & 1.5 & 3 \\
    E141 & W & 9 & $2 \times 10^{15}$ & 0.12 & 35 & 4.5 & 0.0375 & 3419 \\
    E774 & W & 275 & $5.2 \times 10^9$ & 0.3 & 2 & 27.5 & 0.1 & 18 \\
    KEK & W & 2.5 & $1.69 \times 10^{17}$ & 2.4 & 2.2 & 0.1 & 0.047 & 3\\
    Orsay & W & 1.6 & $2 \times 10^{16}$ & 1 & 2 & 0.75 & 0.15 & 3 \\
    JLab & Al & 12 & $10^{20}$ & 10 && & 1\\
    \hline
  \end{tabular}
  \caption{Parameters of the electron beam dump experiments we consider.  The parameters $E_0$ and $E_{thr}$ are in GeV, while $L_{sh}$, $L_{dec}$ and $r_{Acc}$ are in metres.}\label{tab:ebeamexp}
\end{table}

\paragraph{Present and Future Visible Searches} A number of experiments are currently running or have recently been proposed to search for low energy hidden sectors.  This includes MAMI~\cite{Merkel:2011ze}, APEX~\cite{Essig:2010xa,Abrahamyan:2011gv}, HPS~\cite{HPS,2013arXiv1301.1103H}, a proposal at the CERN SPS~\cite{Gninenko:2013rka,Andreas:2013lya}, and DarkLight~\cite{Freytsis:2009bh}.  Regretably, these experiments do not set any limits on our model beyond those that apply in Case A.  For APEX, MAMI and HPS, a crucial tool used to discriminate signal from background is the requirement that the observed electron-positron pair have combined energy equal to the beam energy.  This is based on the sharp peak in the cross sections at $x_e \approx 1$ in \modeqref{eq:dsigdxdth}.  However, if the hidden vector decays to hidden sector particles---or is not produced on-shell---then the initial beam energy will be distributed among a larger number of particles.  Hidden sector events in our model will then not pass the signal cuts.

The proposed CERN SPS search uses different discriminators, but with a similar philosophy.  Calorimeters will be positioned before and after the decay volume, and the total energy in them must sum to the incident beam energy.  For the small scale of the decay volume ($\sim 5$~m) the hidden Higgs is effectively stable, so all non-trivial hidden sector events will involve irreducible missing energy from either \hx\ or $\chi_1^x$.  We thus expect that no events will pass the experimental cuts.  It is also planned to search for invisible decays~\cite{Andreas:2013lya}; this likely would have some sensitivity to cases B--D, but we defer a full investigation of this for a future paper.

For DarkLight, the situation is slightly different.  DarkLight uses a peak in the electron-positron invariant mass spectrum to discriminate the signal.  In our model, the pseudoscalar $A^x$ and fermion $\chi_2^x$ have three-body decays, so will not pass this cut.  The scalar \hx\ \emph{will} pass the cut; however, DarkLight requires a relatively prompt decay.  In the mass range to which DarkLight is senstive (less than 100~MeV), the scalar has a decay length of $\gtrsim 10$~m for $\epsilon \lesssim 0.1$.  It follows that there will be no sensitivity below this kinetic mixing.

\paragraph{JLab Inivisible Search} The proposed search at JLab~\cite{Izaguirre:2013uxa} would have the experimental parameters listed in \tabref{tab:ebeamexp}.  Note that for this experiment, the proposed detector would have a square cross section so $r_{Acc}$ should be interpreted as the side of that square.  We follow the original proposal in taking three different choices for sensitivity, corresponding to different possibilities for background subtraction.  In order from least to most optimistic, we consider sensitivity to 20000, 1000 and 40 events respectively.

\subsection{Limits}\label{sec:limits}

\begin{figure}
  \centering
  \includegraphics[width=0.49\textwidth]{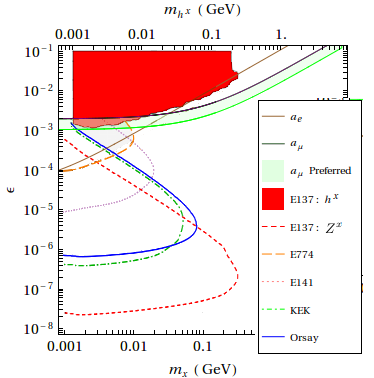}\includegraphics[width=0.5\textwidth]{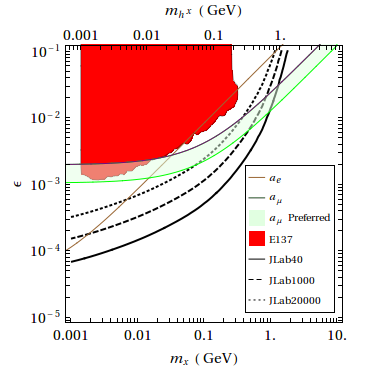}
  \caption{Exclusion plots from electron beam dumps for benchmark slopes A (left) and B (right).  The constraints and preferred region from the electron and muon $g-2$ are shown for scale.    For A, the unshaded contours are prior limits from $Z^x \to e^+ e^-$, and the solid red region is the new limits from $\hx \to e^+ e^-$ at the E137 experiment.  For B, the solid region shows the same $\hx \to e^+ e^-$ limit, and the thicker contours mark 40, 1000 and 20000 expected $\chi_1^x$ scattering events at the JLab search.  See the text for more details.}\label{fig:elec-AB}
\end{figure}

\begin{figure}
  \centering
  \includegraphics[width=0.5\textwidth]{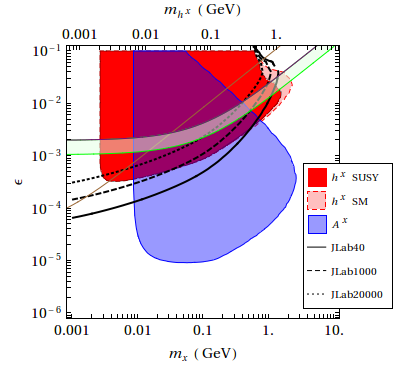}\includegraphics[width=0.48\textwidth]{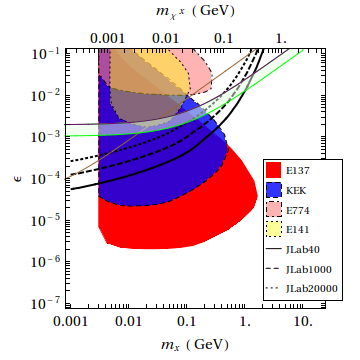}
  \caption{Exclusion plots from electron beam dumps for benchmark slopes C (left) and D (right).  The limits from the electron and muon $g-2$ are shown with the same notation as in \figref{fig:elec-AB}, but are omitted from the keys for space.  For C, the red regions denote E137 limits from $\hx \to e^+ e^-$ and the blue region from $A^x \to \hx \, e^+ e^-$; the solid contours denote the expected number of $h_1^x$ scattering events at JLab.  For D, the shaded regions are limits from $\chi_2^x \to \chi_1^x\, e^+ e^-$ at different experiments, and the solid contours the total number of expected $\chi_1^x$ and $\chi_2^x$ scattering events at JLab.  See the text for more details.}\label{fig:elec-CD}
\end{figure}

We now present current and prospective limits for our four benchmark scenarios.  For clarity, we show only limits from electron beam dump experiments, as well as from the electron and muon anomalous magnetic moments (to aid in comparisons with other results).  For composite plots showing all limits we find, see \secref{sec:conc}.  We show limits in the $m_x$--$\epsilon$ plane for benchmarks A and B in \figref{fig:elec-AB}, and benchmarks C and D in \figref{fig:elec-CD}.  For benchmarks A through C, we also show the mass of the lightest scalar \hx\ on the upper horizontal axis; for benchmark D, we show the mass of the lightest fermion $\chi_1^x$.

The general features of all exclusion plots from visible searches are straightforward.  At large kinetic mixing, the exclusion contours are set when the lifetime of the metastable particle becomes too short.  They are approximately contours of $\epsilon^2 m_x$, except near particle thresholds.  At small kinetic mixing, the boundary is mainly set when the production cross section becomes too small, which is only weakly dependent on mass.  Finally, there is a lower limit on all these searches when the long-lived particle can no longer decay to electrons.  While decays to photons might fake electrons in the experimental detectors, the lifetimes are highly suppressed with the result that no limits from these searches can be set.

\paragraph{Case A} This benchmark corresponds to the previously-studied case, where the hidden vector may only decay to the visible sector.  We therefore include past limits for electron beam dumps taken from \refcite{Andreas:2012mt} in \figref{fig:elec-AB}.  We find no new limits beyond those.  Additional constraints could in principle arise from the production of hidden sector scalars or fermions through an off-shell vector; but  a combination of kinematic and coupling suppressions result in this not being the case.  The strongest such limits we find come from long-lived \hx\ decays at the E137 experiment.  These are shown as the red shaded region in \figref{fig:elec-AB}, but are weaker than the current constraints from $a_e$.\footnote{This was not the case for the previous limits~\cite{Pospelov:2008zw,Odom:2006zz}.}

\paragraph{Case B} Similarly to Case A, there are no new limits from previous experiments.  The limits from long-lived scalars that would exist are comparable to those in Case A, and inferior to those from $a_e$.  However, we now have promising prospective limits from the JLab search.  Even in the most pessimistic background scenario, we can extend the exclusion contours to exclude much of the region preferred by $a_\mu$.   This is unsurprising; the vector here decays invisibly to the stable $\chi_1^x \chi_1^x$ final state, precisely the motivating case for that search.  The limits here are not directly comparable to the plots of \refscite{Izaguirre:2013uxa,Essig:2013vha}: that work fixed the mass of the stable hidden particle and varied the vector mass, while we vary both with a fixed mass ratio.

\paragraph{Case C} This benchmark offers a number of interesting new and prospective limits, that go beyond the constraints from $a_e$ and $a_\mu$.  The improved reach is because we can produce unstable hidden sector states through an \emph{on-shell} hidden vector.  The phenomenology is richer because the vector decays to two particles, \hx\ and $A^x$, that are both potentially long-lived.  

We first address the limits from searches for visible decays.  In \figref{fig:elec-CD} we only show the limits from E137; the limits from the other experiments listed in \tabref{tab:ebeamexp} are strictly inferior.  We show separately the limits from the scalar (red) and pseudoscalar (blue) decays.  The scalar limits are sensitive to the Higgs mixing induced from $D$-terms as discussed in \secref{sec:hxdecays}.  The limits with (without) this mixing are shown in darker red with solid boundary (lighter red with dashed boundary), and labelled ``SUSY'' (``SM'') in the key.  The two cases are equivalent below the muon threshold, as the mass mixing is unimportant there (see \figref{fig:h1ctau}).  Above the muon threshold, the limits with the mixing are weaker due both to the hidden scalar decaying too quickly and the branching ratio to electrons is being more strongly suppressed.

The limits from pseudoscalar decays extend to lower kinetic mixings than those from the scalar decays.  This is because the $A^x$ width is less suppressed than the \hx, so smaller values of $\epsilon$ are needed to have macroscopic decay lengths.  There are two regions where the scalar limits are superior.  At low mass, the threshold for the decay $A^x \to \hx e^+ e^-$ is crossed before that for $\hx \to e^+ e^-$; while at high mass and moderate $\epsilon$, the pseudoscalar is too short-lived to produce an observable signal.  Additionally, the limits from the pseudoscalar are less generic for two reasons.  First, it is possible that the decay $A^x \to \chi^x_1 \chi^x_1$ may be allowed and the corresponding scalar decay forbidden.  Second, if CP is violated in the hidden sector and we choose a value of $\tan\zeta$ slightly closer to 1, the decay $A^x \to \hx\hx$ may become relevant.

Finally, we have the prospective limits from the JLab invisible search.  These limits are dominated from the scattering of the scalar \hx\ in the detector.  As such, while these limits are generically weaker than the pseudoscalar limits from E137, they might still be relevant if the latter are evaded as discussed above.

\paragraph{Case D} This benchmark has non-trivial limits from both visible and invisible searches.  Both come from production of an on-shell vector that decays to $\chi_2^x \chi_1^x$.  Visible searches look for the decay $\chi_2^x \to \chi_1^x e^+ e^-$, suppressed for the same reasons (phase space and $\epsilon^2$) as the pseudoscalar in Case C.  This slope is the only one of our four where visible searches \emph{other} than E137 set limits, though that experiment still gives the strongest constraints.  Of the five experiments in \tabref{tab:ebeamexp}, the limits from the Orsay experiment are strictly inferior to those from the KEK experiment, so we do not show them in \figref{fig:elec-CD}.

Invisible searches for this benchmark are most sensitive to scattering of the $\chi_1^x$ in the detector.  Of the two vector decay modes to the lightest fermion, the $\chi_2^x \chi_1^x$ channel dominates due to its larger branching ratio.  While there remains a region of parameter space where $a_\mu$ can be explained that is not excluded by visible searches, it can be excluded by the JLab search.

\section{Hadronic Fixed Target Experiments}\label{sec:hbeam}
Hadronic fixed target experiments can also be used to constrain low-energy hidden sectors~\cite{Batell:2009di,Schuster:2009au,deNiverville:2011it,Gninenko:2012eq,deNiverville:2012ij,Blumlein:2013cua,Blumlein:2011mv,Gninenko:2011uv}.  This is particularly relevant for hidden sectors with GeV-scale masses, where most constraints from electron experiments are limited by kinematics.  As with electron beam dumps, most previous studies assumed the hidden vector either decays directly to the SM or to stable invisble particles.  When long-lived scalars \emph{were} considered~\cite{Batell:2009di,Schuster:2009au}, they were required to decay only through hidden vector loops.  The possible limits from hidden sector pseudoscalars and fermions have not been studied.

\begin{figure}[ttt]
  \centering
  \includegraphics[width=0.7\textwidth]{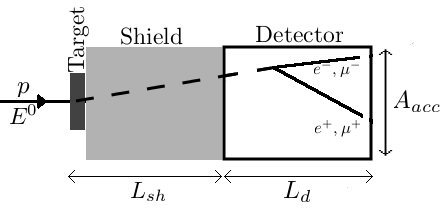}
  \caption{Experimental configuration for hadronic beam dump searches for visible decays.  Note that the detector may be on the beam axis (as shown here) or off it.}\label{fig:hbvis}
\end{figure}

The searches for visible particles discussed in this section have the general configuration shown in \figref{fig:hbvis}.  This is similar to \figref{fig:beamdump}, but the decay volume and detector are combined.  Searches for the scattering of invisible particles have the same configuration of \figref{fig:beaminvis}, with an incident beam of protons rather than electrons, and some of the experiments we consider placed the detector off the beam axis.

We discuss production of hidden sectors at these experiments in \secref{sec:hadprod} and the calculation of acceptances in \secref{sec:hadacc}.  This includes review of material covered elsewhere, both for completeness, and to extend some expressions for our more general hidden sector.  In \secref{sec:hadexpt} we discuss relevant experiments.  The reader who is 
mainly interested in results should skip to \secref{sec:hadlimits}, where we present and discuss the exclusions we find for our four benchmark slopes.

\subsection{Production}\label{sec:hadprod}

Hidden sector production in these experiments can roughly be divided into two domains~\cite{Batell:2009di}, according to the hidden sector mass scale.  If the hidden states have masses below $\Lambda_{QCD}$, then production will mainly occur through the decay of mesons and baryons produced in the initial collision; we term this the ``low mass'' region.  In the complementary ``high mass'' region, we can resolve the quark content of the proton and produce an $s$-channel hidden vector through $q\bar{q}$ fusion.  
\begin{figure}[ttt]
  \begin{center}
    \begin{fmffile}{mesHS}
      \begin{fmfgraph*}(80,50)
        \fmfstraight
        \fmfleft{i1} \fmfright{o1,o2}
        \fmf{dbl_plain,tension=2}{i1,v1}
        \fmfblob{.08w}{v1}
        \fmf{dashes}{v1,o2}
        \fmf{boson,tension=2,label=$\gamma$}{v1,v2}
        \fmf{boson,tension=2}{v2,o1}
        \fmfv{decoration.shape=cross,decoration.angle=60,decoration.size=8,label=$\epsilon$,label.angle=60}{v2}
        \fmflabel{$M$}{i1}
        \fmflabel{$M'$}{o2}
        \fmflabel{$Z^{x(\ast)}$}{o1}
      \end{fmfgraph*}
    \end{fmffile}
    \qquad
    \begin{fmffile}{rhoHS}
      \begin{fmfgraph*}(130,50)
        \fmfstraight
        \fmfleft{i1} \fmfright{o1,o2}
        \fmf{dbl_wiggly,tension=2}{i1,v1}
        \fmf{boson,label=$\gamma$,tension=2}{v1,v2}
        \fmf{boson,label=$Z^x$,tension=2}{v2,v3}
        \fmf{plain}{o1,v3,o2}
        \fmfv{decoration.shape=tetragram,decoration.size=8,filled=full,decoration.angle=45}{v1}
        \fmfv{decoration.shape=cross,decoration.size=8,label=$\epsilon$,label.angle=90}{v2}
        \fmflabel{$V$}{i1}
        \fmflabel{$\chi^x,h^x$}{o1}
        \fmflabel{$\chi^x,A^x,Z^x$}{o2}
      \end{fmfgraph*}
    \end{fmffile}
  \end{center}
  \caption{Hadron decays to the hidden sector.  (Left): for any SM decay $M \to M' + \gamma$, there is a hidden sector companion decay $M \to M' + Z^{x(\ast)}$ through the vector kinetic mixing.  (Right): a vector meson mixes with the hidden vector through their mutual mixing with the photon, allowing it to decay to two hidden sector particles.}\label{fig:mes}
\end{figure}

The low mass regime was previously considered in \refcite{Batell:2009di}.  We extend some of their results to include new decay modes, in particular \modeqref{eq:rhodec}.  
Production of hidden sector states occurs primarily
through the secondary decay of hadrons produced at the interaction point.  These decays can proceed either through kinetic mixing with the hadron decay product or with the hadron itself.  In the former class, for any hadron $M$ with an electromagnetic decay of the form $M \to \gamma + \ldots$, there will be a companion decay $M \to Z^{x(\ast)} + \ldots$.  In the second, a vector meson can mix with the $Z^x$ \emph{e.g.} through a mutual mixing with the photon, allowing it to decay into the hidden sector.  These possibilities are illustrated in \figref{fig:mes}. 

The most important example of the first category is the diphoton decay of a neutral pion.  This has maximal branching ratio, and pions are produced in abundance in any hadron experiment.  To extend the reach above the pion mass, we also consider the diphoton decays of the heavier pseudoscalars $\eta$, $\eta'$; and the 
baryonic decay $\Delta \to N \gamma$.  When the hidden vector is on-shell, its production has branching ratio
\begin{equation}\label{eq:onbr}
  \begin{split}
    \text{Br} (M \to \gamma Z^x) & = 2 \, \epsilon^2 \, \biggl( 1 - \frac{m_x^2}{m_M^2}\biggr)^3 \, \text{Br} (M \to \gamma\gamma) , \\
    \c(\Delta \to N Z^x) & = \epsilon^2 \, \biggl( 1 - \frac{m_x^2}{(m_\Delta - m_N)^2}\biggr)^{3/2} \, \text{Br} (\Delta \to N \gamma) \,.
  \end{split}
\end{equation}
This assumes that the additional decay does not change the total width, a good approximation for $\epsilon \lesssim 0.1$.  We also have the possibility of decays involving an off-shell hidden vector; these have branching ratio
\begin{equation}
  \text{Br}_{HS} (M \to M' Z^{x\ast}) = \frac{1}{\pi} \, \epsilon^2 \int_{m_{HS}^2}^{m_M^2} dq^2 \, \frac{\sqrt{q^2} \, \Gamma_{HS} (q^2)}{(q^2 - m_x^2)^2} \, \text{Br}_2 (m_x^2 = q^2) ,
\end{equation}
where $\Gamma_{HS}$ is the partial width from \appref{app:vecwidth} appropriate to the hidden final states; $m_{HS}$ is the sum of those masses; and Br$_2$ refers to the on-shell expression from \modeqref{eq:onbr} with $m_x^2$ replaced by $q^2$.

The second class of hadronic decays into the hidden sector involve the mixing of a vector meson with the hidden vector.  This is exemplified by the $\rho$ meson.  As is well-known, the $\rho$ mixes with the photon; successive insertions of the $\rho$--$\gamma$ and $\gamma$--$X^\mu$ mixings then lead to a $\rho$--hidden vector mixing, which allows $\rho$ to decay to hidden states.  This leads to a $\rho$ branching ratio
\begin{equation}
  \text{Br}(\rho \to HS) = \frac{12\epsilon^2}{\alpha} \, \text{Br}(\rho \to e^+ e^-) \, \frac{m_\rho^3 \, \Gamma_{HS} (m_\rho^2) }{(m_\rho^2 - m_x^2)^2 + m_x^2 \Gamma_x^2} \, .
\label{eq:rhodec}\end{equation}
The propagator is regulated by the total hidden vector width $\Gamma_x$, which is relevant as $m_x \to m_\rho$.  Analogous results hold for other neutral vector mesons.

The total number of hidden particles produced is proportional to the number of different hadron species produced.  When possible, we use measurements or estimates of this from the experimental collaborations themselves.  Otherwise, we model the production as dominantly in the first interaction length at full beam energy.  This lets us relate the hadron production cross sections and yields using a simplified version of \modeqref{eq:diffNe}:
\begin{align}
  N_M & \approx \sigma_N \, \frac{N_0 X^\text{nuc}}{A} \, N_p \,,
\label{eq:Nhad}\end{align}
where $N_p$ is the total number of protons on target, $X^{\text{nuc}}$ is the nuclear interaction length, and $\sigma_N$ is the per-nucleus cross section.  We then have $N_X = \text{Br}(M \to X) \times N_M$, summed over hadron flavour $M$.

\begin{figure}[ttt]
  \centering
  \begin{fmffile}{qqhA}
    \begin{fmfgraph*}(85,45)
      \fmfleft{i1,i2} \fmfright{o1,o2} \fmfbottom{b1}
      \fmf{fermion}{i1,v1,i2}
      \fmf{boson}{v1,v2,v3}
      \fmf{dashes}{o1,v3,o2}
      \fmflabel{(a)}{b1}
      \fmflabel{$q$}{i1}
      \fmflabel{$\bar{q}$}{i2}
      \fmflabel{$A^x$}{o1}
      \fmflabel{$h^x$}{o2}
      \fmfblob{0.04w}{v1}
      \fmfv{decoration.shape=cross,decoration.size=8,label.angle=90,label=$\epsilon$}{v2}
      \fmfblob{0.04w}{v3}
    \end{fmfgraph*}
  \end{fmffile}
  \quad
  \begin{fmffile}{qqxx}
    \begin{fmfgraph*}(85,45)
      \fmfleft{i1,i2} \fmfright{o1,o2} \fmfbottom{b1}
      \fmf{fermion}{i1,v1,i2}
      \fmf{boson}{v1,v2,v3}
      \fmf{plain}{o1,v3,o2}
      \fmflabel{(b)}{b1}
      \fmflabel{$q$}{i1}
      \fmflabel{$\bar{q}$}{i2}
      \fmflabel{$\chi^x$}{o1}
      \fmflabel{$\chi^x$}{o2}
      \fmfblob{0.04w}{v1}
      \fmfv{decoration.shape=cross,decoration.size=8,label.angle=90,label=$\epsilon$}{v2}
      \fmfblob{0.04w}{v3}
    \end{fmfgraph*}
  \end{fmffile}
  \quad 
  \begin{fmffile}{qqhZ}
    \begin{fmfgraph*}(85,45)
      \fmfleft{i1,i2} \fmfright{o1,o2} \fmfbottom{b1}
      \fmf{fermion}{i1,v1,i2}
      \fmf{boson}{v1,v2,v3,o1}
      \fmf{dashes}{v3,o2}
      \fmflabel{(c)}{b1}
      \fmflabel{$q$}{i1}
      \fmflabel{$\bar{q}$}{i2}
      \fmflabel{$Z^x$}{o1}
      \fmflabel{$h^x$}{o2}
      \fmfblob{0.04w}{v1}
      \fmfv{decoration.shape=cross,decoration.size=8,label.angle=90,label=$\epsilon$}{v2}
      \fmfblob{0.04w}{v3}
    \end{fmfgraph*}
  \end{fmffile}\\
  \vspace{9pt}
  \caption{Feynman diagrams for the three parton-level processes that produce two hidden sector particles at hadronic beam dumps in the high mass regime.  All proceed through an $s$-channel hidden vector to produce hidden scalars (a), fermions (b) or a scalar-vector pair (c).\label{fig:prod_feyn}}
\end{figure}

In the high mass regime, we can produce two-body hidden sector final states through either an on- or off-shell vector, through the processes shown in \figref{fig:prod_feyn}.  We have the scalar, fermion and (for the intermediate vector off-shell) scalar-vector final states.  If the vector is on-shell, the tree-level partonic cross section is
\begin{equation}
  \hat{\sigma}_q (\hat{s}) = \frac{4\pi^2\alpha}{3} \, (\epsilon Q_q)^2 \, \delta (\hat{s} - m_x^2) \, \text{Br}_{x} \,,
\label{eq:hadxsecon}\end{equation}
with Br$_x$ the hidden vector branching ratio to the final state of interest.  If the vector is off-shell, the equivalent expression may be written in terms of the vector partial widths from \appref{app:vecwidth}:
\begin{equation}
  \hat{\sigma}_q (\hat{s}) = \frac{4\pi\alpha}{3} \, (\epsilon Q_q)^2 \, \frac{\sqrt{\hat{s}} \, \Gamma_{HS} (\hat{s})}{(\hat{s} - m_x^2)^2} \,,
\label{eq:hadxsecoff}\end{equation}
with $\hat{s}$ the partonic center of mass energy.  We find the per-nucleon cross sections by convolving \twoeqref{eq:hadxsecon}{eq:hadxsecoff} with the CTEQ~10 parton density functions (pdfs)~\cite{Lai:2010vv} in the usual way,
\begin{equation}
  \sigma_{p,n} = \int dx_1 \, dx_2 \, \sum_{q} f_{q|p,n} (x_1) \, f_{\bar{q}|p,n} (x_2) \, \hat{\sigma}_q (x_1 x_2 s) \,.
\end{equation}
We convert to a per-nucleus cross section by assuming a simple incoherent scaling, $\sigma_N \approx Z \, \sigma_p + (A - Z) \sigma_n$.

\begin{figure}
  \centering
  \includegraphics[width=0.48\textwidth]{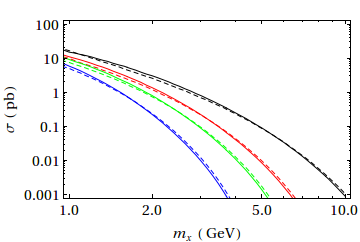} 
  \includegraphics[width=0.48\textwidth]{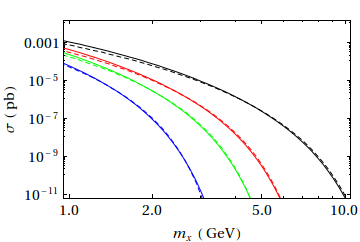}
  \caption{Cross sections $\sigma_p$ at hadronic beam dumps.  (Left): on-shell production in case C.  (Right): Higgsstrahlung production in case A.  Solid (dashed) lines are at tree level ($\alpha_S$).  From top to bottom, the black (red, green, blue) lines are for the CHARM (MINOS, U70, T2K) experiments (see \secref{sec:hadexpt}).  The cross sections $\sigma_n$ are not appreciably different.\label{fig:xsecplots}}
\end{figure}

Though we have named this the high-mass regime, we are concerned with particle masses of one to a few GeV.  This sets the energy scale for the partonic cross sections, and it is still close to $\Lambda_{QCD}$.  One might then be concerned about possible higher-order corrections in $\alpha_S$.  
We computed these terms, and show show the tree and one-loop cross sections for on- and off-shell processes in \figref{fig:xsecplots}.  These plots assume a factorisation scale $\mu_F = m_V$ (on-shell) or $\sqrt{\hat{s}}$ (off-shell); varying this represents an additional theoretical uncertainty.  This can be as large as an $\order{1}$ effect on the overall production rate~\cite{deNiverville:2012ij}.
We use the NLO cross sections in computing all limits, and note that our limits are self-consistently comparable where the low- and high-mass regimes meet.

As with hadron production, we obtain the total number of hidden states produced by assuming all interactions take place in the first interaction length with full beam energy.  The equivalent of \twoeqref{eq:diffNe}{eq:Nhad} is
\begin{align}
  N_X & \approx \sigma_N \, \frac{N_0 X^\text{nuc}}{A} \, N_p \,.
\end{align}

\subsection{Acceptances}\label{sec:hadacc}

As for electron beam dump experiments, the experimental acceptance is a product of three terms: an angular acceptance factor that the (meta-)stable particle hits the detector; a survival probability; and the probability of either decaying or scattering within the detector, as appropriate.  We operate with samples of hidden sector events rather than the kinematic distributions directly.

In the low mass region, we first construct a set of hadron events, then decay them through the hidden sector spectrum using the narrow width approximation.  Our samples are built on uniform grids in the parent hadron phase space, then weighted by the hadronic kinematic distributions.  Our choice of weighting function differs by experiment.  For LSND and MiniBooNE, we use parameterisations defined in the lab frame as functions of the meson momentum $p_{lab}$ and $\theta_{lab}$.  For LSND we use the Burman and Smith fit $f_{BS}$~\cite{Burman:1989ds}, appropriately weighted to account for the use of different target materials over the experiment's lifetime.  For MiniBooNE, we use the Sanford-Wang distribution$f_{SW}$~\cite{AguilarArevalo:2008yp}.  For all other experiments, we use the modified Bourquin-Gaillard parameterisation $f_{BG}$~\cite{Tel-Zur:1996gua,Bourquin:1975fx,Bourquin:1976fe}.  This is defined in terms of the CoM frame rapidity $y_{com}$ and transverse momentum $p_{T,com}$, so after constructing our initial grid we must boost to the lab frame.  Lastly, we decay each parent hadron to the hidden sector multiple times, so as to avoid a rare decay at a high-weight point unduly affecting our results.

In the high mass region, we use an unweighted sample of events construced with standard tools.  We implemented our model in FeynRules~2.0~\cite{Alloul:2013bka,Duhr:2011se,Degrande:2011ua}, and generate events using MadGraph~5.1.5.13~\cite{Alwall:2011uj} for the processes of \figref{fig:prod_feyn}.  Since we can produce any hidden sector particle directly in a $2\to 2$ process, we do not consider secondary production through the decay of heavier hidden sector particles.  This should be a small correction due both to suppressed production of heavier states, and the secondaries having larger production angles relative to the beam axis.

Once we have generated a sample of hidden sector events, the remaining analysis is analogous to \secref{sec:sigmatoN}.  For visible searches, we need modify the decay probability $P_{dec}$ of \modeqref{eq:pdec} by replacing $L_{dec}$ with $L_{d}$.  The angular acceptance is also simplified, as we only need the metastable state to cross the detector.  In the high mass region, the numer of signal events is given by
\begin{equation}
  N_{sig} = N \times \frac{1}{EV} \sum_{i \in HIT} \text{Br}_{tot}^i \times P_{dec}^i \,,
\end{equation}
with $EV$ the number of generated events and $HIT$ the subset of hidden sector particles that hit the detector, with daughter particles that pass the experimental cuts.  In the low mass region, the expression is slightly modified:
\begin{equation}
  N_{sig} = N \times \frac{1}{w_{tot}} \sum_{i \in HIT} w^i \times \text{Br}_{tot}^i \times P_{dec}^i \,,
\end{equation}
with $w^i$ the weighting associated with the event containing the particle $i$ and $w_{tot}$ the total weighting summed over all events.  The weights in turn are given by
\begin{equation}
  w^i = 
  \begin{cases}
    f_{BS} (p_{lab}^i, \theta_{lab}^i) \, \delta p_{lab} \, \delta \theta_{lab} & \text{LSND}, \\
    f_{SW} (p_{lab}^i, \theta_{lab}^i) \, \delta p_{lab} \, \delta \theta_{lab} & \text{MiniBooNE}, \\
    f_{GW} (y_{com}^i , p_{T,com}^i) \, \delta y_{com} \, \delta p_{T,com} & \text{Other experiments,}
  \end{cases}
\label{eq:weight}\end{equation}
with $p_{lab}^i$ \emph{etc.} the parent hadron kinematics, and $\delta p_{lab}$ \emph{etc.} the sampling interval in phase space.

For searches for dark matter scattering in the detector, the analysis here is almost identical to to the electron beam dump case.  There only modification of note refers to the LSND experiment, which searched for electron scattering with the final state electron in a given energy range.  This demands that we use the full scattering cross sections of \appref{app:scatter}, with \modeqref{eq:scatxsec} a poor approximation.  It is also useful to rewrite the scattering probability as
\begin{equation}
  P_{scat} = n_{e,N} \, L_d \, \sigma_{\chi e,N} \,.
\end{equation}
With this correction, the total signal yield in the high mass region is given by \modeqref{eq:neinv}.  In the low-mass region, we modify it to include the weighting,
\begin{equation}
  N_{sig} = N \times \frac{1}{w_{tot}} \sum_{i \in HIT} w^i \times \text{Br}_{tot}^i \times P_{scat}^i \times P_{surv}^i \,,
\end{equation}
with $w^i$ as given in \modeqref{eq:weight}.

\subsection{Experiments}\label{sec:hadexpt}

\begin{table}
  \centering
  \begin{tabular}{ccrrrrr}
    Experiment & Target & $E_p$ & {$N_p$} & $L_{sh}$ & $L_{d}$ & $A_{acc}$ \\
    \hline
    CHARM~\cite{Bergsma:1985qz,Bergsma:1985is} & Cu & 400 & $2.4 \times 10^{18}$ & 480 & 35 & 4.8 \\
    MINOS~\cite{Ambats:1998aa,Adamson:2013whj} & C & 120 & $1.407 \times 10^{21}$ & 1040 & 1.3 & 3.1\\
    $\nu$-Cal I~\cite{Blumlein:1990ay,Blumlein:1991xh} & Fe & 70 & $1.71 \times 10^{18}$ & 64 & 23 & 6.76 \\
    INGRID~\cite{Abe:2011xv} & C & 30 & $5 \times 10^{21}$ & 280 & 0.585 & 21.5 \\
    LSND~\cite{Aguilar:2001ty,Auerbach:2001wg} & See text & 0.798 & See text & 30 & 8.3 & 25.5 \\
    \hline
  \end{tabular}
  \caption{Experimental parameters describing the beam and detector position and size.  Energies $E_p$ are in GeV, lengths $L_{sh}$ and $L_{d}$ are in m, and the area $A_{acc}$ is in m$^2$.  $N_p$ is the number of protons on target recorded or expected.}\label{tab:genhbeam}
\end{table}

\begin{table}
  \centering
  \begin{tabular}{ccccccc}
    Experiment & $\sigma_\pi$ & $\sigma_\eta/\sigma_\pi$ & $\sigma_{\eta'}/\sigma_\pi$ & $\sigma_\rho/\sigma_\pi$ & $\sigma_\omega/\sigma_\pi$ & $\sigma_\Delta/\sigma_\pi$ \\
    \hline
    CHARM & See Text & 0.078 & 0.024 & 0.11 & 0.11 & 0.03 \\
    MINOS & 229.3~mb & 0.035 & 0.0035 & 0.047 & 0.048 & 0.013 \\
    $\nu$-Cal I & 677~mb & 0.035 & 0.0035 & 0.049 & 0.049 & 0.014 \\
    INGRID & 229.3~mb & 0.035 & 0.0035 & 0.046 & 0.047 & 0.017 \\
    \hline
  \end{tabular}
  \caption{Hadron cross sections we take for low mass hidden sectors.  Except as mentioned in the text, we take $\sigma_\pi \approx 25$--$30\,\sigma_\eta$ from \refcite{Teis:1996kx,Flam:1900ijf}, and the multiplicities of other hadrons from Pythia~8.180~\cite{Sjostrand:2007gs,Sjostrand:2006za}.}\label{tab:hadxsec}
\end{table}

We list general properties of the experiments we find set non-trivial limits in \tabref{tab:genhbeam}, and hadron production cross sections in \tabref{tab:hadxsec}.  Properties relevant to visible searches are in \tabref{tab:vishbeam}, and to invisible scattering searches in \tabref{tab:invhbeam}.  Because there is a relatively large variation in the configuration and searches for each experiment, we discuss them individually below.

\paragraph{CHARM} The CHARM experiment featured a 400~GeV proton beam impacting a copper target.  Searches where performed for an axion (pseudoscalar) decaying to $\gamma\gamma$, $e^+e^-$ or $\mu^+\mu^-$ in \refcite{Bergsma:1985qz}; and for heavy neutrinos decaying as $\nu_h \to \nu l^+ l^-$ in \refcite{Bergsma:1985is}.  These cover the relevant topologies of two- and three-body decays with and without additional unobserved states.  No events where observed in either search.

The CHARM detector was positioned 5~m (10~mrad) off the beam axis and had a square transverse cross section.  For low mass hidden resonances, we normalise our results to $N_\pi^{det} = 2.9\times10^{17}$ from the experimental collaboration~\cite{Bergsma:1985qz}, where $N_\pi^{det}$ is the number of pions produced on a trajectory that would intersect the detector.  For other pseudoscalar mesons, we use the cross sections quoted in \refcite{Gninenko:2012eq}; for vectors, the measurements of \refcite{AguilarBenitez:1991yy,Agakishiev:1998mw}; and for fermions, we estimate $N_\Delta \approx 0.03 N_\pi$ from the relative multiplicities in Pythia~8.180~\cite{Sjostrand:2007gs,Sjostrand:2006za}.

Limits from these CHARM searches have previously been interpreted in terms of vector kinetic mixing.  \refcite{Gninenko:2012eq} only considered the decay $Z^x \to l^+l^-$, with the hidden vector produced in psuedoscalar meson decay.  \refcite{Schuster:2009au} did consider production and decay of hidden scalars, but in a non-super\-symmetric context and at high masses.  Our results are then different due to the mass mixing, as discussed in \appref{app:hidden}.

\begin{table}
  \centering
  \begin{tabular}{crrrrr}
    Experiment & $E_{thr}^{e}$ & $E_{thr}^{\mu}$ & $\kappa_{Eff}^{e}$ & $\kappa_{Eff}^{\mu}$ & $N_{up}$ \\
    \hline
    CHARM & 5 & 5 & 0.51 & 0.85 & 3 \\
    MINOS & - & 1 & - & 0.8 & 10 \\
    $\nu$-Cal I & 3 & 3 & 0.7 & 0.9 & 7.76 \\
    LSND & 0.015 & - & 0.19 & - & $10^3$ \\
    \hline
  \end{tabular}
  \caption{Detector parameters relevant for visible searches.  Energies $E_{thr}^e$ and $E_{thr}^\mu$ are in GeV, while $\kappa$ are detection efficiencies.  Dashes signify a column does not apply for that experiment.  See \tabref{tab:genhbeam} or the text for references.}\label{tab:vishbeam}
\end{table}

\paragraph{MINOS} The MINOS experiment~\cite{Ambats:1998aa} uses the 120~GeV NuMI beam on a graphite target.  We consider possible signals in the near detector, positioned 1.04~km away.  We model the detector as a cylinder of radius 1~m and take a recent value~\cite{Adamson:2013whj} for the total number of protons on target.  For hadron production we take the cross section for $pC \to \pi + \ldots$ from \refcite{Abgrall:2011ae}; using \modeqref{eq:Nhad} gives $N_\pi \approx 0.99 N_p$.  

We consider the possibility of signals at MINOS from both hidden sector scattering and decay.  For decays, we follow \refcite{Batell:2009yf} and consider a sensitivity to $\order{10}$ muon pairs with a detection efficiency comparable to that for scattering~\cite{Adamson:2010wi}.  Searches for invisible particle scattering must compete against a large neutrino background.  We follow \refcite{deNiverville:2012ij} and show prospective limits for $10^3$ and $10^4$ events in our plots, corresponding to different estimates for background rejection.

\begin{table}
  \centering
  \begin{tabular}{ccccc}
    Experiment & $n_{e}$ & $n_N$ & $\kappa_{eff}$ & $N_{up}$ \\
    \hline
    MINOS & - & $5 \times 10^{24}$ & 0.8 & $10^3$--$10^4$ \\
    INGRID & - & $5 \times 10^{24}$ & 0.8 & $10^3$--$10^4$ \\
    LSND & $5.1\times 10^{23}$ & - & 0.19 & $10^3$ \\
    \hline
  \end{tabular}
  \caption{Detector parameters for invisible searches.  Dashes signify a column is not relevant for that experiment.  See \tabref{tab:genhbeam} or the text for references.}\label{tab:invhbeam}
\end{table}

\paragraph{$\nu$-Cal I}  The $\nu$ Calorimeter I experiment took data from a beam dump at the U70 accelerator, where a 70~GeV proton beam was delivered to an iron target.  Searches for axions and light Higgs bosons~\cite{Blumlein:1990ay,Blumlein:1991xh} were reinterpreted in terms of hidden vectors decaying to leptons in \refscite{Blumlein:2011mv,Blumlein:2013cua}.  We find the pion production cross section by rescaling the measurement of \refcite{Boratav:1976wx,Blumenfeld:1973ww} as discussed in~\cite{Blumlein:2011mv}:
\begin{equation}
  \sigma (p\,Fe \to \pi + \ldots) \approx A^{0.55} \times \sigma (pp \to \pi + \ldots) \,.
\end{equation}
From \modeqref{eq:Nhad} we find $N_\pi \approx 0.96 N_p$.  Also, our choice for the $\eta$ production cross section is consistent with its non-observation in~\refcite{Boratav:1976wx,Blumenfeld:1973ww}.  

The combination of a relatively small distance between the target and the detector and a moderately large energy are useful in setting limits when the mass mixing of \appref{app:hidden} dominates the \hx\ decay.  However, we find that the limits from meson decays are almost always inferior to the corresponding limits from CHARM.

\paragraph{INGRID} INGRID~\cite{Abe:2011xv} is the near on-axis detector in the T2K experiment~\cite{Abe:2011ks}.  We use an estimate for the prospective total lifetime $N_p$, and approximate the detector efficiencies as similar to MINOS.  The detector has a non-trivial cross shape transverse to the beam direction.  As with MINOS, a large neutrino background would demand a dedicated study to reject backgrounds, which we approximate by considering limits for $10^3$ and $10^4$ events.

\paragraph{LSND} The LSND experiment used a 800~MeV proton beam on two different targets over its lifetime, water and a high-Z metal~\cite{Aguilar:2001ty}.  When computing the acceptance efficiencies, we use a weighted mean of the two distribution functions $f_{BS}$ for the different target materials.  We model the detector as a cylinder of diameter 5.7~m.  Because of the low beam energy, we only consider limits from hidden states produced in pion decays; it is assumed that the production of heavier hadrons is highly kinematically suppressed.  We use an estimate $N_\pi \approx 10^{22}$ of the number of pions produced from~\cite{deNiverville:2011it}.

A search for neutral current scattering on the full LSND data set observed $\order{300}$ events above background, of which $\order{200}$ where expected from neutrinos~\cite{Auerbach:2001wg}.  We conservatively place limits for 1000 hidden sector events as unambiguously ruled out~\cite{Batell:2009di,deNiverville:2011it}.  Along with MINOS, we consider the possibility of both hidden sector decay (to electrons) and scattering, as shown in \tabrefs{tab:vishbeam}{tab:invhbeam}.

\paragraph{Other Experiments} Searches for heavy neutrinos decaying to leptons were carried out at NOMAD~\cite{oai:arXiv.org:hep-ex/0101041} and PS-191~\cite{Bernardi:1985ny,Bernardi:1987ek}, and reinterpreted as searches for hidden vectors in \refcite{Gninenko:2011uv}.  We find that the limits from other hidden decays then $Z^x \to l^+l^-$ are inferior to those from CHARM.  
\refscite{deNiverville:2012ij,deNiverville:2011it,Batell:2009di} proposed looking for hidden particles scattering at MiniBooNE~\cite{AguilarArevalo:2008qa} and in the ND280 detector at T2K~\cite{Assylbekov:2011sh}.  For our benchmarks, the former is always inferior to MINOS and the latter to INGRID.  Finally, we note that future experiments such as Project X~\cite{Kronfeld:2013uoa,Holmes:2013hfa} or AFTER@LHC~\cite{Brodsky:2012vg,Lansberg:2012wj} would likely place further exclusions on our paramter space.  The nature of those limits is beyond the scope of this work.

\subsection{Limits}\label{sec:hadlimits}

\begin{figure}
  \centering
  \includegraphics[width=0.95\textwidth]{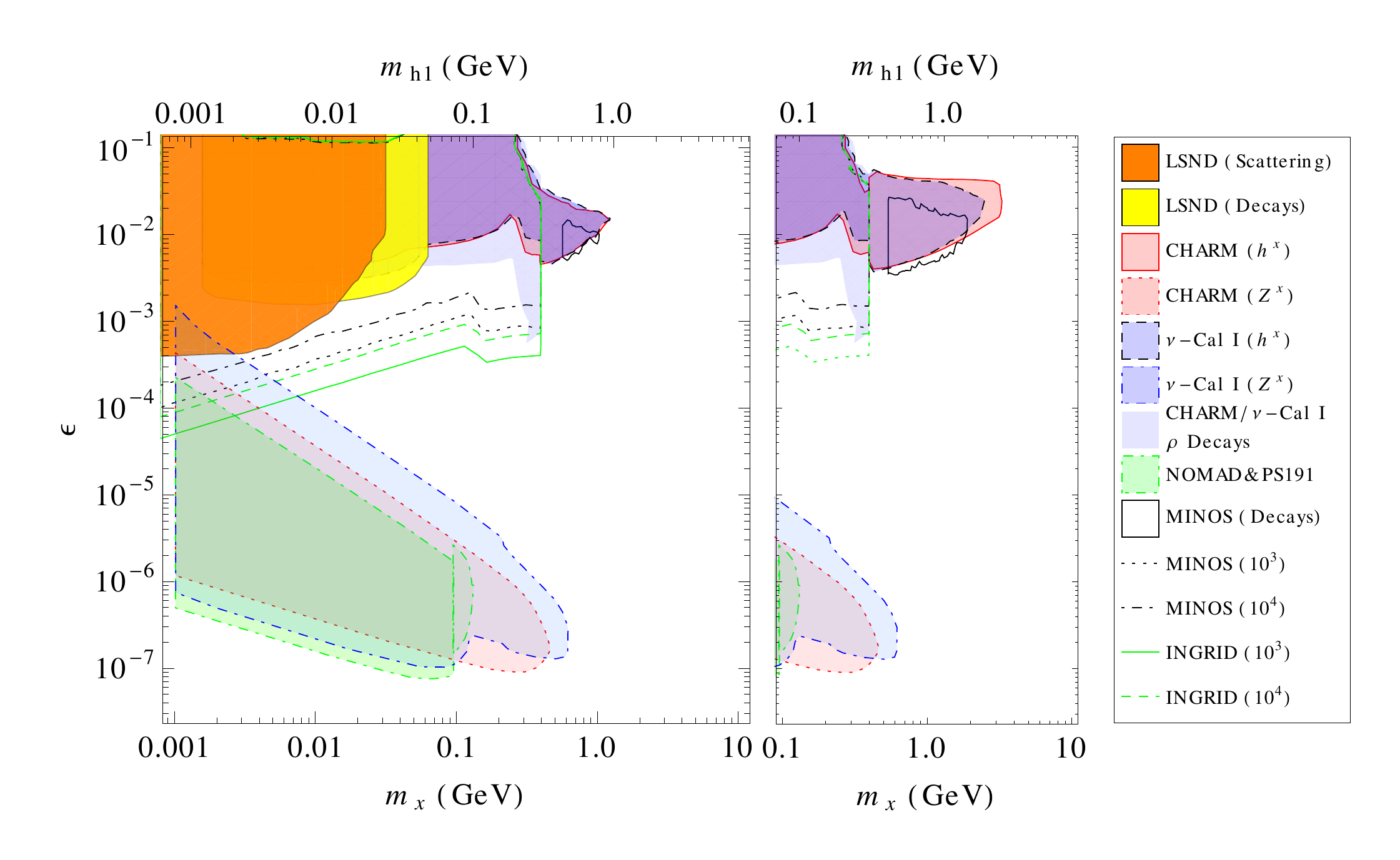}
  \caption{Case A, with (left) and without (right) the Higgs mass mixing induced by the SUSY D-terms.  The shaded contours in the lower left show prioir limits from $Z^x$ decays for CHARM (red, dotted), $\nu$-Cal~I (blue, dot-dashed), and NOMAD and PS-191 (green, dot-dashed).  Shaded upper regions show new limits: from \hx\ decays at CHARM (red and light blue), $\nu$-Cal~I (blue and light blue), and LSND (yellow); and from \hx\ scattering at LSND (orange).  The solid black contour shows prospective limits from \hx\ decays at MINOS.
  The remaining contours show prospective limits from \hx\ scattering at MINOS (black dotted or dot-dashed) and INGRID (green solid or dashed), with different contours for sensitivity to $10^3$ or $10^4$ events.  See the text for more details.  
  }\label{fig:Ahad}
\end{figure}

We now present current and prospective limits for our four benchmarks from hadronic beam dumps.  Composite plots showing all the limits we find are shown in \secref{sec:conc}.  We give limits for Cases A, B and D in \figfigrefs{fig:Ahad}{fig:Bhad}{fig:Dhad}.  For Case C, we show limits separately for constraints from scalars (\figref{fig:Chad}) and pseudoscalars (\figref{fig:CAxHad}) for clarity.  In \figfigrefs{fig:Ahad}{fig:Bhad}{fig:Chad}, we split the constraints for $m_x > 0.1\,\gev$ 
with (left) and without (right) the inclusion of D-term Higgs portal
mass mixing.  As discussed in \appref{app:hidden}, the presence of this mass mixing heavily alters the \hx\ lifetime above the muon threshold.  Hence all experimental searches based on a long-lived scalar will differ between these two possibilities.

We note some general features of our results before discussing each case in detail.  Limits from the hidden Higgs decays are alomst always upper bounds on $\epsilon$.  This is because for kinetic mixings not ruled out by \emph{e.g.} EWPTs, the scalar decay length is large.  Signals are then limited by production ($\propto \epsilon^2$) and, when relevant, scaterring probability ($\propto \epsilon^2$).  The exception comes only for large masses $m_{\hx} > 2m_\mu$, when the lifetime can become short enough for the scalar to decay within the shield.

At low masses, limits from LSND tend to dominate due to its large luminosity and the absence of kinematic suppression.  At intermediate masses, limits from CHARM are stronger; and at high masses above the muon threshold, the $\nu$-cal~I search can also be relevant.  
We show current limits with shaded regions, and use unshaded contours to denote prospective limits.  For MINOS, we label possible limits from visible searches as ``MINOS(VIS)''.  For both MINOS and INGRID, we use ``$10^3$'' and ``$10^4$'' for contours of that many expected scattering events.  We omit any results whose exclusions would be entirely within those of another experiment; this means that not all experiments appear in all plots.

\paragraph{Case A} In this benchmark the hidden vector will decay to the SM, so we include limits for this from NOMAD and PS-191~\cite{Gninenko:2012eq},  $\nu$-calorimeter~I~\cite{Blumlein:2013cua,Blumlein:2011mv}, and CHARM~\cite{Gninenko:2011uv}.  These extend to lower values of the kinetic mixing than the channels we consider here, but the exclusions cannot reach large values of $\epsilon$ as the $Z^x$ lifetime becomes too short.

We find four new limits, shown as the shaded regions in \figref{fig:Ahad} for $\epsilon \gtrsim 10^{-3}$.  At low masses, LSND sets limits from \hx\ scattering, where it is stable over the relevant length scales.  At slightly higher masses, the scalar is metastable and LSND sets limits from visible \hx\ decays.  Above the pion mass, we have limits from CHARM and $\nu$-Cal~I from \hx\ decays.  We split these limits into two regions.  The darker regions with borders correspond to \hx\ production through pseudoscalar decay or $q\bar{q}$ fusion, and we show the two experiments separately.  The light blue region without border shows the limit from the decay $\rho \to Z^x \hx$; the limits in this case are indistinguishable between the two experiments.  The resonance at $m_x \approx m_\rho$ is clearly visible.  These limits are stronger as they arise from a two-body decay, which compensates for the smaller $\rho$ multiplicity.  However, there is a larger systematic uncertainty in the number of $\rho$ produced.  


The limits from INGRID, and from MINOS due to dark particle scattering, also come from the $\rho$.  While the decay $\rho \to \chi_1\chi_1$ is possible, the large fermion mass means our limits are set by $\rho \to Z^x \hx$, with the \hx\ effectively stable in this mass range. 

At high masses, we can clearly see the effect of the Higgs mass mixing by comparing the left (with) and right (without) sides of \figref{fig:Ahad}.  With the mass mixing, CHARM and $\nu$-Cal~I set very similar and much weaker limits.  This is caused by the shorter scalar decay length, leading to lower survival probabilities.  Without the mass mixing, we can exclude higher masses and the higher centre of mass energy of CHARM becomes important.


\begin{figure}
  \centering
  \includegraphics[width=0.95\textwidth]{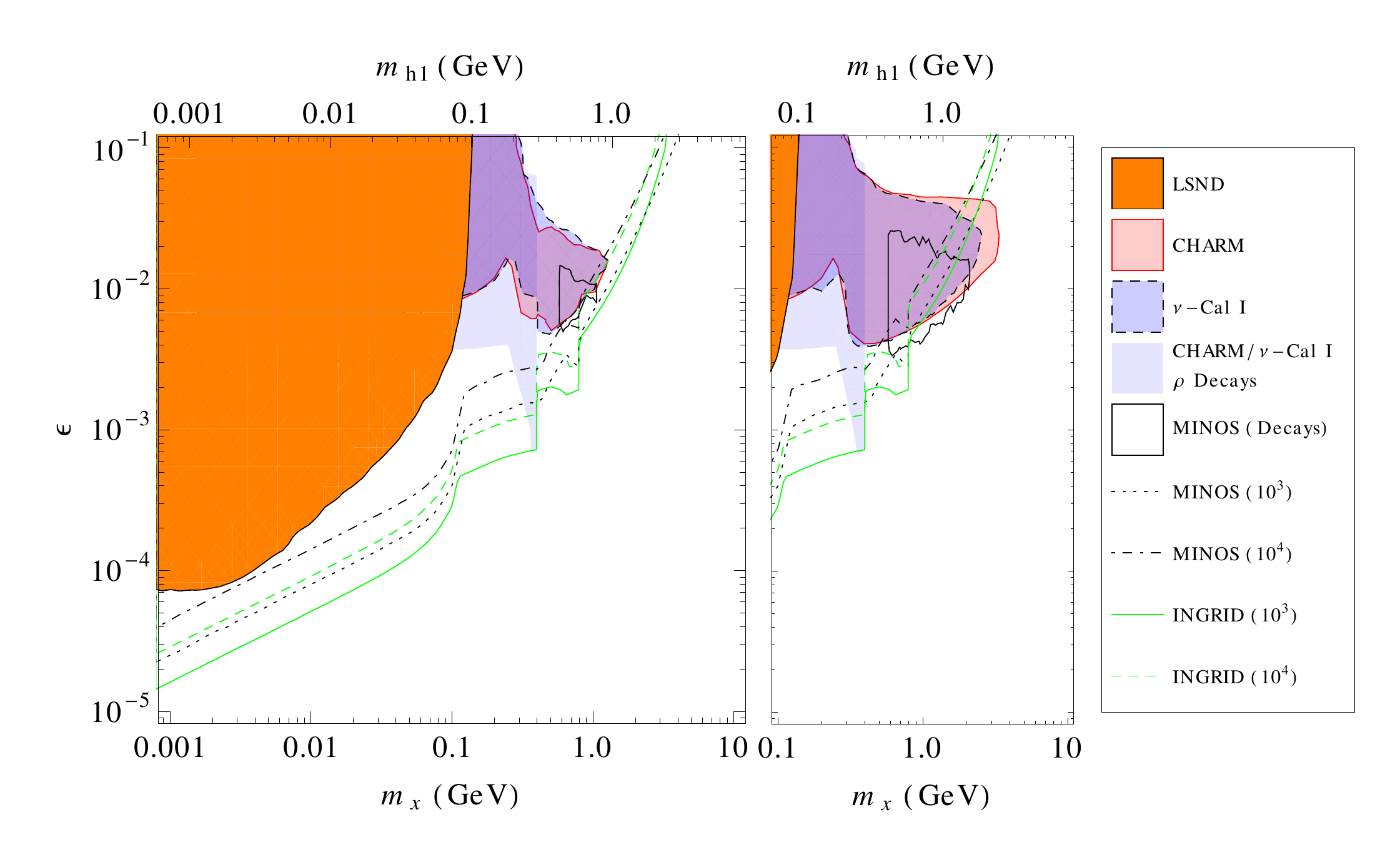}
  \caption{Case B, with (left) and without (right) the Higgs mass mixing.  Shaded regions show new limits: from \hx\ decays at CHARM (red and light blue) and $\nu$-Cal~I (blue and light blue), and from $\chi_1^x$ scattering at LSND (orange).  The solid black contour shows prospective limits from \hx\ decays at MINOS.
  The remaining contours show prospective limits from $\chi_1^x$ scattering at MINOS (black dotted or dot-dashed) and INGRID (green solid or dashed), with different contours for sensitivity to $10^3$ or $10^4$ events.  See the text for more details. 
  %
  }\label{fig:Bhad}
\end{figure}

\paragraph{Case B} For visible searches, this benchmark point differs from Case A in two regards.  The main difference is that there are no longer any limits from $Z^x \to l^+l^-$, as previously discussed.  For limits from other hidden particle decays, apart from small differences in the ratio $m_{\hx}/m_x$ and the mixing matrix $R$, the two sectors are effectively the same.  It follows that the associated limits are nearly identical.  These are the limits from CHARM and $\nu$-Cal~I, and the visible decay prospects from MINOS, shown in \figref{fig:Bhad}.

However, the possibility for the hidden vector to decay to two hidden fermions, $Z^x \to \chi^x_1\chi^x_1$, substantially enhances the limits from dark sector scattering.  We can easily see this by comparing the LSND, MINOS and INGRID exclusions and prospects between \figrefs{fig:Ahad}{fig:Bhad}.  In particular, note that the LSND limits from $\chi^x_1$ scattering now dominate those from \hx\ decay.  At low masses, hidden states at MINOS and INGRID limits are dominantly produced from pions, as this can now produce hidden sectors through successive two-body decays.  Above the pion mass the contributions from the $\rho$ and $\eta$ are comparable, and at high masses quark fusion is the dominant channel.  Finally, we note that the limits here are not directly comparable to those shown in \emph{e.g.}~\refscite{Essig:2013vha,Essig:2013lka}.  Those limits assumed a fixed $\chi_1^x$ mass and varied the vector mass, while we fix the mass ratio and vary them together.

\begin{figure}
  \centering
  \includegraphics[width=0.95\textwidth]{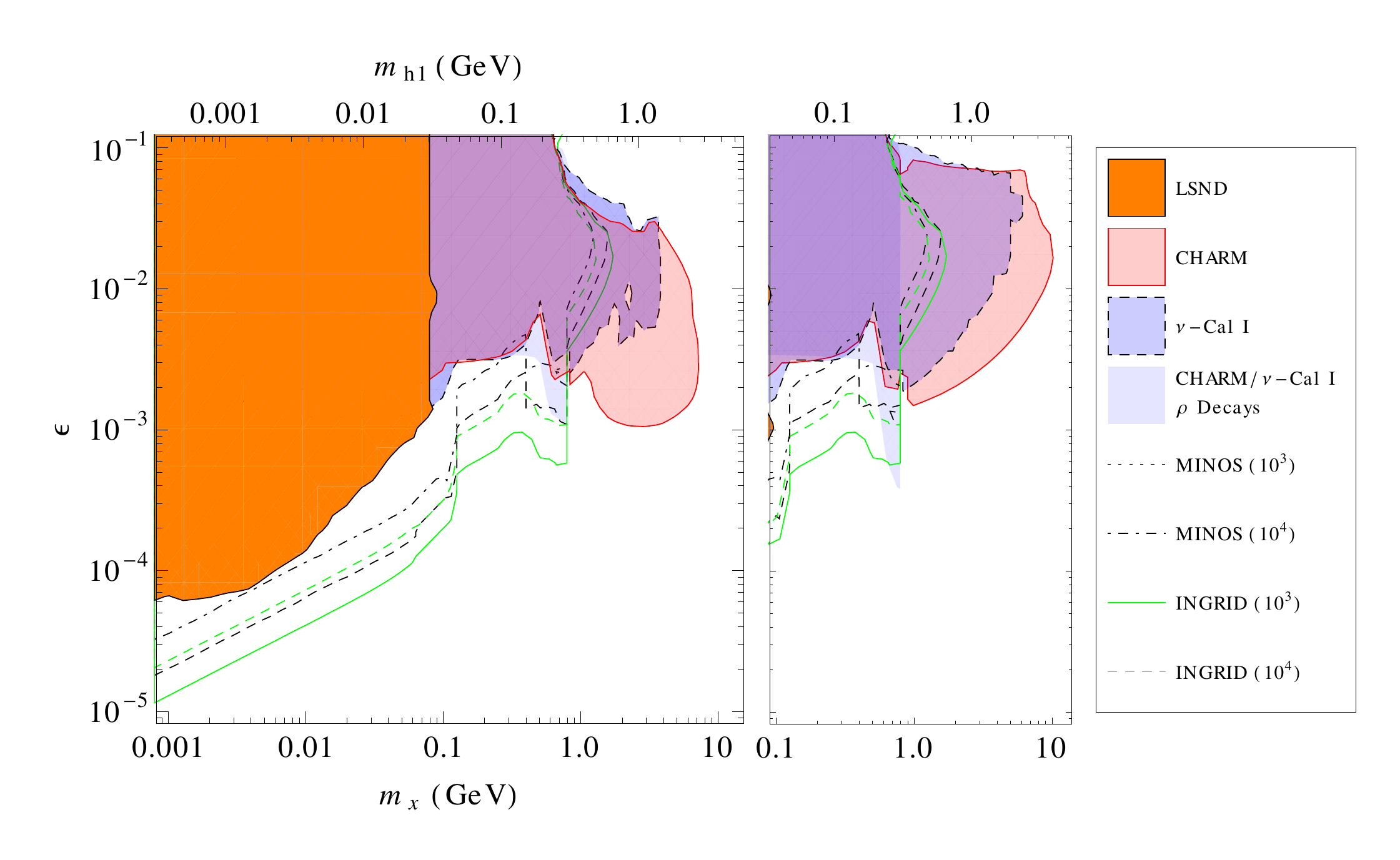}
  \caption{Case C, with (left) and without (right) the Higgs mass mixing.  We show here only limits from hidden sector scattering and from \hx\ decays.  The decay limits at CHARM and $\nu$-Cal~I have the same notation as in \figrefs{fig:Ahad}{fig:Bhad}.  The limits at LSND, and prospective limits at MINOS and INGRID, receive roughly equal contributions from \hx\ and $A^x$ scattering.}\label{fig:Chad}
\end{figure}

\paragraph{Case C}  The ability to produce the scalars through an on-shell vector substantially enhances the limits we derive.  Additionally, the $A^x$ can be long-lived and is also produced in the on-shell decay of a hidden vector, so we also find limits from its decays.  To avoid our plots becoming too cluttered to read, and because they are independent of the Higgs mass mixing, we separate the limits from $A^x$ decays into \figref{fig:CAxHad}.

In \figref{fig:Chad} we show all other limits in this benchmark.  In particular, the exclusions from \hx\ decays at both CHARM and $\nu$-Cal I extend to higher masses and lower mixings than in cases A and B.  We note that the limits we find at CHARM without the mass mixing and at high masses are comparable to those derived in \refcite{Schuster:2009au}.  The limits at low masses at CHARM are dominated by pseudoscalar decays, which can now produce hidden states through two-body channels, \emph{e.g.} $\pi^0 \to \gamma Z^x$ followed by $Z^x \to h^x_1 A^x$.

The limits from scattering experiments come from both \hx\ and $A^x$, which give roughly equal contributions.  This is to be expected from \twoeqref{eq:scatxsec}{eq:scatmixmat}.  Hidden states are again dominantly produced by pions when allowed, and from $\eta$ and $\rho$ decays when not.

\begin{figure}
  \centering
  \includegraphics[width=0.8\textwidth]{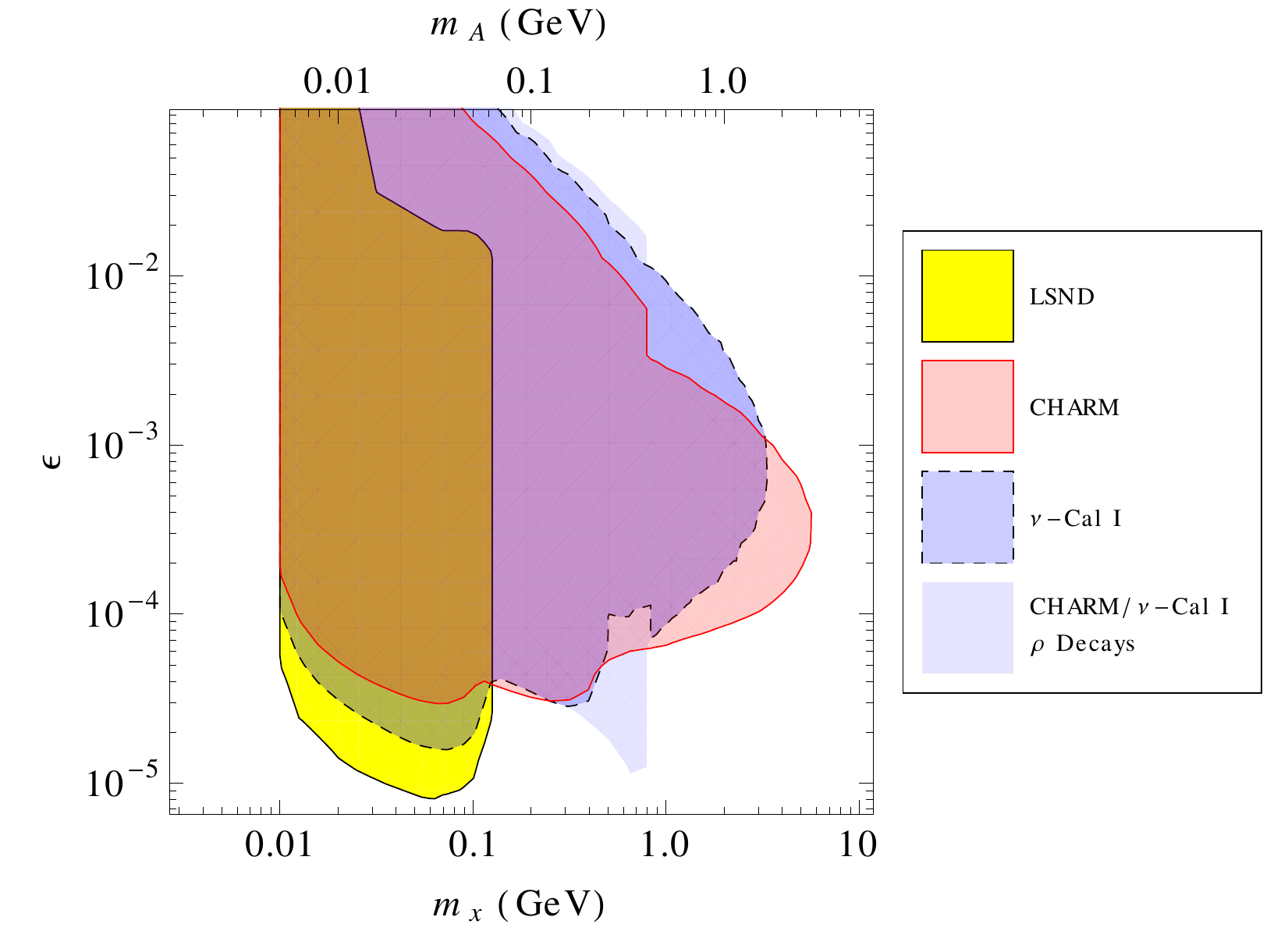}
  \caption{Case C, limits from pseudoscalar decays $A^x \to \hx \, e^+ e^-$ only.  
  }\label{fig:CAxHad}
\end{figure}

The limits from the pseudoscalar in \figref{fig:CAxHad} all come from the visible decay $A^x \to \hx \, e^+ e^-$.  They are interesting in that the exclusion contours have relevant upper bounds.  The $A^x$ decay is less suppressed than the hidden Higgs, and so it can be sufficiently short-lived to decay before reaching the detector.  This will happen at large mass and kinetic mixing.  The $\nu$-Cal I experiment sets better limits in this domain for the simple reason that the detector was closer to the interaction point.  
At low mixings and high masses, the limits from CHARM dominate due to its higher beam energy.  At low masses, limits from LSND dominate due to that experiment's high luminosity.

\begin{figure}
  \centering
  \includegraphics[width=0.75\textwidth]{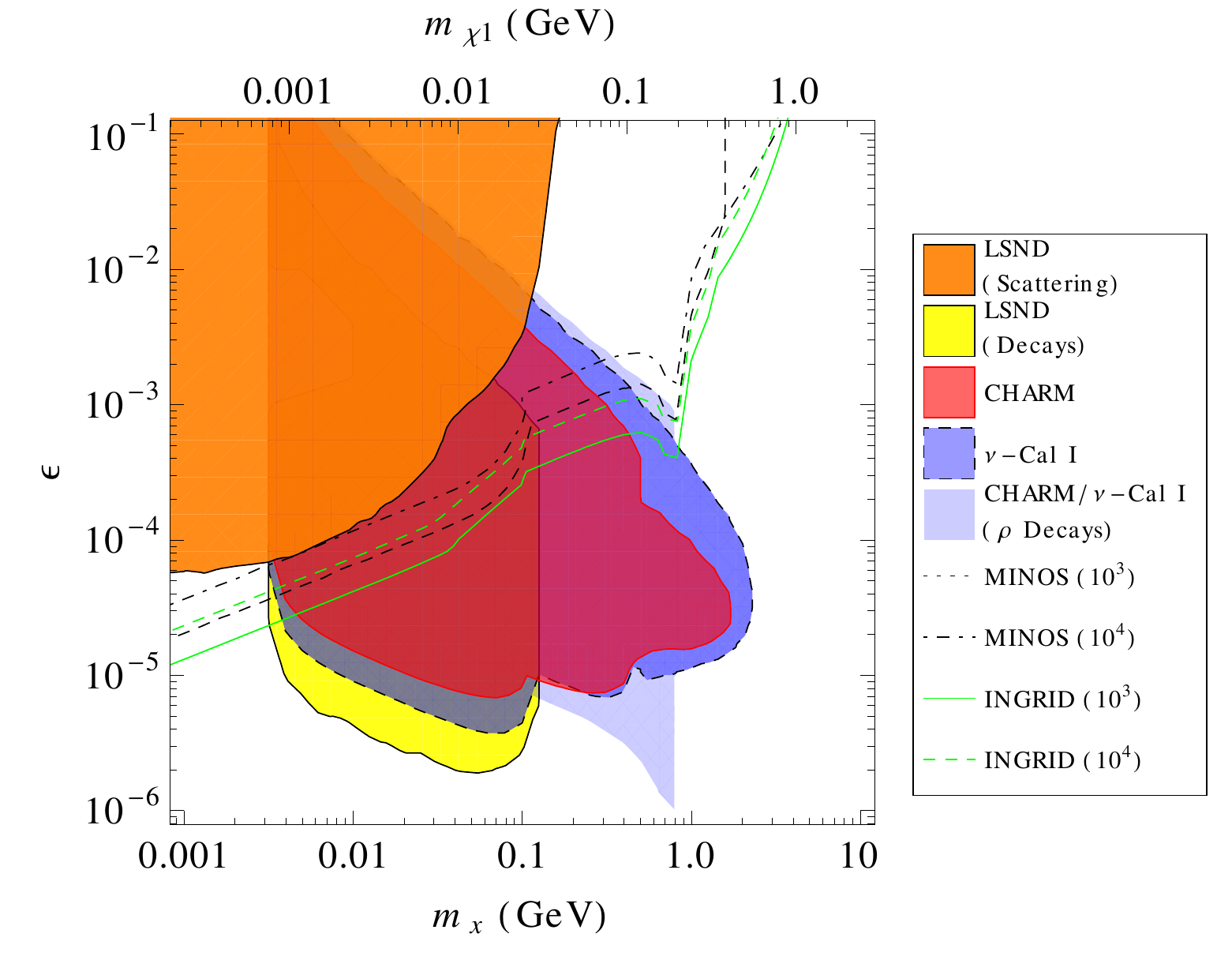}
  \caption{Limits in Case D.  Limits from $\chi_x^2 \to \chi_x^1 \, e^+ e^-$ or $\chi_1^x \, \mu^+ \mu^-$ are shown in red or pink (CHARM) and blue ($\nu$-Cal~I).  Limits from $\chi_{1,2}^x$ scattering from LSND are shown in orange.  Prospective scattering limits from MINOS and INGRID are shown as black and green contours, respectively.
  }\label{fig:Dhad}
\end{figure}

\paragraph{Case D}  Unlike the three previous cases, there are no limits in this benchmark from metastable scalars.  All limits come from the production of hidden sector fermions.  We have significant limits both from $\chi_2^x$ decays to leptons, and from $\chi_1^x$ scattering ($\chi_2^x$ scattering is a subleading effect).  Finally, because we have no limits from \hx\ we have no sensitivity to the Higgs mass mixing, and hence show only a single plot in \figref{fig:Dhad}.

The constraints from visible searches are dominated by LSND at low mass and $\nu$-Cal I at high mass.  Hidden fermion production at CHARM and $\nu$-Cal~I is dominated by pion decay when kinematically accessible, and by $\rho$ decay when not.  The $\rho$ wins out over the $\eta$ due to energy thresholds.  Hidden fermion production from pseudoscalars requires two two-body decays, and so they tend to have lower energies.  Like the pseudoscalar in Case C, the fermion decays are sufficiently less suppressed than the \hx\ for our exclusion contours to have an upper bound.  This occurs when the $\chi_2^x$ lifetime is so short, it decays before reaching the detector.  The superiority of the $\nu$-Cal I experiment at high masses is due to its location closer to the detector.

Limits from scattering experiments are straightforward.  Of particular note is their sensitivity at high mass, where the $\chi_2^x$ decays too promptly.  In this region the MINOS and INGRID limits can be very important.  Otherwise, limits from visible searches win so long as we are above the electron threshold.

\section{Implications at the LHC}\label{sec:lhc}
  Signatures of this theory at the LHC were investigated 
in Ref.~\cite{Chan:2011aa}.  When the MSSM superpartners
are near the TeV scale, hidden states can be produced
readily in supersymmetric cascades~\refscite{
Strassler:2006qa,ArkaniHamed:2008qp,Baumgart:2009tn,Cheung:2009su}.  
In general, the creation of a pair of MSSM superpartners will initiate
a pair of visible cascades down to the lightest SM superpartner~(LSMP).
Each LSMP will subsequently decay to the hidden sector, thereby
initiating a hidden cascade.  Depending on the decay properties of the
hidden states, this may produce additional visible activity in the event
and reduced missing energy~($\met$).  The extent to which this occurs
is related to the signatures of the hidden sector in low-energy experiments.

  MSSM superpartners connect to the hidden sector through 
the supersymmetric kinetic mixing of Eq.~\eqref{eq:svecport} 
between the Bino and the $U(1)_x$ gaugino.  
When the LSMP is a neutralino with a significant Bino component,
the decays $\chi_1^0 \to \chi_i^xS^x$, where $S^x = Z^x,\,h_1^x,h_2^x,\,A^x$, 
are likely to be prompt ~\cite{Baumgart:2009tn}.
As in Ref.~\cite{Chan:2011aa}, we will focus on this case here.
Other species of LSMPs tend to have delayed decays,
possibly leading to charged tracks~\cite{Baumgart:2009tn}.
The decay of an LSMP neutralino populates the four bosonic
states equally, while the relative fractions of the hidden
neutralino species depends on their hidden Higgsino 
contents~\cite{Chan:2011aa}.  

  Decays of the $\chi_1^0$ LSMP will contribute to missing energy,
but they can also produce visible signals.  The latter occurs when
a vector, on- or off-shell, is created in the cascade,
and decays to SM states.  In contrast, direct decays of $h_1^x$ 
to the SM are usually too slow to occur within the LHC detector.
Since we assume $m_{\chi_1^0} \gg m_{x}$, the boson and the fermion 
produced by a $\chi_1^0$ decay will be highly boosted by an amount 
$\gamma \sim m_{\chi_1^0}/2m_x$.
Any visible products created along the subsequent boson and fermion
decay chains will therefore be highly boosted and collimated.  
Following Ref.~\cite{Chan:2011aa}, we will call these 
hidden-valley~(HV) jets.  An HV jet may contain leptons or hadrons,
and it can be prompt or displaced.  
Each $\chi_1^0$ decay can produce zero, one, or two HV jets.  

  The average number of HV jets per $\chi_1^0$ decay depends 
on the spectrum of the hidden sector, and is related to the signals
expected in low-energy experiments.  In case~A, the HV jets
come mostly from on-shell vectors and can be prompt or delayed, with
\beq
\gamma\,c\tau \sim 
500\,\mu\text{m}\,
\lrf{10^{-3}}{\epsilon}^2
\lrf{0.1\,\gev}{m_x}^2 
\lrf{m_{\chi_1^0}}{100\,\gev}\ .
\eeq
Hidden vectors are produced directly from the Bino decay, or in a two-step decay $\chi_1^0 \to A^x \to Z^x$.  The heavier scalar preferentially decays $h_2^x \to \hx\hx$.  At least one HV jet will be present in $3/4$ of supersymmetric events.

  In case~B, nearly all LSMP decays are completely invisible.
With the exception of $h_1^x$, the dominant decay channel of all
the bosons in the theory is $S^x \to \chi_1^x\chi_1^x$.
While the $h_1^x$ state does decay to the SM, it is typically too slow
to be seen in the LHC detectors.

  Case~C can yield HV jets when $A^x$ pseudoscalars are created in cascades.  
These decay through $A^x\to h_1^x+Z^{x*}$ with $Z^{x*}$ producing 
an HV jet.  However, these HV jets are frequently displaced or delayed
when the boost from the $\chi_1^0$ decay is included.
Case~D is similar to case~C, but now the HV jets come mainly from
the vector in $\chi_2^x\to\chi_1^+Z^{x*}$ decays.  Again, these are likely
to be delayed.  This is especially true given that our limits in these benchmarks tend to force the kinetic mixing to be small, $\epsilon \lesssim 10^{-4}$--$10^{-3}$.

\section{Conclusions}\label{sec:conc}
\begin{figure}[ttt]
  \centering
  \includegraphics[width=\textwidth]{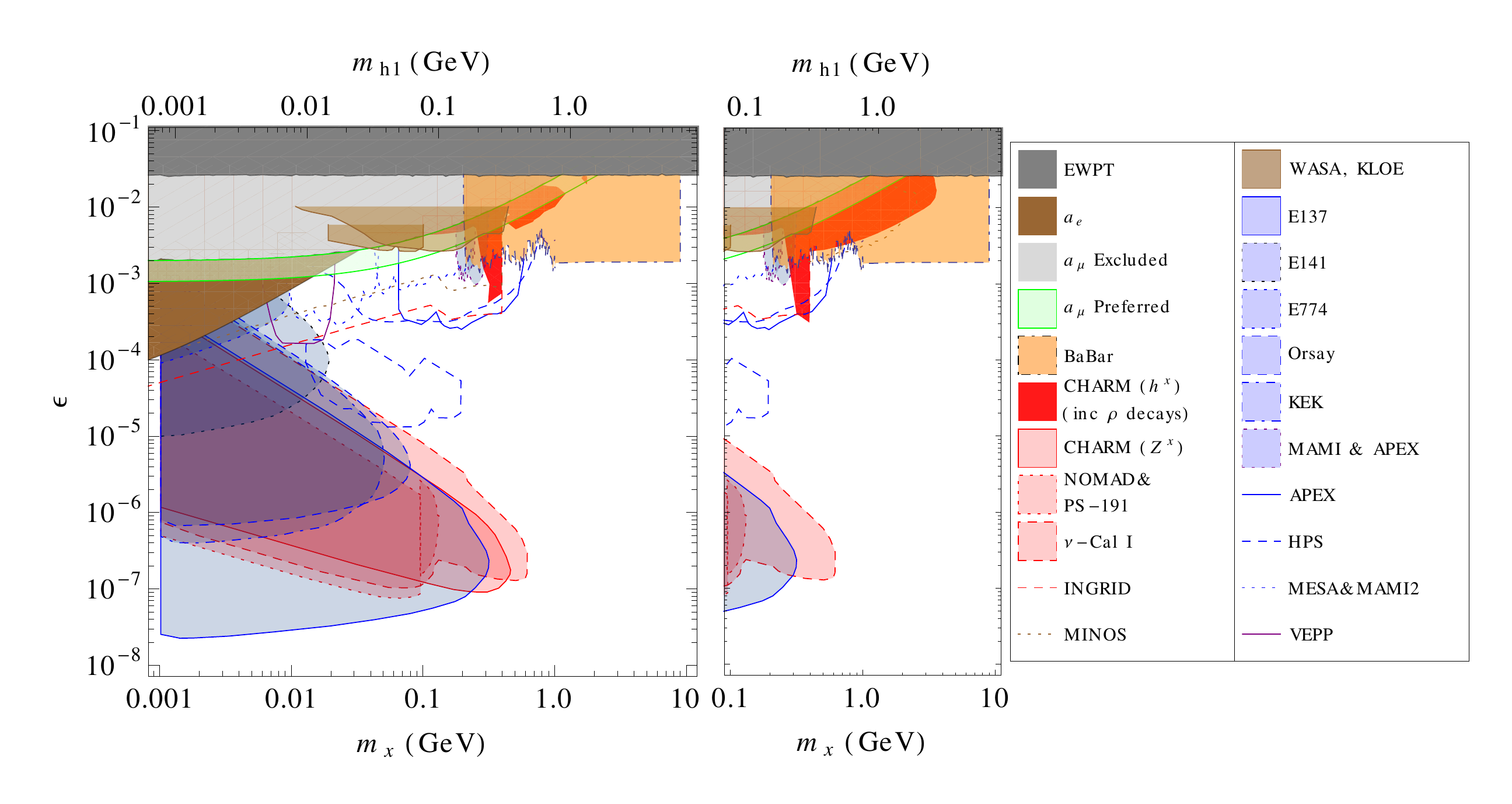}
  \caption{Composite plot of all limits we find for Case A.  Shaded (unshaded) regions are current (prospective) constraints. The exclusions from electron beam dumps, and from proton beam dumps at low $m_x$ and $\epsilon$, come from the decay $Z^x \to l^+l^-$.  These are as cited in \secrefs{sec:ebeam}{sec:hbeam}, plus the exclusions from the MAMI~\cite{Merkel:2011ze} and APEX~\cite{Abrahamyan:2011gv} test runs.  We do not include the new electron limits from \figref{fig:elec-AB}, as they are inferior to those from the anomalous magnetic moment.  Prospective limits from electron beam dumps are from APEX~\cite{Essig:2010xa}, HPS~\cite{HPS}, DarkLight~\cite{Freytsis:2009bh}, MAMI and MESA~\cite{Beranek:2013yqa} projections.  The limits from BaBar are from \refscite{Bjorken:2009mm,Reece:2009un,Aubert:2009cp,Essig:2013lka}.  We also show limits from KLOE~\cite{Babusci:2012cr,Archilli:2011zc} and WASA-at-COSY~\cite{Adlarson:2013eza}.}\label{fig:allA}
\end{figure}

  In this work we have investigated the limits on a minimal 
supersymmetric hidden sector from lower-energy precision experiments.  
This model provides a well-defined and plausible extension of the simplified
hidden-sector theories studied previously.  In certain limits, our model
behaves nearly identically to some of these simplified models.
However, our supersymmetric extension can also lead to richer 
and more complicated signals that have received little attention so far.

  To survey the range experimental signatures of this model,
we have focussed on four specific parameter slopes that cover the bulk
of the most likely possibilities.  
For each slope, the dimensionful parameters
of the theory are varied in fixed ratios to the vector mass $m_x$.
In Case~A, the vector $Z^x$ has no hidden decay channels and goes exclusively
to the SM.  In Case~B, the vector decays almost completely invisibly to
the lightest stable hidden neutralino $\chi_1^x$.  In Case~C, the vector decays
primarily to hidden scalar states which decay in turn to the SM, typically
with a significant delay.  In Case~D, the vector decays to both the
$\chi_1^x$ and $\chi_2^x$ hidden fermions, with $\chi_2^x$ subsequently
decaying $\chi_1^x$ and SM states.  

  Of these, Cases A and B lead to signals that are very similar to the 
commonly-considered scenarios of a hidden vector decaying 
to the SM (Case A)~\cite{Bjorken:2009mm}, 
and invisibly (Case B)~\cite{Izaguirre:2013uxa,Essig:2013vha}.  
Cases C and D have not been studied in nearly as much detail,
and we have substantially extended (Case C) or performed 
the first study of (Case D) the exclusions that apply in these cases.
Additionally, we have investigated how the presence of more states
in the hidden sector can affect the limits even in Cases A and B.  
In all four cases, producing hidden sector states through off-shell vectors 
or through the decay of a vector meson grant access to new detection channels.  
Of particular importance is the production and decay of the hidden 
sector Higgs, which sets new limits in phenomenologically interesting 
regions of parameter space.  We have also studied how limits 
from hidden scalar decays vary depending on whether $D$-term 
mixing with the visible Higgs (as expected in our supersymmetric realization) 
is included or not.

  Summary plots containing the limits derived on all four scenarios
are shown in \thrufigrefs{fig:allA}{fig:allD}, 
together with some additional limits taken from the literature.  
We do not include the limits from \secref{sec:cosmo} from cosmologically 
late decays since they depend on the detailed cosmological history.
We show a single line for prospective limits from current 
and future experiments
we considered: specifically, observation of 1000 scattering events at JLab, 
INGRID and MINOS; and 10 muon pairs at MINOS.  
For Cases A--C, we show separately the high mass ($m_x > 0.1\,\gev$) 
region both with (left) and without (right) the Higgs mass mixing.

In Case A, the most important new feature in \figref{fig:allA} relative to the exclusions for minimal models is the region excluded by CHARM and $\nu$-Cal~I from decays $\rho \to \hx Z^x$.  These are the strongest limits in the vicinity of the $\rho$ resonance, 
$250\,\mev \lesssim m_x \lesssim 400\,\mev$.  
We also note that the prospective limits from INGRID are stronger than those from DarkLight/MESA for $m_x \lesssim 40\,\mev$.

\begin{figure}[ttt]
  \centering
  \includegraphics[width=\textwidth]{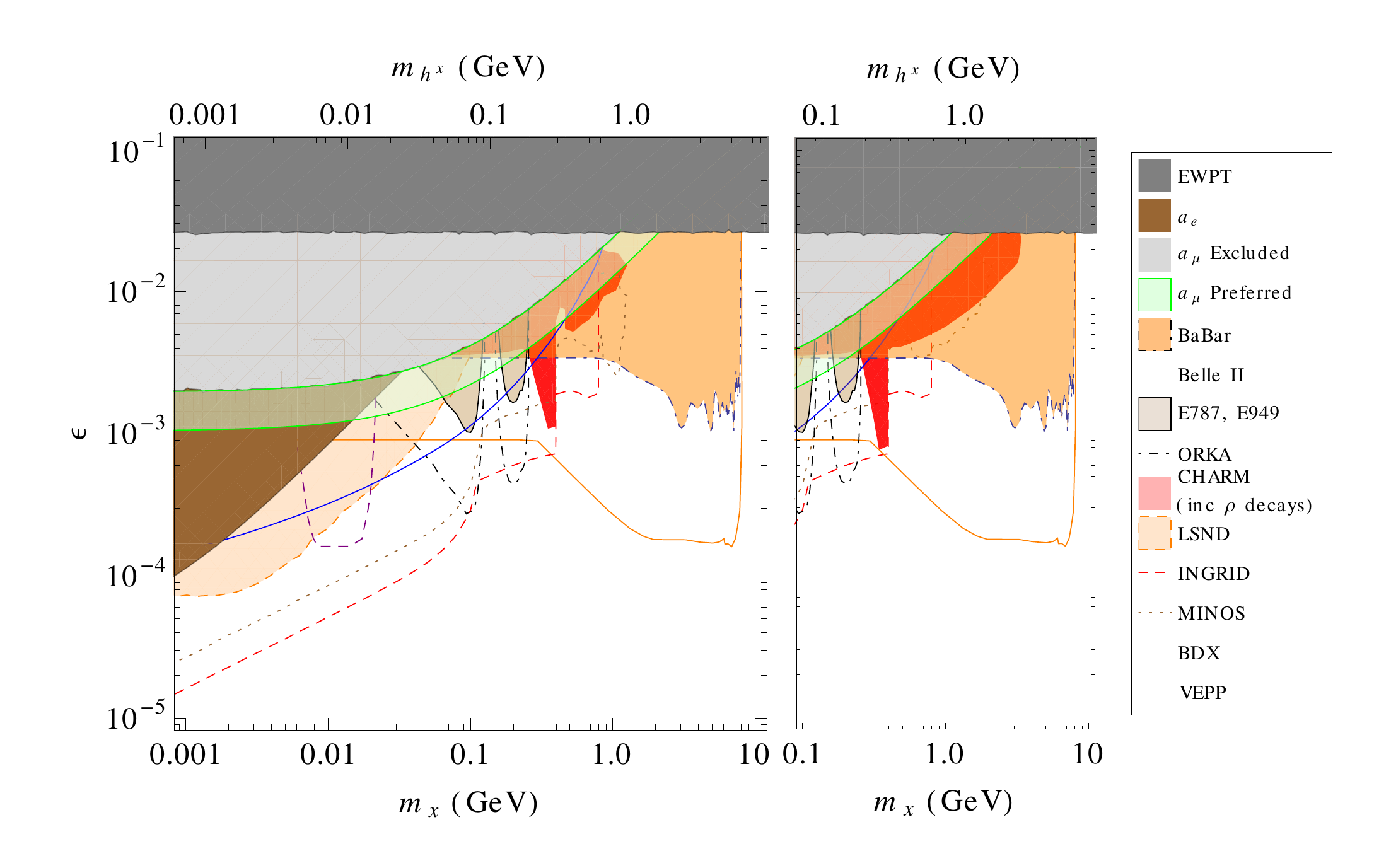}
  \caption{Composite plot of all limits we find for Case B.  The limits from BaBar and prospective exclusions from Belle II are estimated from \refscite{Aubert:2008as,Essig:2013vha,Izaguirre:2013uxa} and~\cite{Essig:2013vha} respectively.  Limits from E787~\cite{Pospelov:2008zw} and E949~\cite{Artamonov:2009sz} are from invisible kaon decays.  Prospective limits from ILC are beyond the scope of this work, but are likely slightly better than those from JLab~\cite{Izaguirre:2013uxa,Andersen:2013rda}.}\label{fig:allB}
\end{figure}

In Case B (\figref{fig:allB}), new exclusions relative to minimal models
are again found from CHARM  and $\nu$-Cal~I between $250\,\mev \lesssim m_x \lesssim 400\,\mev$.  
Unlike in Case A, there is some parameter space sensitive to the Higgs mass mixing that is not excluded by BaBar: a small region around $m_x \approx 1\,\gev$ can be excluded by MINOS with the mass mixing, but not without it.

The additional exclusions can be avoided in generic hidden sectors in two ways: giving the hidden vector a St\"uckelberg mass, or making the hidden Higgs decay to the hidden sector.  If the hidden vector has a St\"uckelberg mass, then there is no hidden Higgs and the limits obviously do not apply.  Higgs decays of the form $\hx \to \chi^x \chi^x$ also trivially sidestep these constraints.  Note, though, that in Case A this is only possible if $m_{\hx} > m_x$, which is not possible in our minimal supersymmetric model.  Finally, a more complex hidden sector might force the hidden states produced in the $\rho$ decay to return to the visible sector only after a complex hidden cascade; such a scenario might suppress the total acceptance enough that no limits can be set.

\begin{figure}[ttt]
  \centering
  \includegraphics[width=\textwidth]{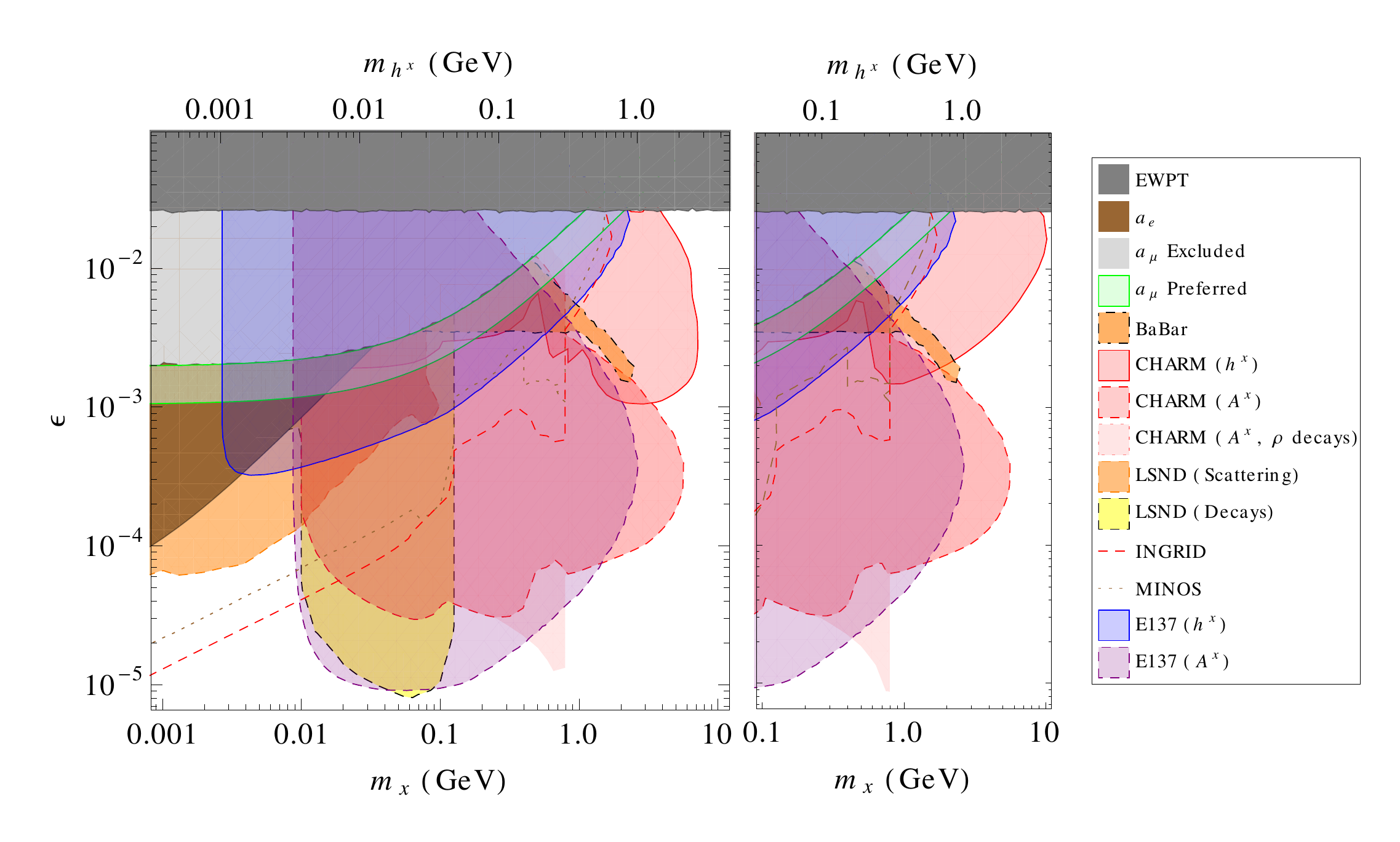}
  \caption{Composite plot of all non-cosmological limits we find for Case C.  
  Limits from invisible kaon decays are much weaker than the limits shown.  If the limits from late decays discussed in \secref{sec:cosmo} are included, they rule out all masses $m_x \lesssim 1\,\gev$ except for a narrow region just above the $\hx \to \mu^+ \mu^-$ threshold.}\label{fig:allC}
\end{figure}

In Case C (\figref{fig:allC}), we have limits that substantially extend 
the reach and type of those of \refcite{Schuster:2009au}
in the supersymmetric hidden sector we consider.  
We show limits from hidden Higgs and pseudoscalar decays separately for each experiment where relevant.  As discussed in \secref{sec:limits}, the latter are more model-dependent and, in particular, can be avoided if CP is violated in the hidden sector.  In either case, we find that both electron and proton experiments comfortably exclude the region of parameter space where this benchmark can explain $a_\mu$.  The effects of the presence or absence of the Higgs mass mixing are most obvious here; note that BaBar only sets independent limits without it. 


\begin{figure}[ttt]
  \centering
  \includegraphics[width=0.8\textwidth]{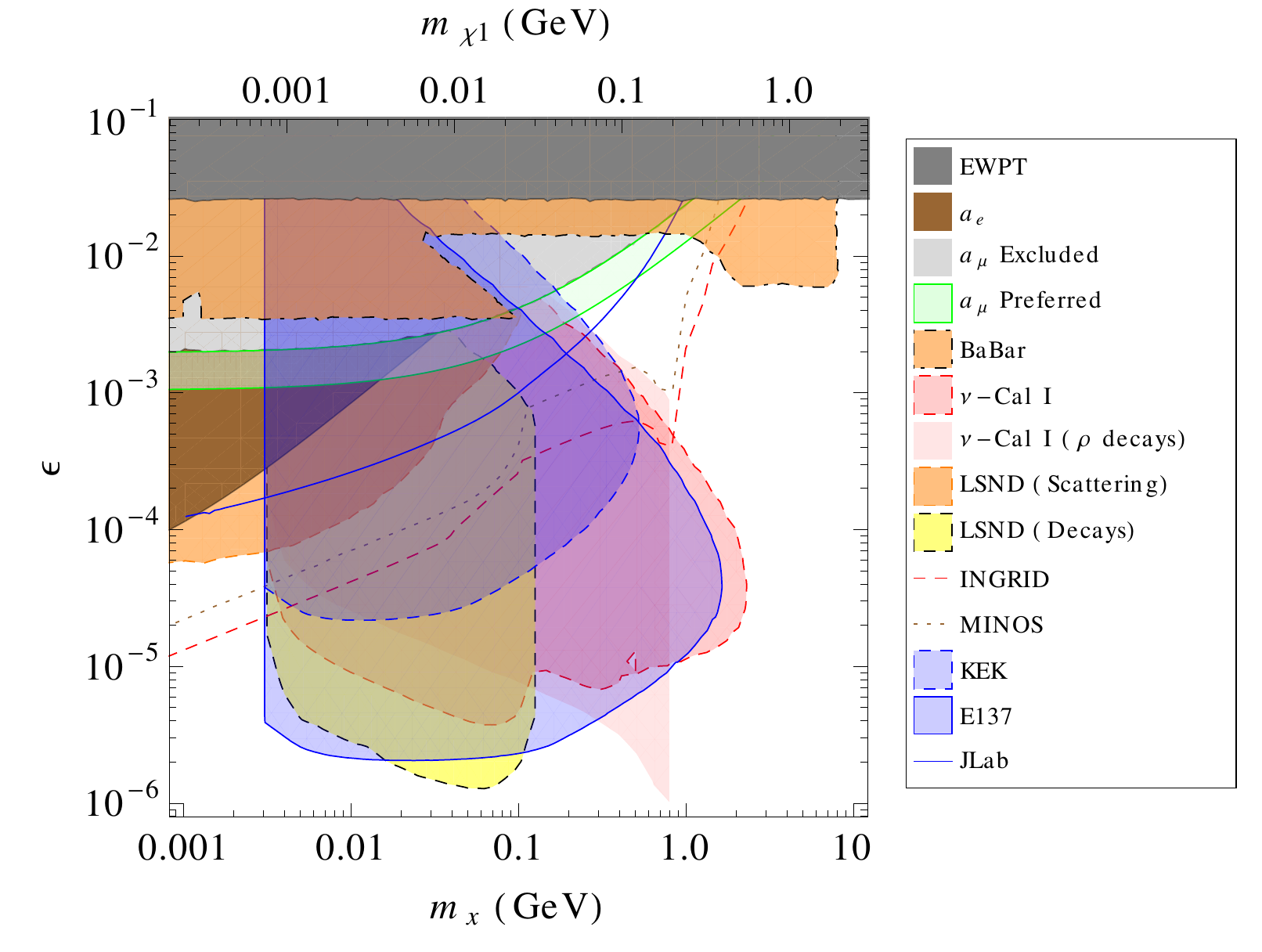}
  \caption{Composite plot of all non-cosmological limits we find for Case D.  Limits from invisible kaon decays are much weaker than the limits shown.  Including the limits from \secref{sec:cosmo} would exclude $m_x \lesssim 3\,\mev$, below which the decay $\chi_2^x \to \chi_1^x e^+ e^-$ is kinematically forbidden.  However, their remains significant allowed parameter space at small $\epsilon$ even after this.}\label{fig:allD}
\end{figure}

Finally, in Case D (\figref{fig:allD}), we find the first limits on hidden sectors where the vector decays to hidden fermions with visible decays.  We have a diverse mix of limits from searches for visible and invisible final states, and reach to low values of $\epsilon$ thanks to the relatively short $\chi_2^x$ lifetime.  Most interestingly, uniquely among our benchmarks this model has substantial parameter space still allowed where the $a_\mu$ discrepancy can be explained; this is due to the hidden Higgs decaying within the hidden sector.  This region of parameter space will be probed by JLab and can be definitively excluded by INGRID.

  The four parameter slopes considered in this paper cover the most likely
experimental ``phases'' of the theory.  Deviating away from these specific
slopes will change the decay lifetimes of metastable states
(such as $A^x$ and $\chi_2^x$), but will not tend to alter the
qualitative signatures.  Thus, a similar set of experimental analyses
are expected to be applicable to the general parameter space of the theory.  
The only exception would occur if the hidden vector had multiple significant
hidden decay modes, \emph{i.e.} to both hidden scalars and fermions.  
In this case, the limits we have found would still apply, but would 
be suppressed by branching ratios.

  In summary, the model presented here strikes a convenient balance between simplicity and flexibility.  From a supersymmetric standpoint, it is minimal.  Despite this, it not only encompasses the most common hidden sectors previously considered, but it also extends those to situations with new and ineteresting phenomenology.  For this reason, it serves as a convenient framework for studying general hidden sectors, including non-supersymmetric ones.  We hope that the new hidden sector phenomenology observed here might encourage our experimental colleagues to perform dedicated searches for them.

\section*{Acknowledgements}

  We would like to thank Sonia Bacca, Nikita Blinov, Christopher Carone, 
Rouven Essig, Eder Izaguirre, Patrick de Niverville, Adam Ritz, 
Philip Schuster, Mark Sher, Michael Spannowsky, Natalia Toro, and Itay Yavin for useful discussions.  We also thank an anonymous referee for their comments.
This work was supported by the Australian Research Council
and the National Sciences and Engineering Research Council of Canada~(NSERC).

\newpage


\appendix

\section{Masses and Mixings in the Hidden Sector}\label{app:hidden}

  In this appendix we describe in more detail the masses and mixings 
within the minimal supersymmetric hidden sector formulated 
in Refs.~\cite{Morrissey:2009ur,Chan:2011aa} 
and investigated in this work.  
We also show how the hidden states connect to the visible MSSM sector,
and we describe the calculation of the decay of $h^x_1$ to SM states.

\subsection{Masses for $\epsilon \to 0$}
  
  In the absence of kinetic mixing, $\epsilon \to 0$, 
the hidden and visible sectors decouple and it 
is straightforward to obtain the mass eigenstates.  
The scalar components of $H$ and $H'$ are assumed to develop real VEVs 
near or below the GeV scale,
\beq
  \langle H\rangle = \eta\,\sin\zeta,~~~~~\langle H'\rangle = \eta\,\cos\zeta \ ,
\eeq
with $\zeta \in [0,\pi/2]$.  As a result, the hidden vector $X_{\mu}$ obtains 
a mass equal to
\beq
  m_x = \sqrt{2}\, g_x\eta \ .
\eeq
The remaining physical bosons are a pair of real scalars $h_1^x$ and $h_2^x$ 
(with $m_{h_1^x} \leq m_{h_2^x}$), and a 
pseudoscalar $A^x$.\footnote{We do not consider CP violation 
in the present work.}  
In terms of the vector and pseudoscalar masses, the scalar mass matrix 
is\footnote{We will make use of the abbreviation, 
$c_\zeta$ ($s_\zeta$) = $\cos \zeta$ ($\sin \zeta$).}
\beq
\mathcal{M}^2_{h^x} = \left(
\begin{array}{cc}
m_x^2s_{\zeta}^2+m_{A^x}^2c_{\zeta}^2&-(m_x^2+m_{A^x}^2)s_{\zeta}c_{\zeta}\\
-(m_x^2+m_{A^x}^2)s_{\zeta}c_{\zeta}&m_x^2c_{\zeta}^2+m_{A^x}^2s_{\zeta}^2
\end{array}
\right) \ .
\label{eq:hmass}
\eeq
This matrix is identical in structure to the MSSM scalar Higgs mass matrix, 
and implies the tree-level relations
\beq
  m_{h_1^x} \leq m_x|\cos\zeta|,~~~\text{and}~~~m_{h_1^x} \leq m_{A^x}|\cos\zeta| \ ,
\eeq
as well as
\beq
  m_{h_1^x}^2+m_{h_2^x}^2 = m_x^2+m_{A^x}^2 \ .
\eeq
We will also write
\begin{equation}\label{eq:hmix}
  \begin{split}
    H &= \eta \, s_{\zeta}+\frac{1}{\sqrt{2}}  \left(R_{1a} \, h_a^x + i c_{\zeta} \, A^x\right) \ , \\
    H' &= \eta \, c_{\zeta}+\frac{1}{\sqrt{2}} \left(R_{2a} \, h_a^x + i s_{\zeta} \, A^x\right) \ ,
  \end{split}
\end{equation}
where $R$ is the real orthogonal matrix that diagonalizes the masses.

  Spontaneous symmetry breaking also mixes the $U(1)_x$ gaugino with 
the fermion components of $H$ and $H'$ to yield three hidden neutralinos 
$\chi_{1}^x,\,\chi_{2}^x,\,\chi_{3}^x$ 
(with $|m_{\chi_1^x}| < |m_{\chi_2^x}| < |m_{\chi_3^x}|$).  
The entire spectrum of the theory can be specified by 
the seven input parameters $\{g_x,\mu',m_x,m_{A^x},M_x,\tan\zeta,\epsilon\}$.

\subsection{Mixing with the MSSM}
 
When the kinetic mixing parameter $\epsilon$ is non-zero, the mixed term in \modeqref{eq:svecport} induces  kinetic mixing between hidden and visible gauge bosons and gauginos, as well as further mixing among the auxiliary fields.  In components, 
\begin{align}
  \lag & \supset -\frac{1}{4} B^{\mu\nu} B_{\mu\nu} - \frac{1}{4} X^{\mu\nu} X_{\mu\nu} - \frac{\epsilon}{2c_W} \, X^{\mu\nu} B_{\mu\nu} \notag \\
  & \quad + \frac{1}{2} i \, \tilde{B}^\dagger \sigma \ccdot \partial \tilde{B} + \frac{1}{2} i \, \tilde{X}^\dagger \sigma \ccdot \partial \tilde{X} + \frac{\epsilon}{2c_W} i \, \tilde{B}^\dagger \sigma \ccdot \partial \tilde{X} + \frac{\epsilon}{2c_W} i \, \tilde{X}^\dagger \sigma \ccdot \partial \tilde{B} \notag\\
  & \quad + \frac{1}{2} D_B^2 + \frac{1}{2} D_X^2 + \frac{\epsilon}{c_W} \, D_X D_B .
\label{eq:kmixcomponent}\end{align}
where $c_W = \cos\theta_W$ refers to the weak mixing angle.
The kinetic terms can be brought to their canonical form by making 
field redefinitions at the expense of generating new interactions 
among the visible and hidden sectors.

The gauge kinetic terms can be diagonalized to leading order in 
$\epsilon$ by the transformations
\begin{align}
  X_\mu & \to X_\mu + \epsilon \, t_W \, Z_\mu \ , \notag \\
  A_\mu & \to A_\mu - \epsilon \, X_\mu \ , \\
  Z_\mu & \to Z_\mu \ , \notag
\end{align}
where the fields $A_\mu$ and $Z_\mu$ are defined as in the SM, 
and $t_W$ is the tangent of the Weinberg angle.  These transformations 
are not unique, but they have the advantage that the mass mixing induced 
between $Z$ and $X$ is very small, suppressed by both 
$\epsilon$ and $m_x^2/m_Z^2$, and has a negligible effect 
on the mass eigenvalues.  

  For the gauginos, neglecting a possible soft mass mixing term, 
the kinetic mixing is most cleanly removed by the 
redefinition\footnote{No mass mixing is induced by gauge mediation, 
while the residual gravity-mediated contribution is expected to be suppressed 
relative to the $\tilde{X}$ mass by a factor of 
$\epsilon$~\cite{Morrissey:2009ur}.}
\begin{equation}\label{eq:inos}
  \begin{split}
    \tilde{B} & \to \tilde{B} - \frac{\epsilon}{c_W} \, \tilde{X} \ , \\
    \tilde{X} & \to \tilde{X} \ ,
  \end{split}
\end{equation}
to leading order in $\epsilon$.  This transformation generates a coupling 
between the visible neutralinos and the hidden sector.  The induced mass mixing 
between the visible and hidden sectors is again negligible, 
suppressed by both $\epsilon$ and the ratio of the hidden to visible 
neutralino masses.  

  Turning next to the scalars, the last term in \modeqref{eq:kmixcomponent} 
mixes the hidden and visible $D$ terms.  Let us represent the visible sector scalars by $\phi_i$ with hypercharge $Y_i$, and the hidden scalars by $\varphi_j = \{ H, H' \}$ with $U(1)_x$ charge $x_j$.    
Upon integrating out the auxiliary fields, the $D$-term potential becomes~\cite{Kumar:2011nj}
\begin{align}
  V_D & = \frac{1}{2} c_\epsilon^2 g_Y^2 \biggl( \sum_i Y_i \abs{\phi_i}^2 \biggr)^2 + \frac{1}{2} c_\epsilon^2 g_x^2 \biggl( \sum_j x_j \abs{\varphi_j}^2 \biggr)^2 \notag \\
  & \quad - s_\epsilon c_\epsilon g_x g_Y \biggl( \sum_i Y_i \abs{\phi_i}^2 \biggr) \biggl( \sum_j x_j \abs{\varphi_j}^2 \biggr) \ , \label{eq:dtermpot}
\end{align}
where $c_\epsilon = 1/\sqrt{1 - \epsY^2}$ and $s_\epsilon = \epsY \, c_\epsilon$.  The last term in this expression is precisely a Higgs portal interaction.

For $\eta \ll v$ (as we assume here) the modified $D$-term potential has a negligible effect on electroweak symmetry breaking, but it plays an important role in inducing symmetry breaking in the hidden sector.  When the MSSM Higgs fields develop VEVs, driven almost entirely by the larger MSSM soft terms, an effective Fayet-Iliopoulos term is generated for $U(1)_x$ which pushes the $H$ and $H'$ fields to develop VEVs of their own.\footnote{In the absence of soft terms in the hidden sector, this would also cause the spontaneous breaking of supersymmetry in that sector~\cite{Morrissey:2009ur}.}  

Expanding the visible and hidden Higgs fields about their VEVs, the Higgs portal interaction also induces a mass mixing between the visible and hidden mass eigenstates.  This has only a very small effect on the mass eigenvalues, with deviations from the $\epsilon \to 0$ values suppressed by at least $\epsilon\,m_x^2/m_z^2$.   For this reason it is an excellent approximation to separately diagonalise the scalars in the visible and hidden sectors, and treat the mixing as a mass insertion.  To be explicit, we will assume that the visible Higgs sector has the form of the MSSM~\cite{Martin:1997ns}, with mass eigenstates $H^{\pm}$, $A^0$, and
\beq
  \left(
  \begin{array}{c}
  h^0\\
  H^0
  \end{array}
  \right)
  = 
  \left(
  \begin{array}{rr}
  \cos{\alpha}&-\sin\alpha\\
  \sin\alpha&\cos\alpha
  \end{array}
  \right)
  \left(
  \begin{array}{c}
  \sqrt{2}\left[Re(H_u^0)-v_u\right]\\
  \sqrt{2}\left[Re(H_d^0)-v_d\right]
  \end{array}
  \right) \ .
\eeq
The mass mixing term is then\footnote{This corrects the original version of \refcite{Chan:2011aa}.}
\beq
  -\mathscr{L}\supset 
  - \epsilon \, c_\epsilon^2 t_W m_x m_Z  
  \left[\sin(\alpha\!+\!\beta)\,h^0 -\cos(\alpha\!+\!\beta)\,H^0 \right]
  \left(s_{\zeta}R_{1a}-c_{\zeta}R_{2a}\right)\,h_a^x \ \ 
\label{eq:massmixing}
\eeq
  In the MSSM Higgs decoupling limit, $\alpha = \beta-\pi/2$, this reduces to
\beq
  -\mathscr{L}\supset 
  \epsilon \, c_\epsilon^2 t_W m_x m_Z 
  \bigl( s_\zeta R_{1a} - c_\zeta R_{2a} \bigr)h_a^x
  \left[\cos(2\beta)h^0 + \sin(2\beta)H^0\right] \ .
\eeq

\subsection{Decays of the Light Scalar}\label{sec:hxdecays}

\begin{figure}
  \centering
  \includegraphics[scale=0.5, bb=0 0 617 151]{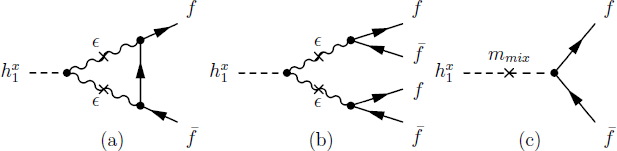}
  \caption{Three main contributions to the decay of the hidden sector Higgs: (a) loop decay through vector kinetic mixing; (b) four-body decay through vector kinetic mixing; (c) two-body decay through $D$-term-induced mass mixing.\label{fig:hdec_feyn}}
\end{figure}

  The $h_1^x$ scalar is the lightest $R$-even state in the hidden sector.
When it is lighter than twice the mass of the lightest neutralino $\chi_1^x$,
it will decay exclusively to the SM.  The three leading contributions 
to the decay width are shown in \figref{fig:hdec_feyn}.  The first two are 
discussed in detail in \cite{Batell:2009yf}, 
and require a pair of insertions of $\epsilon$ in the amplitudes 
as well as the coupling of the hidden scalar to the hidden vector,
\begin{equation}
  \lag \supset 
  g_xm_x\,h^x_a X_\mu X^{\mu} \bigl( s_\zeta R_{1a} + c_\zeta R_{2a} \bigr) \ .
\label{eq:hXXcoupling}\end{equation}
The third contribution to the decay comes from the Higgs portal interaction 
of \modeqref{eq:massmixing}~\cite{Chan:2011aa}.  This piece has only one power 
of $\epsilon$ in the amplitude, but also receives suppression by 
$m_x^2/m_Z^2$ and small SM Yukawa couplings.  In the absence of interference 
with the loop-mediated two-body channel and working in the MSSM decoupling limit, 
the Higgs-portal-induced decay width is
\beq
  \Gamma(h_1^x) = 
  \epsilon^2 t_W^2 c^2_{2\beta}
  \bigl( s_\zeta R_{1a} - c_\zeta R_{2a} \bigr)^2\,\lrf{m_xm_Z}{m_h^2}^2
  \Gamma(h^0_{SM},m_h=m_{h_1^x}) \ ,
\label{eq:hxdecayhiggs}\eeq
where $\Gamma(h^0_{SM},m_h=m_{h_1^x})$ is the width the SM Higgs would 
have if its mass were equal to $m_{h_1^x}$.  
The branching fractions  will also follow those of a SM Higgs 
with $m_h=m_{h_1^x}$.  Decays through mixing with the heavier Higgs 
will generally be suppressed unless $\tan\beta$ is close to unity, 
when the mixing with $h^0$ vanishes.

\begin{figure}
  \centering
  \includegraphics[width=0.8\textwidth]{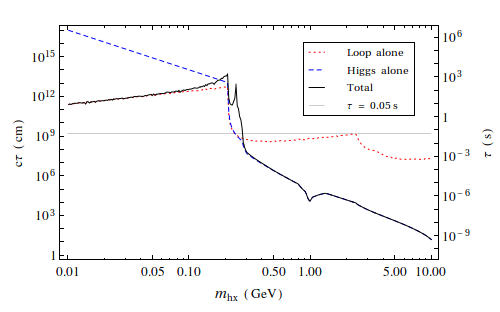}
  \caption{Decay length ($c\tau$) for a hidden sector Higgs \hx\ for different decay mechanisms.  In making this figure, we have fixed $\epsilon = 10^{-3}$, $\tan \beta \to \infty$, and the other parameters as in benchmark C.  The horizontal line shows $\tau = 0.05$~s, relevent for cosmological constraints as discussed in the text.\label{fig:h1ctau}}
\end{figure}

In \figref{fig:h1ctau} we plot the total decay length $c\tau$ of $h_1^x$ as a function of its mass.  For this plot, we use $\epsilon = 10^{-3}$ and the hidden sector benchmark C, together with the decoupling and large $\tan\beta$ limit of the MSSM.  The results for benchmarks A and B are qualitatively similar, while for benchmark D the scalar decays promptly to the hidden sector, $\hx \to \chi_1^x \chi_1^x$. 
This result includes the interference between the loop and Higgs-portal contributions to the amplitude.  We also show in \figref{fig:h1ctau} lines corresponding to what the decay length would be with the Higgs-portal alone or with the loop-induced vector-mediated decay alone.  

For $2m_{\pi} < m_{h_1^x} \lesssim 1\,\gev$ there is a significant uncertainty in the Higgs width to hadronic modes.  For the mass mixing contribution, we use the parametrization of \refcite{Donoghue:1990xh} to compute the total rate and the branching fractions.  A comparison of this and other estimates is given in \refcite{Clarke:2013aya} (see also \cite{Bezrukov:2009yw,Bezrukov:2013fca}).  For the loop diagram, we do not compute any contribution for hadronic decays below $\Lambda_{QCD} \approx 2.5$~\gev; this represents an additional theoretical uncertainty.  

We would like to estimate the size of this uncertainty, and its corresponding effect on the limits we extract in \secrefs{sec:ebeam}{sec:hbeam}.  When we have no mass mixing, our limits are surprisingly robust to the omission of any hadronic decay modes below $\Lambda_{QCD}$.  The upper bounds on $\epsilon$ we find in this region of parameter space come when $c\tau_{\hx} \sim 10^4$~m.  This is large compared to the typical size of experiments, especially once a boost of $\gamma \sim 10~$--100 is included.  The decay probability \modeqref{eq:pdec} is then
\begin{equation}
  P_{dec} \sim \frac{L_{dec}}{\gamma c \tau} \propto \Gamma \,.
\label{eq:pdecapprox}\end{equation}
Increasing the total width will increase the number of hidden states decaying within our detector.  This is offset by the fact that the extra width is to hadrons, while all searches tagged on leptons.  We thus pay a branching ratio penalty $\propto 1/\Gamma$, cancelling the factor from $P_{dec}$.

This argument holds provided that the width does not increase too much, specifically that the approximation \eqref{eq:pdecapprox} is valid.  This is why our exclusions change with the inclusion of mass mixing; those limits are set for parameters where the scalar lifetime $c\tau_{\hx} \sim 10$--100~m.  Consequently, those limits are more sensitive to uncertainties in the hidden Higgs width.  The sensitivity is greatest to increases in the width, which would suppress signals by having more decays within the shield.  Decreasing the width would make the limits stronger, approaching the exclusions without mass mixing.  We note that our choice for the Higgs hadronic width is both the most recent and among the largest of those discussed in \refcite{Clarke:2013aya}, suggesting the main theoretical uncertainty makes our limits conservative.

\section{Details of Hidden Vector Production at Electron Beam Dumps}\label{app:edetails}
In \secref{sec:ebeam}, we sketched the calculation of the number of hidden sector events at an electron beam dump.  Here we provide some of the details omitted there.  

\subsection{Effective Flux}\label{app:effflux}

We begin by considering the effective flux $\chi$, introduced in \modeqref{eq:wwonshell} that contains the form factors describing the details of the target.  This quantity was defined in \refcite{Kim:1973he} as
\begin{equation}
  \chi = \frac{1}{2M_i} \int_{t_{min}}^{t_{max}} dt \int_{M_i^2}^{M_{up}^2} dM_f^2 \, \frac{1}{t^2} \, \bigl( 2 t_{aux} \, W_1 (t, M_f^2) + (t - t_{min}) \, W_2 (t, M_f^2) \bigr) \, .
\end{equation}
Here, $t = -q^2$ ranges between $t_{min} = (m_x/2E_e)^2$ and $t_{max} = m_x^2$; $M_f$ is the mass of the target final state, and ranges between $M_i$ and $M_{up}^2 \approx M_i (M_i + 2E_e)$.  $W_{1,2}$ are form factors describing the target and including both elastic and inelastic behaviour, and $t_{aux} \equiv (p_e \cdot k_X - \frac{1}{2} m_x^2)/(E_e - E_x)$.

We use analytic approximations for the form factors taken from \refscite{Kim:1973he,Tsai:1973py}.  These are given in terms of the partially integrated quantities $G$:
\begin{equation}
  G_{1,2} (t) = \frac{1}{2M_i} \int_{M_i^2}^{M_{up}^2} dM_f^2 \ W_{1,2} (t, M_f^2) \,.
\end{equation}
The form factor $W_1$ has a negligible contribution to hidden vector production~\cite{Bjorken:2009mm}.  We split $G_2$ into a sum of elastic and inelastic contributions, which in turn are the product of atomic and nuclear contributions.  The elastic term is
\begin{equation}
  G_2^{el} (t) = \biggl( \frac{a^2 t}{1 + a^2 t} \biggr)^2 \, \biggl( \frac{1}{1 + t/d} \biggr)^2 \, Z^2 .
\end{equation}
The parameters $a$ and $d$ are well approximated as $a \approx 111 /( m_e\sqrt[3]{Z})$ and $d \approx 0.164$~GeV$^2/\sqrt[3]{A^2}$, with $Z$ and $A$ the usual atomic and mass numbers.  The terms in the first (second) brackets describe the effects of atomic screening (finite nuclear size).

For the inelastic term we take, based on \refcite{Tsai:1973py},
\begin{equation}
  G_2^{inel} =  \biggl( \frac{a'^2 t}{1 + a'^2 t} \biggr)^2 \, C(t) \, \bigl[ Z \, G_{2p}^{el} + (A - Z) \, G_{2n}^{el} \bigr] \,.
\end{equation}
The first set of brackets again describes the atomic screening factor, with $a' \approx 773/(m_e \sqrt[3]{Z^2})$.  The remaining terms describe the inelastic nuclear form factors, in terms of a prefactor $C$ and nucleon functions $G_{2p,n}^{el}$.  The prefactor is
\begin{equation}
  C(t) = \begin{cases}
  1 & \text{if } Q > 2P_F = 0.5 \text{ GeV} \\
  \frac{3Q}{4 P_F} \biggl[ 1 - \frac{Q^2}{12 P_F^2} \biggr] & \text{otherwise,}
  \end{cases}
\end{equation}
with $Q$ defined by
\begin{equation}
  Q = \frac{t^2}{4m_p^2} + t \,;
\end{equation}
while the nucleon functions are
\begin{equation}
  \begin{split}
    G_{2p}^{el} & = \frac{m_p}{M_i} \, \frac{1}{(1 + t/t_0)^4} \, \frac{1 + 2.79^2 \tau}{1 + \tau} \,, \\
    G_{2n}^{el} & = \frac{m_p}{M_i} \, \frac{1}{(1 + t/t_0)^4} \, \frac{1.91^2 \tau}{1 + \tau} \,,
  \end{split}
\end{equation}
where $t_0 = 0.71$~GeV$^2$ and $\tau = t/(4m_p^2)$.

\begin{figure}
  \centering
  \includegraphics[width=0.6\textwidth]{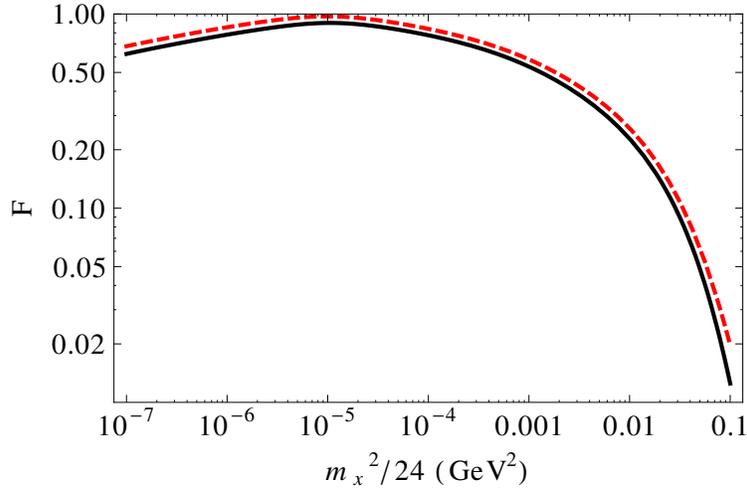}
  \caption{The scaled flux $F$ as a function of $q_{min} = m_x^2/(2E)$ for an electron beam of energy $E = 12$~GeV impacting an aluminium target.  The black, solid line represents our choice of form factors, while the red, dashed line corresponds to that of \emph{e.g.} \refcite{Izaguirre:2013uxa} (compare Fig.~4 of that reference).}\label{fig:normflux}
\end{figure}

Note that \refscite{Bjorken:2009mm,Andreas:2012mt,Izaguirre:2013uxa} used a different form for $G_2^{inel}$, specifically the expressions given for elastic scattering from protons in \refcite{Kim:1973he}.  The difference between the two choices is small, as the elastic contribution to $G_2$ is dominant.  We compare the two in \figref{fig:normflux} using the scaled flux
\begin{equation}
  F = \frac{4}{3} \alpha^3 \, \chi \ \frac{X_0\, N_0}{A\, m_e^2} \,,
\end{equation}
introduced in \refcite{Izaguirre:2013uxa}.  The difference is only a few percent for the mass range of interest, with our parameterisation the smaller of the two.

\subsection{Off-Shell Production}\label{app:offshell}

The general process for hidden sector production through an off-shell vector is shown in \figref{fig:epepXoffgen} (a simple generalisation of \figref{fig:epepXoffshell}).  We can write the matrix element as
\begin{equation}
  \mathcal{M}_{2\to n+2} = \frac{1}{q^2} T_\mu^N(p_N, k_{N'}) \, S_e^{\mu\nu} (p_e, q, k_e) \, N^x_\nu (k_1, \ldots k_n) \,,
\end{equation}
with $T^N$, $S_e$ and $N^x$ respectively the target fermion line trace, the electron line trace, and the terms related to the hidden sector.  Squaring this and integrating over the $k_{N'}$ phase space  lets us introduce form factors:
\begin{multline}
  \int d \Pi_{n+2} \, \frac{1}{2N_i} \, \sum \abs{\mathcal{M}_{2\to n+2}}^2 = \int \Biggl( \prod_{i=1}^n \frac{d^3 k_i}{(2\pi)^3 2E_i} \Biggr) \, \frac{d^3 k_e}{(2\pi)^3 2E_{ef}} \, \frac{e^2}{(q^2)^2} \, 2 \pi M_i \\
  \times \biggl\{ - W_1 \eta_{\mu\nu} + W_2 \, \frac{p_{N\mu} p_{N\nu}}{M_i^2} \biggr\} \sum S_e^{\mu\rho} (S_e^{\nu\sigma})^\ast N^x_\rho (N^x_\sigma)^\ast ,
\end{multline}
with $M_i$ the target mass and $E_{ef}$ the final state electron energy.

\begin{figure}
  \begin{center}
    \begin{fmffile}{epepXgen1}
      \begin{fmfgraph*}(150,60)
        \fmfstraight
        \fmfleft{i1,ia,i2,i3} \fmfright{o0,oa,o1,o2,o3}
        \fmf{fermion,label=$p_e$}{i2,v1}
        \fmf{dbl_plain,label=$\overrightarrow{p_N}$,label.side=right}{i1,v3}
        \fmf{dbl_plain,label=$\overrightarrow{k_{N'}}$,label.side=right}{v3,o0}
        \fmfblob{.08w}{v3}
        \fmf{fermion,label=$q_a$,label.side=left}{v1,v2}
        \fmf{fermion,label=$k_e$,label.side=right}{v2,o1}
        \fmf{boson,label=$q\uparrow$,label.side=left,tension=0.4}{v3,v1}
        \fmffreeze
        \fmf{boson,label=$k_x$,label.side=left}{v2,v4}
        \fmf{plain}{o2,v4,o3}
        \fmfblob{.05w}{v4}
        \fmf{dots}{o2,o3}
      \end{fmfgraph*}
    \end{fmffile}
    \begin{fmffile}{epepXgen2}
      \begin{fmfgraph*}(150,60)
        \fmfstraight
        \fmfleft{i1,ia,i2,i3} \fmfright{o0,oa,o1,o2,o3}
        \fmf{fermion,label=$p_e$}{i2,v1}
        \fmf{fermion,label=$q_b$}{v1,v2}
        \fmf{fermion,label=$k_e$}{v2,o1}
        \fmf{dbl_plain,label=$\overrightarrow{p_N}$,label.side=right}{i1,v3}
        \fmf{dbl_plain,label=$\overrightarrow{k_{N'}}$}{v3,o0}
        \fmfblob{.08w}{v3}
        \fmf{boson,label=$\uparrow q$,label.side=right,tension=0.4}{v3,v2}
        \fmffreeze
        \fmf{boson,label=$k_x$,label.side=left}{v1,v4}
        \fmf{plain}{o2,v4,o3}
        \fmfblob{.05w}{v4}
        \fmf{dots}{o2,o3}
      \end{fmfgraph*}
    \end{fmffile}
  \end{center}
  \caption{The two leading Feynman diagrams for production of off-shell massive vector in electron-fixed target scattering.  The unlabelled lines represent the $n$ hidden sector states ultimately produced through the hidden vector.}\label{fig:epepXoffgen}
\end{figure}

To generalise the WW approximation, we follow \refcite{Kim:1973he} in rewriting the phase space integral over $k_e$ into an integral over $M_f^2$, $t = -q^2$ and an angular degree of freedom $\phi_S$.  This is done by making use of a special frame $S$, defined as the frame where $\vec{u} \equiv \vec{p}_e + \vec{p}_N - \vec{k}_x$ vanishes.  Specifically, $\phi_S$ is the azimuthal angle of $\vec{k}_e$ relative to $\vec{p}_e - \vec{k}_x$ in this frame.  We also note that the hidden sector factors $N$ have no dependence on $k_e$, while the form factors depend only on $t$ and $M_f^2$.  This brings us to the following expression for the integrated matrix element:
\begin{multline}
  \int \frac{d \Pi_{n+2}}{2N_i} \, \sum \abs{\mathcal{M}_{2\to n+2}}^2 = \frac{\alpha}{4} \int \Biggl( \prod_{i=1}^n \frac{d^3 k_i}{(2\pi)^3 2E_i} \Biggr) \sum N^x_\rho (N^x_\sigma)^\ast (k_1, \ldots k_n)\\
  \times \frac{M_i}{\sqrt{(u\cdot p_N)^2 - M_i^2 u^2}} \int dM_f^2 \, dt \, \frac{1}{t^2}\biggl\{ - W_1 \eta_{\mu\nu} + W_2 \, \frac{p_{N\mu} p_{N\nu}}{M_i^2} \biggr\} \\
  \times \int \frac{d\phi_S}{2\pi} \, \sum S_e^{\mu\rho} (S_e^{\nu\sigma})^\ast (p_e, k_x, M_f^2, t, \phi_S) .
\label{eq:intermediate}\end{multline}
Note that the only difference so far with the derivation of \refcite{Kim:1973he} is that in that case, the integral over $k_1 \cdots k_n$ is replaced by a single integral over $k_x$, and the function $N^x_\rho (N^x_\sigma)^\ast$ is replaced by $- \eta_{\rho\sigma}$.

At this point, we sketch the derivation of \modeqref{eq:offshell}.  We introduce a trivial integral over $k_x$ using a delta function.  This lets us rewrite the integral over $N_\rho^x (N^x_\sigma)^\ast$ in terms of the vector partial decay widths:
\begin{align}
  & \int \Biggl( \prod_{i=1}^n \frac{d^3 k_i}{(2\pi)^3 2E_i} \Biggr) \sum N^x_\rho (N^x_\sigma)^\ast (k_1, \ldots k_n) \notag \\
  & = \int \Biggl( \prod_{i=1}^n \frac{d^3 k_i}{(2\pi)^3 2E_i} \Biggr) \frac{d^4k_x}{(2\pi)^4} \, (2\pi)^4 \, \delta^{(4)} (k_x - k_1 - \cdots k_n) \sum N^x_\rho (N^x_\sigma)^\ast \notag \\
  & = \int \frac{d^4k_x}{(2\pi)^4} \frac{1}{(k_x^2 - m_x^2)^2} \bigl( A(k_x^2) \, \eta_{\rho\sigma} + B(k_x^2) \, k_{x\rho} k_{x\sigma} \bigr) \,.
\end{align}
The last term vanishes when contracted with $S_e^{\mu\rho}$ via the QED Ward identity, while
\begin{equation}
  A(k_x^2) = - 2 \sqrt{k_x^2} \, \Gamma_x (k_x^2) \,,
\end{equation}
with $\Gamma_x$ the partial width for a vector of mass $m_x^2 = k_x^2$ to decay to the relevant hidden sector final state.  From this, the usual WW argument leads to \modeqref{eq:offshell}.

The WW approximation is a Taylor expansion of the (integral of the) electron line trace about the minimal momentum transfer $t_{min}$, leading to the two approximations
\begin{equation}
  \int \frac{d\phi_S}{2\pi} \, \sum S_e^{\mu\nu} (S_{e\mu\nu})^\ast (p_e, k_x, M_f^2, t, \phi_S) \approx \sum S_e^{\mu\nu} (S_{e\mu\nu})^\ast (p_e, k_x, M_i^2, t_{min}) \,,
\label{eq:approx1}\end{equation}
and
\begin{multline}
  - \int \frac{d\phi_S}{2\pi} \, \frac{p_{N\mu}p_{N\nu}}{M_i^2} \sum S_e^{\mu\rho} (S^\nu_{e\rho})^\ast (p_e, k_x, M_f^2, t, \phi_S) \\
  \approx \frac{t - t_{min}}{2 t_{aux}} \, \sum S_e^{\mu\nu} (S_{e\mu\nu})^\ast (p_e, k_x, M_i^2, t_{min}) \,.
\label{eq:approx2}\end{multline}
As noted before, $\vec{q^\ast}$ is parallel to $\vec{p}_e - \vec{k}_x$.  It follows that $\vec{k}_e$ is also parallel to $\vec{p}_e - \vec{k}_x$, allowing us to omit $\phi_S$ from the right-hand side of \twoeqref{eq:approx1}{eq:approx2}.  This approximation can be checked using the explicit expressions for $S_e^{\mu\nu}$.  It follows from the presence of the collinear pole when the electron is nearly on-shell, as well as the photon propagator that results in the dominance of processes where $t \approx t_{min}$.

The equivalent approximations for the off-shell case are easy to guess:
\begin{equation}
  \int \frac{d\phi_S}{2\pi} \, \sum S_e^{\rho\mu} (S^{\phantom{e\rho}\nu}_{e\rho})^\ast (p_e, k_x, M_f^2, t, \phi_S) \approx \sum S_e^{\rho\mu} (S^{\phantom{e\rho}\nu}_{e\rho})^\ast (p_e, k_x, M_i^2, t_{min}) \,,
\label{eq:apnew1}\end{equation}
and
\begin{multline}
  - \int \frac{d\phi_S}{2\pi} \, \frac{p_{N\mu}p_{N\nu}}{M_i^2} \sum S_e^{\mu\rho} (S^{\nu\sigma}_{e})^\ast (p_e, k_x, M_f^2, t, \phi_S) \\
  \approx \frac{t - t_{min}}{2 t_{aux}} \, \sum S_e^{\rho\mu} (S^{\phantom{e\rho}\nu}_{e\rho})^\ast (p_e, k_x, M_i^2, t_{min}) \,.
\label{eq:apnew2}\end{multline}
Once these are put into \modeqref{eq:intermediate}, \modeqref{eq:WWgenoff} is immediate.  The validity of these approximations can also be checked by direct computation.  We comment on a few points in the derivation.  The easiest way to analyse these expressions is to expand in a tensor basis that respects the QED Ward identity.  For example, the tensor in \modeqref{eq:apnew1} can only depend on $p_e$, $q^\ast$, $k_x$ and the metric, so we may define four functions $f_i$ through
\begin{equation}
  \int \frac{d\phi_S}{2\pi} \, \sum S_e^{\rho\mu} (S^{\phantom{e\rho}\nu}_{e\rho})^\ast (p_e, k_x, M_f^2, t, \phi_S) = \sum_i f_i (p_e, k_x, M_f^2, t) \, L^{\mu\nu}_i \,,
\label{eq:Lexptr}\end{equation}
with (for example)
\begin{align*}
  L_1^{\mu\nu} & = \frac{k_x^2}{p_e \cdot k_x} \, p_e^\mu p_e^\nu + p_e \cdot k_x \, \eta^{\mu\nu} - p_e^\mu k_x^\nu - k_x^\mu p_e^\nu \,, \displaybreak[0]\\
  L_2^{\mu\nu} & = \frac{k_x^2}{q^\ast \cdot k_x} \, q^{\ast\mu} q^{\ast\nu} + q^\ast \cdot k_x \, \eta^{\mu\nu} - q^{\ast\mu} k_x^\nu - k_x^\mu q^{\ast\nu} \,, \displaybreak[0]\\
  L_3^{\mu\nu} & = \bigl( p_e \cdot k_x \, q^{\ast\mu} - q^\ast \cdot k_x \, p_e^\mu \bigr) \, (\mu \to \nu) \,, \displaybreak[0]\\
  L_4^{\mu\nu} & = k_x^2 \, \eta^{\mu\nu} - k_x^\mu k_x^\nu \,.
\end{align*}
The approximation of \modeqref{eq:apnew1} is then no more than replacing the functions $f_i$ of \modeqref{eq:Lexptr} by their values at the minimal kinematics $t = t_{min}$, $M_f^2 = M_i^2$.  We know that these functions are finite and non-zero at this point, as it corresponds to the physical point for photon-electron scattering; while the factor of $t^{-2}$ in \modeqref{eq:intermediate} leads the contribution from this region to generically dominate.

The approximation of \modeqref{eq:apnew2} is slightly more complicated.  The tensor expansion is na\"\i vely more involved, as we also have a possible dependence on $p_N$.  However, because $\vec{q^\ast}$ is parallel to $\vec{p}_e - \vec{k}_x$, we can rewrite terms involving $p_N^{\mu}$ in terms of $p_e^\mu - k_x^\mu$ and $q^{\ast\mu}$; the basis $L_i^{\mu\nu}$ is still sufficient.  So we have the expansion
\begin{equation}
  \int \frac{d\phi_S}{2\pi} \, \frac{p_{N\mu}p_{N\nu}}{M_i^2} \sum S_e^{\mu\rho} (S^{\nu\sigma}_{e})^\ast (p_e, k_x, M_f^2, t, \phi_S) = \sum_i g_i (p_e, k_x, M_f^2, t) \, L^{\mu\nu}_i \,.
\label{eq:Lexpen}\end{equation}
The WW approximation involves using not the first term in the Taylor expansion of the $g_i$, but the second.  The leading term is suppressed; specifically, it is not singular when the intermediate electron goes on-shell.  The term linear in $t$ \emph{is} singular in that limit, so is the first term we keep.

The final step is to relate the linear terms in \modeqref{eq:Lexpen} and the leading terms in \modeqref{eq:Lexptr}.  It is clear that \modeqref{eq:apnew2} will hold if
\begin{equation}
  g_i (t) = g_i(t_{min}) + \frac{t - t_{min}}{2t_{aux}} f_i(t) + \order{t - t_{min}}^2 \,.
\label{eq:keyassum}\end{equation}
This identification only holds approximately, to leading order in small parameters $\delta/E$ where 
\begin{equation}
  E \in \{ E_e, E_x, E_e - E_x\} \quad \text{and} \quad \delta \in \biggl\{ \sqrt{t_{aux}}, \frac{M_f^2 - M_i^2}{2M_i}, \frac{m_e^2}{\sqrt{t_{aux}}}, \frac{k_x^2}{\sqrt{t_{aux}}} \biggr\} \,.
\end{equation}
The assumption that $\delta/E$ is small is also part of the WW approximation.

There is one further non-trivial point, which is that in the basis we have chosen the function $g_3$ does not actually satisfy \modeqref{eq:keyassum}.  This is not an issue because when this term is contracted with $N_\rho^x (N^x_\sigma)^\ast$, it always gives to a subleading contribution in $\delta/E$.  Since $N_\rho^x (N^x_\sigma)^\ast$ can be expanded in terms of the metric and the hidden state momenta $k_i$, we need consider two contractions.  The trace terms are
\begin{align}
  \eta_{\mu\nu} \, g_1 \, L_1^{\mu\nu} & \sim \frac{1}{t_{aux}} & \eta_{\mu\nu} \, g_2 \, L_2^{\mu\nu} &\sim \frac{1}{t_{aux}} \,, \notag \\
  \eta_{\mu\nu} \, g_3 \, L_3^{\mu\nu} & \sim \frac{1}{\sqrt{t_{aux}}} & \eta_{\mu\nu} \, g_4 \, L_4^{\mu\nu} & \sim \frac{k_x^2}{t_{aux}^{3/2}} \,,
\end{align}
and the third term is suppressed relative to the first two.  For the contraction with terms of the form $k_{i\mu} k_{j\nu}$, note that in general the angular separation of the hidden state terms is large compared to the separation between $\vec{p}_e$ and $\vec{k}_x$; so terms of the form $\{ p_e, q^\ast, k_x \} \cdot k_i$ will not be small.  Representing such scalar products by $\Lambda$, we have
\begin{align}
  k_{i\mu} \, k_{j\nu} \, g_1 \, L_1^{\mu\nu} & \sim \Lambda^2 & k_{i\mu} \, k_{j\nu} \, g_2 \, L_2^{\mu\nu} &\sim \Lambda^2 \,, \notag \\
  k_{i\mu} \, k_{j\nu} \, g_3 \, L_3^{\mu\nu} & \sim \Lambda^2 t_{aux} & k_{i\mu} \, k_{j\nu} \, g_4 \, L_4^{\mu\nu} & \sim \Lambda^2 \,.
\end{align}
Again, the third term is suppressed compared to the others.  It follows that the approximation of \modeqref{eq:apnew2} is valid within the WW assumptions.

\subsection{Vector Decay Widths}\label{app:vecwidth}

The partial widths for a vector to decay to hidden states, used in \secrefs{sec:ebeam}{sec:hbeam}, are given here.  For the Higgsstrahlung ``decay'', we have
\begin{equation}
  \Gamma_{Ha} (k_x^2) = \frac{\alpha_x \, N_{H,a}^2}{12} \, \sqrt{k_x^2} \, \sqrt{I \bigl( \mu_a^2 , \mu_{x}^2 \bigr)} \bigl[ I \bigl( \mu_a^2 , \mu_{x}^2 \bigr) + 12 \mu_x^2 \bigr] \,;
\end{equation}
for the scalar final states
\begin{equation}
  \Gamma_{Sa} (k_x^2) = \frac{\alpha_x \, N_{S,a}^2}{12} \, \sqrt{k_x^2} \, \bigl\{ I \bigl( \mu_a^2 , \mu_{A^x}^2 \bigr) \bigr\}^{3/2} \,;
\label{eq:vecscadec}\end{equation}
and for decays to fermions,
\begin{multline}
  \Gamma_{Fij} (k_x^2) = \frac{1}{3} \, \alpha_x \, \abs{N_{F,ij}}^2 \bigl(1 - \tfrac{1}{2} \delta_{ij} \bigr) \, \sqrt{k_x^2} \, \sqrt{I \bigl( \mu_i^2 , \mu_j^2 \bigr)} \\
  \times \biggl\{ 1 - \frac{1}{2} \bigl( \mu_i^2 + \mu_j^2 \bigr) - \frac{1}{2} \bigl( \mu_i^2 - \mu_j^2 \bigr)^2 - 3 \, \mu_i \, \mu_j \, \frac{\Re \, N_{F,ij}^2}{\abs{N_{F,ij}}^2} \biggr\} \,.
\label{eq:vecferdec}\end{multline}
The expressions for the actual decay widths derive from \twoeqref{eq:vecscadec}{eq:vecferdec} with the replacement $k_x^2 \to m_x^2$.  The triangle function is
\begin{equation}
  I (a, b) = 1 - 2a - 2b + a^2 + b^2 - 2ab \,;
\end{equation}
the kinematic factors are
\begin{align}
  \mu_x^2 & = \frac{m_x^2}{k_x^2} \,, &
  \mu_a^2 & = \frac{m_{h^x_a}^2}{k_x^2} \,, &
  \mu_{A^x}^2 & = \frac{m_{A^x}^2}{k_x^2} \,, &
  \mu_i^2 & = \frac{m_{\chi_i^x}^2}{k_x^2} \,;
\end{align}
and the mixing matrix functions are as defined in \modeqref{eq:couplings}.

\section{Hidden Sector-Visible Sector Scattering}\label{app:scatter}
In this appendix we provide relevant details on hidden sector-visible sector scattering.  We are interested in the $2\to 2$ processes shown in \figref{fig:scatter}, with the two sectors coupled through a $t$-channel $Z^x$.  The dotted lines represent one of a number of possible initial and final hidden states: $\chi^x_i \to \chi^x_j$ (with both $i=j$ and $i\neq j$); $A^x \to h^x_a$; and $h^x_a \to A^x$ or $Z^x$.  The visible sector particles can be either nucleons or electrons.

\begin{figure}
  \begin{center}
    \begin{fmffile}{xNxN}
      \begin{fmfgraph*}(150,40)
        \fmfstraight
        \fmfleft{i1,i2} \fmfright{o1,o2}
        \fmf{dbl_plain,label=$\overrightarrow{p_N}$,label.side=right}{i1,v1}
        \fmf{dbl_plain,label=$\overrightarrow{k_{N'}}$,label.side=right}{v1,o1}
        \fmfblob{.08w}{v1}
        \fmf{dots,label=$\vec{p}_i$,label.side=left}{i2,v2}
        \fmf{dots,label=$\vec{k}_j$,label.side=left}{v2,o2}
        \fmf{boson,label=$q_x\downarrow$,label.side=left,tension=0.4}{v1,v2}
        \fmffreeze
      \end{fmfgraph*}
    \end{fmffile}
  \end{center}
  \caption{Dominant scattering process between visible and hidden sectors.  The dotted lines represent various possible hidden sector states as discussed in the text, while the solid lines are either nucleons or electrons.}\label{fig:scatter}
\end{figure}

We define $Q^2 \equiv -q_x^2 > 0$.  Additionally, $m_i$ ($m_f$) are the initial (final) hidden sector masses; and $E_i$, $p_i$ the initial hidden sector energy and momentum.  The differential cross sections are
\begin{equation}
  \frac{d \sigma_{HS}}{d Q^2} = \frac{4 \pi \alpha \alpha_x \epsilon^2}{p_i^2} \, \frac{1}{(Q^2 + m_x^2)^2} \, \frac{1}{4m_N^2} \, \mathcal{I}_{HS}(Q^2) \,.
\end{equation}
The functions $\mathcal{I}_{HS}$ are, for fermion scattering:
\begin{equation}
  \begin{split}
    \mathcal{I}_{F, ij} & = \abs{N_{F, ij}}^2 \biggl[ \Bigl( 2 E_i m_N - \frac{1}{2} (Q^2 - m_i^2 + m_f^2) \Bigr)^2 \biggl( F_{1,N}^2 + \frac{Q^2}{4 m_N^2} \, F_{2,N}^2 \biggr) \\
    & \quad - \Bigl( m_i^2 Q^2 + \frac{1}{4} (Q^2 - m_i^2 + m_f^2)^2 \Bigr) (F_{1,N} + F_{2, N})^2 \biggr] \\
    & \quad - \bigl( \abs{N_{F, ij}}^2 Q^2 + \abs{N_{F, ij} m_i + N_{F, ij}^\ast m_f}^2 \bigr) \biggl[ -3 Q^2 (F_{1,N} + F_{2, N})^2 \\
    & \quad + (Q^2 + 4m_N^2) \biggl( F_{1,N}^2 + \frac{Q^2}{4 m_N^2} \, F_{2,N}^2 \biggr) \biggr] \,,
  \end{split}
\end{equation}
for a pseudoscalar initial or final state:
\begin{equation}
  \begin{split}
    \mathcal{I}_{S, a} & =  N_{S, a}^2 \biggl[ \Bigl( 2 E_i m_N - \frac{1}{2} (Q^2 - m_i^2 + m_f^2) \Bigr)^2 \biggl( F_{1,N}^2 + \frac{Q^2}{4 m_N^2} \, F_{2,N}^2 \biggr) \\
    & \quad - \Bigl( m_i^2 Q^2 + \frac{1}{4} (Q^2 - m_i^2 + m_f^2)^2 \Bigr) (F_{1,N} + F_{2, N})^2 \biggr] \,,
  \end{split}
\end{equation}
and for the $h_x^a \to Z^x$ channel:
\begin{multline}
    \mathcal{I}_{H, a}  = N_{H, a}^2 \biggl[ - \Bigl( m_i^2 Q^2 + \frac{1}{4} (Q^2 - m_i^2 + m_f^2)^2 + 3 m_x^2 Q^2 \bigr) (F_{1,N} + F_{2, N})^2 \\
    + \Bigl\{ \Bigl( 2 E_i m_N - \frac{1}{2} (Q^2 - m_i^2 + m_f^2) \Bigr)^2 - m_x^2 (Q^2 + 4m_N^2) \Bigr\} \biggl( F_{1,N}^2 + \frac{Q^2}{4 m_N^2} \, F_{2,N}^2 \biggr) \biggr] \,.
\end{multline}
In the limit $E_i m_N \gg m_i^2$, $m_f^2$, $m_N^2$ and $Q^2$, all these expressions have the same approximate limit:
\begin{equation}
  \mathcal{I}_{HS} \approx \abs{N_{HS}}^2 \, 4 E_i^2 m_N^2 F_{1,N}^2 (Q^2) \,.
\end{equation}
This approximation is required to derive \modeqref{eq:scatxsec}.

For the nucleon form factors, we take~\cite{Izaguirre:2013uxa}
\begin{align}
  F_{1, N} (Q^2) & \frac{q_N}{(1 + Q^2/m_N^2)^2} \,, & F_{2,N} (Q^2) & = \frac{\kappa_N}{(1 + Q^2/m_N^2)^2} \,,
\end{align}
with $q_{p, n} = 1, 0$ and $\kappa_{p, n} = 1.79, -1.9$.  For electron scattering, we can use the same expressions with $F_{1,e} = 1$, $F_{2,e} = 0$ and consistently replacing $m_N \to m_e$.


\bibliography{CHARM1,CHARM2}{}

\providecommand{\href}[2]{#2}\begingroup\raggedright\begin{thebibliography}{100}

\bibitem{Essig:2013lka}
R.~Essig, J.~A. Jaros, W.~Wester, P.~H. Adrian, S.~Andreas, {\em et.~al.}, {\it
  {Dark Sectors and New, Light, Weakly-Coupled Particles}},
  \href{http://xxx.lanl.gov/abs/1311.0029}{{\tt arXiv:1311.0029}}.

\bibitem{Jaeckel:2010ni}
J.~Jaeckel and A.~Ringwald, {\it {The Low-Energy Frontier of Particle
  Physics}},  {\em Ann.Rev.Nucl.Part.Sci.} {\bf 60} (2010) 405--437,
  [\href{http://xxx.lanl.gov/abs/1002.0329}{{\tt arXiv:1002.0329}}].

\bibitem{Hewett:2012ns}
J.~Hewett, H.~Weerts, R.~Brock, J.~Butler, B.~Casey, {\em et.~al.}, {\it
  {Fundamental Physics at the Intensity Frontier}},
  \href{http://xxx.lanl.gov/abs/1205.2671}{{\tt arXiv:1205.2671}}.

\bibitem{Holdom:1985ag}
B.~Holdom, {\it {Two U(1)'s and Epsilon Charge Shifts}},  {\em Phys.Lett.} {\bf
  B166} (1986) 196.

\bibitem{Foot:1991kb}
R.~Foot and X.-G. He, {\it {Comment on Z Z-prime mixing in extended gauge
  theories}},  {\em Phys.Lett.} {\bf B267} (1991) 509--512.

\bibitem{Davidson:2000hf}
S.~Davidson, S.~Hannestad, and G.~Raffelt, {\it {Updated bounds on millicharged
  particles}},  {\em JHEP} {\bf 0005} (2000) 003,
  [\href{http://xxx.lanl.gov/abs/hep-ph/0001179}{{\tt hep-ph/0001179}}].

\bibitem{Arias:2012az}
P.~Arias, D.~Cadamuro, M.~Goodsell, J.~Jaeckel, J.~Redondo, {\em et.~al.}, {\it
  {WISPy Cold Dark Matter}},  {\em JCAP} {\bf 1206} (2012) 013,
  [\href{http://xxx.lanl.gov/abs/1201.5902}{{\tt arXiv:1201.5902}}].

\bibitem{ArkaniHamed:2008qn}
N.~Arkani-Hamed, D.~P. Finkbeiner, T.~R. Slatyer, and N.~Weiner, {\it {A Theory
  of Dark Matter}},  {\em Phys.Rev.} {\bf D79} (2009) 015014,
  [\href{http://xxx.lanl.gov/abs/0810.0713}{{\tt arXiv:0810.0713}}].

\bibitem{ArkaniHamed:2008qp}
N.~Arkani-Hamed and N.~Weiner, {\it {LHC Signals for a SuperUnified Theory of
  Dark Matter}},  {\em JHEP} {\bf 0812} (2008) 104,
  [\href{http://xxx.lanl.gov/abs/0810.0714}{{\tt arXiv:0810.0714}}].

\bibitem{Cheung:2009qd}
C.~Cheung, J.~T. Ruderman, L.-T. Wang, and I.~Yavin, {\it {Kinetic Mixing as
  the Origin of Light Dark Scales}},  {\em Phys.Rev.} {\bf D80} (2009) 035008,
  [\href{http://xxx.lanl.gov/abs/0902.3246}{{\tt arXiv:0902.3246}}].

\bibitem{Katz:2009qq}
A.~Katz and R.~Sundrum, {\it {Breaking the Dark Force}},  {\em JHEP} {\bf 0906}
  (2009) 003, [\href{http://xxx.lanl.gov/abs/0902.3271}{{\tt
  arXiv:0902.3271}}].

\bibitem{Morrissey:2009ur}
D.~E. Morrissey, D.~Poland, and K.~M. Zurek, {\it {Abelian Hidden Sectors at a
  GeV}},  {\em JHEP} {\bf 0907} (2009) 050,
  [\href{http://xxx.lanl.gov/abs/0904.2567}{{\tt arXiv:0904.2567}}].

\bibitem{Goodsell:2009xc}
M.~Goodsell, J.~Jaeckel, J.~Redondo, and A.~Ringwald, {\it {Naturally Light
  Hidden Photons in LARGE Volume String Compactifications}},  {\em JHEP} {\bf
  0911} (2009) 027, [\href{http://xxx.lanl.gov/abs/0909.0515}{{\tt
  arXiv:0909.0515}}].

\bibitem{Essig:2010ye}
R.~Essig, J.~Kaplan, P.~Schuster, and N.~Toro, {\it {On the Origin of Light
  Dark Matter Species}},  {\em Submitted to Physical Review D} (2010)
  [\href{http://xxx.lanl.gov/abs/1004.0691}{{\tt arXiv:1004.0691}}].

\bibitem{Andreas:2011in}
S.~Andreas, M.~Goodsell, and A.~Ringwald, {\it {Dark Matter and Dark Forces
  from a Supersymmetric Hidden Sector}},  {\em Phys.Rev.} {\bf D87} (2013)
  025007, [\href{http://xxx.lanl.gov/abs/1109.2869}{{\tt arXiv:1109.2869}}].

\bibitem{Pospelov:2008zw}
M.~Pospelov, {\it {Secluded U(1) below the weak scale}},  {\em Phys.Rev.} {\bf
  D80} (2009) 095002, [\href{http://xxx.lanl.gov/abs/0811.1030}{{\tt
  arXiv:0811.1030}}].

\bibitem{Batell:2009yf}
B.~Batell, M.~Pospelov, and A.~Ritz, {\it {Probing a Secluded U(1) at
  B-factories}},  {\em Phys.Rev.} {\bf D79} (2009) 115008,
  [\href{http://xxx.lanl.gov/abs/0903.0363}{{\tt arXiv:0903.0363}}].

\bibitem{Reece:2009un}
M.~Reece and L.-T. Wang, {\it {Searching for the light dark gauge boson in
  GeV-scale experiments}},  {\em JHEP} {\bf 0907} (2009) 051,
  [\href{http://xxx.lanl.gov/abs/0904.1743}{{\tt arXiv:0904.1743}}].

\bibitem{Bjorken:2009mm}
J.~D. Bjorken, R.~Essig, P.~Schuster, and N.~Toro, {\it {New Fixed-Target
  Experiments to Search for Dark Gauge Forces}},  {\em Phys.Rev.} {\bf D80}
  (2009) 075018, [\href{http://xxx.lanl.gov/abs/0906.0580}{{\tt
  arXiv:0906.0580}}].

\bibitem{Batell:2009di}
B.~Batell, M.~Pospelov, and A.~Ritz, {\it {Exploring Portals to a Hidden Sector
  Through Fixed Targets}},  {\em Phys.Rev.} {\bf D80} (2009) 095024,
  [\href{http://xxx.lanl.gov/abs/0906.5614}{{\tt arXiv:0906.5614}}].

\bibitem{deNiverville:2011it}
P.~deNiverville, M.~Pospelov, and A.~Ritz, {\it {Observing a light dark matter
  beam with neutrino experiments}},  {\em Phys.Rev.} {\bf D84} (2011) 075020,
  [\href{http://xxx.lanl.gov/abs/1107.4580}{{\tt arXiv:1107.4580}}].

\bibitem{Izaguirre:2013uxa}
E.~Izaguirre, G.~Krnjaic, P.~Schuster, and N.~Toro, {\it {New Electron
  Beam-Dump Experiments to Search for MeV to few-GeV Dark Matter}},
  \href{http://xxx.lanl.gov/abs/1307.6554}{{\tt arXiv:1307.6554}}.

\bibitem{Diamond:2013oda}
M.~D. Diamond and P.~Schuster, {\it {Searching for Light Dark Matter with the
  SLAC Millicharge Experiment}},  {\em Phys.Rev.Lett.} {\bf 111} (2013) 221803,
  [\href{http://xxx.lanl.gov/abs/1307.6861}{{\tt arXiv:1307.6861}}].

\bibitem{Essig:2013vha}
R.~Essig, J.~Mardon, M.~Papucci, T.~Volansky, and Y.-M. Zhong, {\it
  {Constraining Light Dark Matter with Low-Energy $e^+e^-$ Colliders}},  {\em
  JHEP} {\bf 1311} (2013) 167, [\href{http://xxx.lanl.gov/abs/1309.5084}{{\tt
  arXiv:1309.5084}}].

\bibitem{Schuster:2009au}
P.~Schuster, N.~Toro, and I.~Yavin, {\it {Terrestrial and Solar Limits on
  Long-Lived Particles in a Dark Sector}},  {\em Phys.Rev.} {\bf D81} (2010)
  016002, [\href{http://xxx.lanl.gov/abs/0910.1602}{{\tt arXiv:0910.1602}}].

\bibitem{Chan:2011aa}
Y.~F. Chan, M.~Low, D.~E. Morrissey, and A.~P. Spray, {\it {LHC Signatures of a
  Minimal Supersymmetric Hidden Valley}},  {\em JHEP} {\bf 1205} (2012) 155,
  [\href{http://xxx.lanl.gov/abs/1112.2705}{{\tt arXiv:1112.2705}}].

\bibitem{Martin:1997ns}
S.~P. Martin, {\it {A Supersymmetry primer}},
  \href{http://xxx.lanl.gov/abs/hep-ph/9709356}{{\tt hep-ph/9709356}}.

\bibitem{Chang:2006fp}
W.-F. Chang, J.~N. Ng, and J.~M. Wu, {\it {A Very Narrow Shadow Extra Z-boson
  at Colliders}},  {\em Phys.Rev.} {\bf D74} (2006) 095005,
  [\href{http://xxx.lanl.gov/abs/hep-ph/0608068}{{\tt hep-ph/0608068}}].

\bibitem{Chun:2010ve}
E.~J. Chun, J.-C. Park, and S.~Scopel, {\it {Dark matter and a new gauge boson
  through kinetic mixing}},  {\em JHEP} {\bf 1102} (2011) 100,
  [\href{http://xxx.lanl.gov/abs/1011.3300}{{\tt arXiv:1011.3300}}].

\bibitem{Hook:2010tw}
A.~Hook, E.~Izaguirre, and J.~G. Wacker, {\it {Model Independent Bounds on
  Kinetic Mixing}},  {\em Adv.High Energy Phys.} {\bf 2011} (2011) 859762,
  [\href{http://xxx.lanl.gov/abs/1006.0973}{{\tt arXiv:1006.0973}}].

\bibitem{Beringer:1900zz}
{\bf Particle Data Group} Collaboration, J.~Beringer {\em et.~al.}, {\it
  {Review of Particle Physics (RPP)}},  {\em Phys.Rev.} {\bf D86} (2012)
  010001.

\bibitem{Davoudiasl:2012ig}
H.~Davoudiasl, H.-S. Lee, and W.~J. Marciano, {\it {Dark Side of Higgs Diphoton
  Decays and Muon g-2}},  {\em Phys.Rev.} {\bf D86} (2012) 095009,
  [\href{http://xxx.lanl.gov/abs/1208.2973}{{\tt arXiv:1208.2973}}].

\bibitem{Giudice:2012ms}
G.~Giudice, P.~Paradisi, and M.~Passera, {\it {Testing new physics with the
  electron g-2}},  {\em JHEP} {\bf 1211} (2012) 113,
  [\href{http://xxx.lanl.gov/abs/1208.6583}{{\tt arXiv:1208.6583}}].

\bibitem{Endo:2012hp}
M.~Endo, K.~Hamaguchi, and G.~Mishima, {\it {Constraints on Hidden Photon
  Models from Electron g-2 and Hydrogen Spectroscopy}},  {\em Phys.Rev.} {\bf
  D86} (2012) 095029, [\href{http://xxx.lanl.gov/abs/1209.2558}{{\tt
  arXiv:1209.2558}}].

\bibitem{Hanneke:2008tm}
D.~Hanneke, S.~Fogwell, and G.~Gabrielse, {\it {New Measurement of the Electron
  Magnetic Moment and the Fine Structure Constant}},  {\em Phys.Rev.Lett.} {\bf
  100} (2008) 120801, [\href{http://xxx.lanl.gov/abs/0801.1134}{{\tt
  arXiv:0801.1134}}].

\bibitem{Aoyama:2012wj}
T.~Aoyama, M.~Hayakawa, T.~Kinoshita, and M.~Nio, {\it {Tenth-Order QED
  Contribution to the Electron g-2 and an Improved Value of the Fine Structure
  Constant}},  {\em Phys.Rev.Lett.} {\bf 109} (2012) 111807,
  [\href{http://xxx.lanl.gov/abs/1205.5368}{{\tt arXiv:1205.5368}}].

\bibitem{Bouchendira:2010es}
R.~Bouchendira, P.~Clade, S.~Guellati-Khelifa, F.~Nez, and F.~Biraben, {\it
  {New determination of the fine structure constant and test of the quantum
  electrodynamics}},  {\em Phys.Rev.Lett.} {\bf 106} (2011) 080801,
  [\href{http://xxx.lanl.gov/abs/1012.3627}{{\tt arXiv:1012.3627}}].

\bibitem{Aubert:2009au}
{\bf BaBar Collaboration} Collaboration, B.~Aubert {\em et.~al.}, {\it {Search
  for Dimuon Decays of a Light Scalar in Radiative Transitions tau(3S) --->
  gamma A0}},  \href{http://xxx.lanl.gov/abs/0902.2176}{{\tt arXiv:0902.2176}}.

\bibitem{Aubert:2009cp}
{\bf BaBar Collaboration} Collaboration, B.~Aubert {\em et.~al.}, {\it {Search
  for Dimuon Decays of a Light Scalar Boson in Radiative Transitions Upsilon
  ---> gamma A0}},  {\em Phys.Rev.Lett.} {\bf 103} (2009) 081803,
  [\href{http://xxx.lanl.gov/abs/0905.4539}{{\tt arXiv:0905.4539}}].

\bibitem{Borodatchenkova:2005ct}
N.~Borodatchenkova, D.~Choudhury, and M.~Drees, {\it {Probing MeV dark matter
  at low-energy e+e- colliders}},  {\em Phys.Rev.Lett.} {\bf 96} (2006) 141802,
  [\href{http://xxx.lanl.gov/abs/hep-ph/0510147}{{\tt hep-ph/0510147}}].

\bibitem{Archilli:2011zc}
F.~Archilli, D.~Babusci, D.~Badoni, I.~Balwierz, G.~Bencivenni, {\em et.~al.},
  {\it {Search for a vector gauge boson in phi meson decays with the KLOE
  detector}},  {\em Phys.Lett.} {\bf B706} (2012) 251--255,
  [\href{http://xxx.lanl.gov/abs/1110.0411}{{\tt arXiv:1110.0411}}].

\bibitem{Babusci:2012cr}
{\bf KLOE-2 Collaboration} Collaboration, D.~Babusci {\em et.~al.}, {\it {Limit
  on the production of a light vector gauge boson in phi meson decays with the
  KLOE detector}},  {\em Phys.Lett.} {\bf B720} (2013) 111--115,
  [\href{http://xxx.lanl.gov/abs/1210.3927}{{\tt arXiv:1210.3927}}].

\bibitem{Adlarson:2013eza}
{\bf WASA-at-COSY Collaboration} Collaboration, P.~Adlarson {\em et.~al.}, {\it
  {Search for a dark photon in the $\pi^0 \to e^+e^-\gamma$ decay}},  {\em
  Phys.Lett.} {\bf B726} (2013) 187--193,
  [\href{http://xxx.lanl.gov/abs/1304.0671}{{\tt arXiv:1304.0671}}].

\bibitem{Batell:2009jf}
B.~Batell, M.~Pospelov, and A.~Ritz, {\it {Multi-lepton Signatures of a Hidden
  Sector in Rare B Decays}},  {\em Phys.Rev.} {\bf D83} (2011) 054005,
  [\href{http://xxx.lanl.gov/abs/0911.4938}{{\tt arXiv:0911.4938}}].

\bibitem{Clarke:2013aya}
J.~D. Clarke, R.~Foot, and R.~R. Volkas, {\it {Phenomenology of a very light
  scalar (100 MeV $<m_h<$ 10 GeV) mixing with the SM Higgs}},
  \href{http://xxx.lanl.gov/abs/1310.8042}{{\tt arXiv:1310.8042}}.

\bibitem{Aditya:2012ay}
Y.~Aditya, K.~J. Healey, and A.~A. Petrov, {\it {Searching for super-WIMPs in
  leptonic heavy meson decays}},  {\em Phys.Lett.} {\bf B710} (2012) 118--124,
  [\href{http://xxx.lanl.gov/abs/1201.1007}{{\tt arXiv:1201.1007}}].

\bibitem{Aubert:2008as}
{\bf BaBar Collaboration} Collaboration, B.~Aubert {\em et.~al.}, {\it {Search
  for Invisible Decays of a Light Scalar in Radiative Transitions
  $\upsilon_{3S} \to \gamma$ A0}},
  \href{http://xxx.lanl.gov/abs/0808.0017}{{\tt arXiv:0808.0017}}.

\bibitem{Adler:2004hp}
{\bf E787 Collaboration} Collaboration, S.~Adler {\em et.~al.}, {\it {Further
  search for the decay K+ --->; pi+ nu anti-nu in the momentum region P <
  195-MeV/c}},  {\em Phys.Rev.} {\bf D70} (2004) 037102,
  [\href{http://xxx.lanl.gov/abs/hep-ex/0403034}{{\tt hep-ex/0403034}}].

\bibitem{Artamonov:2009sz}
{\bf BNL-E949 Collaboration} Collaboration, A.~Artamonov {\em et.~al.}, {\it
  {Study of the decay K+ ---> pi+ nu anti-nu in the momentum region 140 < P(pi)
  < 199-MeV/c}},  {\em Phys.Rev.} {\bf D79} (2009) 092004,
  [\href{http://xxx.lanl.gov/abs/0903.0030}{{\tt arXiv:0903.0030}}].

\bibitem{Harrison:1998yr}
{\bf BaBar Collaboration} Collaboration, P.~Harrison and H.~R. Quinn, {\it {The
  BABAR physics book: Physics at an asymmetric $B$ factory}}, .

\bibitem{Essig:2009nc}
R.~Essig, P.~Schuster, and N.~Toro, {\it {Probing Dark Forces and Light Hidden
  Sectors at Low-Energy e+e- Colliders}},  {\em Phys.Rev.} {\bf D80} (2009)
  015003, [\href{http://xxx.lanl.gov/abs/0903.3941}{{\tt arXiv:0903.3941}}].

\bibitem{Aubert:2009af}
{\bf BaBar Collaboration} Collaboration, B.~Aubert {\em et.~al.}, {\it {Search
  for a Narrow Resonance in e+e- to Four Lepton Final States}},
  \href{http://xxx.lanl.gov/abs/0908.2821}{{\tt arXiv:0908.2821}}.

\bibitem{Lees:2012ra}
{\bf BaBar Collaboration} Collaboration, J.~Lees {\em et.~al.}, {\it {Search
  for Low-Mass Dark-Sector Higgs Bosons}},  {\em Phys.Rev.Lett.} {\bf 108}
  (2012) 211801, [\href{http://xxx.lanl.gov/abs/1202.1313}{{\tt
  arXiv:1202.1313}}].

\bibitem{Pospelov:2007mp}
M.~Pospelov, A.~Ritz, and M.~B. Voloshin, {\it {Secluded WIMP Dark Matter}},
  {\em Phys.Lett.} {\bf B662} (2008) 53--61,
  [\href{http://xxx.lanl.gov/abs/0711.4866}{{\tt arXiv:0711.4866}}].

\bibitem{Pospelov:2008jd}
M.~Pospelov and A.~Ritz, {\it {Astrophysical Signatures of Secluded Dark
  Matter}},  {\em Phys.Lett.} {\bf B671} (2009) 391--397,
  [\href{http://xxx.lanl.gov/abs/0810.1502}{{\tt arXiv:0810.1502}}].

\bibitem{Chang:2009yt}
S.~Chang, A.~Pierce, and N.~Weiner, {\it {Momentum Dependent Dark Matter
  Scattering}},  {\em JCAP} {\bf 1001} (2010) 006,
  [\href{http://xxx.lanl.gov/abs/0908.3192}{{\tt arXiv:0908.3192}}].

\bibitem{Cushman:2013zza}
P.~Cushman, C.~Galbiati, D.~McKinsey, H.~Robertson, T.~Tait, {\em et.~al.},
  {\it {Snowmass CF1 Summary: WIMP Dark Matter Direct Detection}},
  \href{http://xxx.lanl.gov/abs/1310.8327}{{\tt arXiv:1310.8327}}.

\bibitem{Essig:2011nj}
R.~Essig, J.~Mardon, and T.~Volansky, {\it {Direct Detection of Sub-GeV Dark
  Matter}},  {\em Phys.Rev.} {\bf D85} (2012) 076007,
  [\href{http://xxx.lanl.gov/abs/1108.5383}{{\tt arXiv:1108.5383}}].

\bibitem{Essig:2012yx}
R.~Essig, A.~Manalaysay, J.~Mardon, P.~Sorensen, and T.~Volansky, {\it {First
  Direct Detection Limits on sub-GeV Dark Matter from XENON10}},  {\em
  Phys.Rev.Lett.} {\bf 109} (2012) 021301,
  [\href{http://xxx.lanl.gov/abs/1206.2644}{{\tt arXiv:1206.2644}}].

\bibitem{Kawasaki:2004qu}
M.~Kawasaki, K.~Kohri, and T.~Moroi, {\it {Big-Bang nucleosynthesis and
  hadronic decay of long-lived massive particles}},  {\em Phys.Rev.} {\bf D71}
  (2005) 083502, [\href{http://xxx.lanl.gov/abs/astro-ph/0408426}{{\tt
  astro-ph/0408426}}].

\bibitem{Jedamzik:2006xz}
K.~Jedamzik, {\it {Big bang nucleosynthesis constraints on hadronically and
  electromagnetically decaying relic neutral particles}},  {\em Phys.Rev.} {\bf
  D74} (2006) 103509, [\href{http://xxx.lanl.gov/abs/hep-ph/0604251}{{\tt
  hep-ph/0604251}}].

\bibitem{Hu:1993gc}
W.~Hu and J.~Silk, {\it {Thermalization constraints and spectral distortions
  for massive unstable relic particles}},  {\em Phys.Rev.Lett.} {\bf 70} (1993)
  2661--2664.

\bibitem{Essig:2013goa}
R.~Essig, E.~Kuflik, S.~D. Mcdermott, T.~Volansky, and K.~M. Zurek, {\it
  {Constraining Light Dark Matter with Diffuse X-Ray and Gamma-Ray
  Observations}},  {\em JHEP} {\bf 1311} (2013) 193,
  [\href{http://xxx.lanl.gov/abs/1309.4091}{{\tt arXiv:1309.4091}}].

\bibitem{Pospelov:2010cw}
M.~Pospelov and J.~Pradler, {\it {Metastable GeV-scale particles as a solution
  to the cosmological lithium problem}},  {\em Phys.Rev.} {\bf D82} (2010)
  103514, [\href{http://xxx.lanl.gov/abs/1006.4172}{{\tt arXiv:1006.4172}}].

\bibitem{Dent:2012mx}
J.~B. Dent, F.~Ferrer, and L.~M. Krauss, {\it {Constraints on Light Hidden
  Sector Gauge Bosons from Supernova Cooling}},
  \href{http://xxx.lanl.gov/abs/1201.2683}{{\tt arXiv:1201.2683}}.

\bibitem{Dreiner:2013mua}
H.~K. Dreiner, J.-F. Fortin, C.~Hanhart, and L.~Ubaldi, {\it {Supernova
  Constraints on MeV Dark Sectors from e+ e- Annihilations}},
  \href{http://xxx.lanl.gov/abs/1310.3826}{{\tt arXiv:1310.3826}}.

\bibitem{Andreas:2012mt}
S.~Andreas, C.~Niebuhr, and A.~Ringwald, {\it {New Limits on Hidden Photons
  from Past Electron Beam Dumps}},  {\em Phys.Rev.} {\bf D86} (2012) 095019,
  [\href{http://xxx.lanl.gov/abs/1209.6083}{{\tt arXiv:1209.6083}}].

\bibitem{Merkel:2011ze}
{\bf A1 Collaboration} Collaboration, H.~Merkel {\em et.~al.}, {\it {Search for
  Light Gauge Bosons of the Dark Sector at the Mainz Microtron}},  {\em
  Phys.Rev.Lett.} {\bf 106} (2011) 251802,
  [\href{http://xxx.lanl.gov/abs/1101.4091}{{\tt arXiv:1101.4091}}].

\bibitem{Abrahamyan:2011gv}
{\bf APEX Collaboration} Collaboration, S.~Abrahamyan {\em et.~al.}, {\it
  {Search for a New Gauge Boson in Electron-Nucleus Fixed-Target Scattering by
  the APEX Experiment}},  {\em Phys.Rev.Lett.} {\bf 107} (2011) 191804,
  [\href{http://xxx.lanl.gov/abs/1108.2750}{{\tt arXiv:1108.2750}}].

\bibitem{Andreas:2012rm}
S.~Andreas, {\it {Hidden Photons in beam dump experiments and in connection
  with Dark Matter}},  {\em Frascati Phys.Ser.} {\bf 56} (2012) 23--32,
  [\href{http://xxx.lanl.gov/abs/1212.4520}{{\tt arXiv:1212.4520}}].

\bibitem{Essig:2010xa}
R.~Essig, P.~Schuster, N.~Toro, and B.~Wojtsekhowski, {\it {An Electron Fixed
  Target Experiment to Search for a New Vector Boson A' Decaying to e+e-}},
  {\em JHEP} {\bf 1102} (2011) 009,
  [\href{http://xxx.lanl.gov/abs/1001.2557}{{\tt arXiv:1001.2557}}].

\bibitem{Freytsis:2009bh}
M.~Freytsis, G.~Ovanesyan, and J.~Thaler, {\it {Dark Force Detection in Low
  Energy e-p Collisions}},  {\em JHEP} {\bf 1001} (2010) 111,
  [\href{http://xxx.lanl.gov/abs/0909.2862}{{\tt arXiv:0909.2862}}].

\bibitem{HPS}
{The Heavy Photon Search Collaboration (HPS),
  \texttt{https://confluence.slac.stanford.edu/display/hpsg}}.

\bibitem{2013arXiv1301.1103H}
P.~{Hansson Adrian}, {\it {The Heavy Photon Search Experiment}},  {\em ArXiv
  e-prints} (Jan., 2013) [\href{http://xxx.lanl.gov/abs/1301.1103}{{\tt
  arXiv:1301.1103}}].

\bibitem{Kim:1973he}
K.~J. Kim and Y.-S. Tsai, {\it {IMPROVED WEIZSACKER-WILLIAMS METHOD AND ITS
  APPLICATION TO LEPTON AND W BOSON PAIR PRODUCTION}},  {\em Phys.Rev.} {\bf
  D8} (1973) 3109.

\bibitem{Beranek:2013nqa}
T.~Beranek and M.~Vanderhaeghen, {\it {Study of the discovery potential for
  hidden photon emission at future electron scattering fixed target
  experiments}},  \href{http://xxx.lanl.gov/abs/1311.5104}{{\tt
  arXiv:1311.5104}}.

\bibitem{Riordan:1987aw}
E.~Riordan, M.~Krasny, K.~Lang, P.~De~Barbaro, A.~Bodek, {\em et.~al.}, {\it {A
  Search for Short Lived Axions in an Electron Beam Dump Experiment}},  {\em
  Phys.Rev.Lett.} {\bf 59} (1987) 755.

\bibitem{Tsai:1986tx}
Y.-S. Tsai, {\it {AXION BREMSSTRAHLUNG BY AN ELECTRON BEAM}},  {\em Phys.Rev.}
  {\bf D34} (1986) 1326.

\bibitem{Bjorken:1988as}
J.~Bjorken, S.~Ecklund, W.~Nelson, A.~Abashian, C.~Church, {\em et.~al.}, {\it
  {Search for Neutral Metastable Penetrating Particles Produced in the SLAC
  Beam Dump}},  {\em Phys.Rev.} {\bf D38} (1988) 3375.

\bibitem{Bross:1989mp}
A.~Bross, M.~Crisler, S.~H. Pordes, J.~Volk, S.~Errede, {\em et.~al.}, {\it {A
  Search for Shortlived Particles Produced in an Electron Beam Dump}},  {\em
  Phys.Rev.Lett.} {\bf 67} (1991) 2942--2945.

\bibitem{Konaka:1986cb}
A.~Konaka, K.~Imai, H.~Kobayashi, A.~Masaike, K.~Miyake, {\em et.~al.}, {\it
  {Search for Neutral Particles in Electron Beam Dump Experiment}},  {\em
  Phys.Rev.Lett.} {\bf 57} (1986) 659.

\bibitem{Davier:1989wz}
M.~Davier and H.~Nguyen~Ngoc, {\it {An Unambiguous Search for a Light Higgs
  Boson}},  {\em Phys.Lett.} {\bf B229} (1989) 150.

\bibitem{Gninenko:2013rka}
S.~Gninenko, {\it {Search for MeV dark photons in a light-shining-through-walls
  experiment at CERN}},  \href{http://xxx.lanl.gov/abs/1308.6521}{{\tt
  arXiv:1308.6521}}.

\bibitem{Andreas:2013lya}
S.~Andreas, S.~Donskov, P.~Crivelli, A.~Gardikiotis, S.~Gninenko, {\em
  et.~al.}, {\it {Proposal for an Experiment to Search for Light Dark Matter at
  the SPS}},  \href{http://xxx.lanl.gov/abs/1312.3309}{{\tt arXiv:1312.3309}}.

\bibitem{Odom:2006zz}
B.~C. Odom, D.~Hanneke, B.~D'Urso, and G.~Gabrielse, {\it {New Measurement of
  the Electron Magnetic Moment Using a One-Electron Quantum Cyclotron}},  {\em
  Phys.Rev.Lett.} {\bf 97} (2006) 030801.

\bibitem{Gninenko:2012eq}
S.~Gninenko, {\it {Constraints on sub-GeV hidden sector gauge bosons from a
  search for heavy neutrino decays}},  {\em Phys.Lett.} {\bf B713} (2012)
  244--248, [\href{http://xxx.lanl.gov/abs/1204.3583}{{\tt arXiv:1204.3583}}].

\bibitem{deNiverville:2012ij}
P.~deNiverville, D.~McKeen, and A.~Ritz, {\it {Signatures of sub-GeV dark
  matter beams at neutrino experiments}},  {\em Phys.Rev.} {\bf D86} (2012)
  035022, [\href{http://xxx.lanl.gov/abs/1205.3499}{{\tt arXiv:1205.3499}}].

\bibitem{Blumlein:2013cua}
J.~Bluemlein and J.~Brunner, {\it {New Exclusion Limits on Dark Gauge Forces
  from Proton Bremsstrahlung in Beam-Dump Data}},
  \href{http://xxx.lanl.gov/abs/1311.3870}{{\tt arXiv:1311.3870}}.

\bibitem{Blumlein:2011mv}
J.~Blumlein and J.~Brunner, {\it {New Exclusion Limits for Dark Gauge Forces
  from Beam-Dump Data}},  {\em Phys.Lett.} {\bf B701} (2011) 155--159,
  [\href{http://xxx.lanl.gov/abs/1104.2747}{{\tt arXiv:1104.2747}}].

\bibitem{Gninenko:2011uv}
S.~Gninenko, {\it {Stringent limits on the $\pi^0 \to \gamma X, X \to e+e-$
  decay from neutrino experiments and constraints on new light gauge bosons}},
  {\em Phys.Rev.} {\bf D85} (2012) 055027,
  [\href{http://xxx.lanl.gov/abs/1112.5438}{{\tt arXiv:1112.5438}}].

\bibitem{Lai:2010vv}
H.-L. Lai, M.~Guzzi, J.~Huston, Z.~Li, P.~M. Nadolsky, {\em et.~al.}, {\it {New
  parton distributions for collider physics}},  {\em Phys.Rev.} {\bf D82}
  (2010) 074024, [\href{http://xxx.lanl.gov/abs/1007.2241}{{\tt
  arXiv:1007.2241}}].

\bibitem{Burman:1989ds}
R.~Burman and E.~Smith, {\it {PARAMETRIZATION OF PION PRODUCTION AND REACTION
  CROSS-SECTIONS AT LAMPF ENERGIES}}, .

\bibitem{AguilarArevalo:2008yp}
{\bf MiniBooNE Collaboration} Collaboration, A.~Aguilar-Arevalo {\em et.~al.},
  {\it {The Neutrino Flux prediction at MiniBooNE}},  {\em Phys.Rev.} {\bf D79}
  (2009) 072002, [\href{http://xxx.lanl.gov/abs/0806.1449}{{\tt
  arXiv:0806.1449}}].

\bibitem{Tel-Zur:1996gua}
G.~Tel-Zur, {\it {Electron pair production in p-Be and p-Au collisions at 450
  GeV/c}}, .

\bibitem{Bourquin:1975fx}
M.~Bourquin and J.-M. Gaillard, {\it {Vector Meson and psi Contributions to
  Single Lepton Spectra}},  {\em Phys.Lett.} {\bf B59} (1975) 191.

\bibitem{Bourquin:1976fe}
M.~Bourquin and J.-M. Gaillard, {\it {A Simple Phenomenological Description of
  Hadron Production}},  {\em Nucl.Phys.} {\bf B114} (1976) 334.

\bibitem{Alloul:2013bka}
A.~Alloul, N.~D. Christensen, C.~Degrande, C.~Duhr, and B.~Fuks, {\it
  {FeynRules 2.0 - A complete toolbox for tree-level phenomenology}},
  \href{http://xxx.lanl.gov/abs/1310.1921}{{\tt arXiv:1310.1921}}.

\bibitem{Duhr:2011se}
C.~Duhr and B.~Fuks, {\it {A superspace module for the FeynRules package}},
  {\em Comput.Phys.Commun.} {\bf 182} (2011) 2404--2426,
  [\href{http://xxx.lanl.gov/abs/1102.4191}{{\tt arXiv:1102.4191}}].

\bibitem{Degrande:2011ua}
C.~Degrande, C.~Duhr, B.~Fuks, D.~Grellscheid, O.~Mattelaer, {\em et.~al.},
  {\it {UFO - The Universal FeynRules Output}},  {\em Comput.Phys.Commun.} {\bf
  183} (2012) 1201--1214, [\href{http://xxx.lanl.gov/abs/1108.2040}{{\tt
  arXiv:1108.2040}}].

\bibitem{Alwall:2011uj}
J.~Alwall, M.~Herquet, F.~Maltoni, O.~Mattelaer, and T.~Stelzer, {\it {MadGraph
  5 : Going Beyond}},  {\em JHEP} {\bf 1106} (2011) 128,
  [\href{http://xxx.lanl.gov/abs/1106.0522}{{\tt arXiv:1106.0522}}].

\bibitem{Bergsma:1985qz}
{\bf CHARM Collaboration} Collaboration, F.~Bergsma {\em et.~al.}, {\it {Search
  for Axion Like Particle Production in 400-{GeV} Proton - Copper
  Interactions}},  {\em Phys.Lett.} {\bf B157} (1985) 458.

\bibitem{Bergsma:1985is}
{\bf CHARM Collaboration} Collaboration, F.~Bergsma {\em et.~al.}, {\it {A
  Search for Decays of Heavy Neutrinos in the Mass Range 0.5-{GeV} to
  2.8-{GeV}}},  {\em Phys.Lett.} {\bf B166} (1986) 473.

\bibitem{Ambats:1998aa}
{\bf MINOS Collaboration} Collaboration, I.~Ambats {\em et.~al.}, {\it {The
  MINOS Detectors Technical Design Report}}, .

\bibitem{Adamson:2013whj}
{\bf MINOS Collaboration} Collaboration, P.~Adamson {\em et.~al.}, {\it
  {Measurement of Neutrino and Antineutrino Oscillations Using Beam and
  Atmospheric Data in MINOS}},  {\em Phys.Rev.Lett.} {\bf 110} (2013) 251801,
  [\href{http://xxx.lanl.gov/abs/1304.6335}{{\tt arXiv:1304.6335}}].

\bibitem{Blumlein:1990ay}
J.~Blumlein, J.~Brunner, H.~Grabosch, P.~Lanius, S.~Nowak, {\em et.~al.}, {\it
  {Limits on neutral light scalar and pseudoscalar particles in a proton beam
  dump experiment}},  {\em Z.Phys.} {\bf C51} (1991) 341--350.

\bibitem{Blumlein:1991xh}
J.~Blumlein, J.~Brunner, H.~Grabosch, P.~Lanius, S.~Nowak, {\em et.~al.}, {\it
  {Limits on the mass of light (pseudo)scalar particles from Bethe-Heitler e+
  e- and mu+ mu- pair production in a proton - iron beam dump experiment}},
  {\em Int.J.Mod.Phys.} {\bf A7} (1992) 3835--3850.

\bibitem{Abe:2011xv}
K.~Abe, N.~Abgrall, Y.~Ajima, H.~Aihara, J.~Albert, {\em et.~al.}, {\it
  {Measurements of the T2K neutrino beam properties using the INGRID on-axis
  near detector}},  {\em Nucl.Instrum.Meth.} {\bf A694} (2012) 211--223,
  [\href{http://xxx.lanl.gov/abs/1111.3119}{{\tt arXiv:1111.3119}}].

\bibitem{Aguilar:2001ty}
{\bf LSND Collaboration} Collaboration, A.~Aguilar-Arevalo {\em et.~al.}, {\it
  {Evidence for neutrino oscillations from the observation of
  anti-neutrino(electron) appearance in a anti-neutrino(muon) beam}},  {\em
  Phys.Rev.} {\bf D64} (2001) 112007,
  [\href{http://xxx.lanl.gov/abs/hep-ex/0104049}{{\tt hep-ex/0104049}}].

\bibitem{Auerbach:2001wg}
{\bf LSND Collaboration} Collaboration, L.~Auerbach {\em et.~al.}, {\it
  {Measurement of electron - neutrino - electron elastic scattering}},  {\em
  Phys.Rev.} {\bf D63} (2001) 112001,
  [\href{http://xxx.lanl.gov/abs/hep-ex/0101039}{{\tt hep-ex/0101039}}].

\bibitem{Teis:1996kx}
S.~Teis, W.~Cassing, M.~Effenberger, A.~Hombach, U.~Mosel, {\em et.~al.}, {\it
  {Pion production in heavy ion collisions at SIS energies}},  {\em Z.Phys.}
  {\bf A356} (1997) 421--435,
  [\href{http://xxx.lanl.gov/abs/nucl-th/9609009}{{\tt nucl-th/9609009}}].

\bibitem{Flam:1900ijf}
V.~Flaminio, W.~Moorhead, D.~Morrison, and N.~Rivoire, {\it {Compilation of
  cross-sections. 3. p and anti-p induced reactions}}, .

\bibitem{Sjostrand:2007gs}
T.~Sjostrand, S.~Mrenna, and P.~Z. Skands, {\it {A Brief Introduction to PYTHIA
  8.1}},  {\em Comput.Phys.Commun.} {\bf 178} (2008) 852--867,
  [\href{http://xxx.lanl.gov/abs/0710.3820}{{\tt arXiv:0710.3820}}].

\bibitem{Sjostrand:2006za}
T.~Sjostrand, S.~Mrenna, and P.~Z. Skands, {\it {PYTHIA 6.4 Physics and
  Manual}},  {\em JHEP} {\bf 0605} (2006) 026,
  [\href{http://xxx.lanl.gov/abs/hep-ph/0603175}{{\tt hep-ph/0603175}}].

\bibitem{AguilarBenitez:1991yy}
M.~Aguilar-Benitez, W.~Allison, A.~Batalov, E.~Castelli, P.~Cecchia, {\em
  et.~al.}, {\it {Inclusive particle production in 400-GeV/c p p
  interactions}},  {\em Z.Phys.} {\bf C50} (1991) 405--426.

\bibitem{Agakishiev:1998mw}
G.~Agakishiev, M.~Appenheimer, R.~Averbeck, F.~Ballester, R.~Baur, {\em
  et.~al.}, {\it {Neutral meson production in p Be and p Au collisions at
  450-GeV beam energy}},  {\em Eur.Phys.J.} {\bf C4} (1998) 249--257.

\bibitem{Abgrall:2011ae}
{\bf NA61/SHINE Collaboration} Collaboration, N.~Abgrall {\em et.~al.}, {\it
  {Measurements of Cross Sections and Charged Pion Spectra in Proton-Carbon
  Interactions at 31 GeV/c}},  {\em Phys.Rev.} {\bf C84} (2011) 034604,
  [\href{http://xxx.lanl.gov/abs/1102.0983}{{\tt arXiv:1102.0983}}].

\bibitem{Adamson:2010wi}
{\bf MINOS Collaboration} Collaboration, P.~Adamson {\em et.~al.}, {\it {Search
  for sterile neutrino mixing in the MINOS long baseline experiment}},  {\em
  Phys.Rev.} {\bf D81} (2010) 052004,
  [\href{http://xxx.lanl.gov/abs/1001.0336}{{\tt arXiv:1001.0336}}].

\bibitem{Boratav:1976wx}
{\bf French-Soviet Collaboration} Collaboration, M.~Boratav {\em et.~al.}, {\it
  {Gamma Production and Multiplicity Correlations Between Neutral and Charged
  Pions in p p Interactions at 69-GeV/c}},  {\em Nucl.Phys.} {\bf B111} (1976)
  529.

\bibitem{Blumenfeld:1973ww}
{\bf France-Soviet Union Collaboration} Collaboration, H.~Blumenfeld {\em
  et.~al.}, {\it {Photon production in 69-GeV p p interactions}},  {\em
  Phys.Lett.} {\bf B45} (1973) 525--527.

\bibitem{Abe:2011ks}
{\bf T2K Collaboration} Collaboration, K.~Abe {\em et.~al.}, {\it {The T2K
  Experiment}},  {\em Nucl.Instrum.Meth.} {\bf A659} (2011) 106--135,
  [\href{http://xxx.lanl.gov/abs/1106.1238}{{\tt arXiv:1106.1238}}].

\bibitem{oai:arXiv.org:hep-ex/0101041}
{\bf NOMAD Collaboration} Collaboration, P.~Astier {\em et.~al.}, {\it {Search
  for heavy neutrinos mixing with tau neutrinos}},  {\em Phys.Lett.} {\bf B506}
  (2001) 27--38, [\href{http://xxx.lanl.gov/abs/hep-ex/0101041}{{\tt
  hep-ex/0101041}}].

\bibitem{Bernardi:1985ny}
G.~Bernardi, G.~Carugno, J.~Chauveau, F.~Dicarlo, M.~Dris, {\em et.~al.}, {\it
  {Search for Neutrino Decay}},  {\em Phys.Lett.} {\bf B166} (1986) 479.

\bibitem{Bernardi:1987ek}
G.~Bernardi, G.~Carugno, J.~Chauveau, F.~Dicarlo, M.~Dris, {\em et.~al.}, {\it
  {FURTHER LIMITS ON HEAVY NEUTRINO COUPLINGS}},  {\em Phys.Lett.} {\bf B203}
  (1988) 332.

\bibitem{AguilarArevalo:2008qa}
{\bf MiniBooNE Collaboration} Collaboration, A.~Aguilar-Arevalo {\em et.~al.},
  {\it {The MiniBooNE Detector}},  {\em Nucl.Instrum.Meth.} {\bf A599} (2009)
  28--46, [\href{http://xxx.lanl.gov/abs/0806.4201}{{\tt arXiv:0806.4201}}].

\bibitem{Assylbekov:2011sh}
S.~Assylbekov, G.~Barr, B.~Berger, H.~Berns, D.~Beznosko, {\em et.~al.}, {\it
  {The T2K ND280 Off-Axis Pi-Zero Detector}},  {\em Nucl.Instrum.Meth.} {\bf
  A686} (2012) 48--63, [\href{http://xxx.lanl.gov/abs/1111.5030}{{\tt
  arXiv:1111.5030}}].

\bibitem{Kronfeld:2013uoa}
A.~S. Kronfeld, R.~S. Tschirhart, U.~Al-Binni, W.~Altmannshofer,
  C.~Ankenbrandt, {\em et.~al.}, {\it {Project X: Physics Opportunities}},
  \href{http://xxx.lanl.gov/abs/1306.5009}{{\tt arXiv:1306.5009}}.

\bibitem{Holmes:2013hfa}
S.~Holmes, S.~Nagaitsev, and R.~Tschirhart, {\it {Project X: A Flexible High
  Power Proton Facility}},  \href{http://xxx.lanl.gov/abs/1305.3809}{{\tt
  arXiv:1305.3809}}.

\bibitem{Brodsky:2012vg}
S.~Brodsky, F.~Fleuret, C.~Hadjidakis, and J.~Lansberg, {\it {Physics
  Opportunities of a Fixed-Target Experiment using the LHC Beams}},  {\em
  Phys.Rept.} {\bf 522} (2013) 239--255,
  [\href{http://xxx.lanl.gov/abs/1202.6585}{{\tt arXiv:1202.6585}}].

\bibitem{Lansberg:2012wj}
J.~Lansberg, V.~Chambert, J.~Didelez, B.~Genolini, C.~Hadjidakis, {\em
  et.~al.}, {\it {A Fixed-Target ExpeRiment at the LHC (AFTER@LHC) :
  luminosities, target polarisation and a selection of physics studies}},  {\em
  PoS} {\bf QNP2012} (2012) 049, [\href{http://xxx.lanl.gov/abs/1207.3507}{{\tt
  arXiv:1207.3507}}].

\bibitem{Strassler:2006qa}
M.~J. Strassler, {\it {Possible effects of a hidden valley on supersymmetric
  phenomenology}},  \href{http://xxx.lanl.gov/abs/hep-ph/0607160}{{\tt
  hep-ph/0607160}}.

\bibitem{Baumgart:2009tn}
M.~Baumgart, C.~Cheung, J.~T. Ruderman, L.-T. Wang, and I.~Yavin, {\it
  {Non-Abelian Dark Sectors and Their Collider Signatures}},  {\em JHEP} {\bf
  0904} (2009) 014, [\href{http://xxx.lanl.gov/abs/0901.0283}{{\tt
  arXiv:0901.0283}}].

\bibitem{Cheung:2009su}
C.~Cheung, J.~T. Ruderman, L.-T. Wang, and I.~Yavin, {\it {Lepton Jets in
  (Supersymmetric) Electroweak Processes}},  {\em JHEP} {\bf 1004} (2010) 116,
  [\href{http://xxx.lanl.gov/abs/0909.0290}{{\tt arXiv:0909.0290}}].

\bibitem{Beranek:2013yqa}
T.~Beranek, H.~Merkel, and M.~Vanderhaeghen, {\it {Theoretical framework to
  analyze searches for hidden light gauge bosons in electron scattering fixed
  target experiments}},  \href{http://xxx.lanl.gov/abs/1303.2540}{{\tt
  arXiv:1303.2540}}.

\bibitem{Andersen:2013rda}
J.~R. Andersen, M.~Rauch, and M.~Spannowsky, {\it {Dark Sector spectroscopy at
  the ILC}},  \href{http://xxx.lanl.gov/abs/1308.4588}{{\tt arXiv:1308.4588}}.

\bibitem{Kumar:2011nj}
A.~Kumar, D.~E. Morrissey, and A.~Spray, {\it {Kinetically-Enhanced Anomaly
  Mediation}},  {\em JHEP} {\bf 1112} (2011) 013,
  [\href{http://xxx.lanl.gov/abs/1109.1565}{{\tt arXiv:1109.1565}}].

\bibitem{Donoghue:1990xh}
J.~F. Donoghue, J.~Gasser, and H.~Leutwyler, {\it {The Decay of a Light Higgs
  Boson}},  {\em Nucl.Phys.} {\bf B343} (1990) 341--368.

\bibitem{Bezrukov:2009yw}
F.~Bezrukov and D.~Gorbunov, {\it {Light inflaton Hunter's Guide}},  {\em JHEP}
  {\bf 1005} (2010) 010, [\href{http://xxx.lanl.gov/abs/0912.0390}{{\tt
  arXiv:0912.0390}}].

\bibitem{Bezrukov:2013fca}
F.~Bezrukov and D.~Gorbunov, {\it {Light inflaton after LHC8 and WMAP9
  results}},  {\em JHEP} {\bf 1307} (2013) 140,
  [\href{http://xxx.lanl.gov/abs/1303.4395}{{\tt arXiv:1303.4395}}].

\bibitem{Tsai:1973py}
Y.-S. Tsai, {\it {Pair Production and Bremsstrahlung of Charged Leptons}},
  {\em Rev.Mod.Phys.} {\bf 46} (1974) 815.

\end{thebibliography}\endgroup
\bibliographystyle{JHEP}

\end{document}